\documentclass[useAMS,usenatbib,usegraphicx]{mn2e}
\usepackage{amssymb} 
\usepackage{times} 
\usepackage{aas_macros} %needed to define journal macros generated by 
\usepackage[dvipsnames,usenames]{color}

\usepackage{longtable}

\title[TESS Ap stars]{Rotation and pulsation in Ap stars: first light results from TESS sectors 1 and 2} 
\author[Cunha et al.]{M.~S.~Cunha,$^1$ V.~Antoci,$^2$ D.~L.~Holdsworth,$^3$ D. W. Kurtz,$^3$ 
 L.~A.~Balona,$^4$ \newauthor Zs.~Bogn\'ar,$^{5,6}$ D.~M.~Bowman,$^7$ Z.~Guo,$^{8,9}$ P.~A.~Ko\l{}aczek-Szyma\'nski,$^{10}$ \newauthor 
M.~Lares-Martiz,$^{11}$ E.~Paunzen,$^{12}$  M.~Skarka,$^{12,13}$ B.~Smalley,$^{14}$ \'A.~S\'odor,$^{5,6}$ \newauthor 
 O.~Kochukhov,$^{15}$ J. Pepper,$^{16}$ T. Richey-Yowell,$^{17}$ G.~R. Ricker,$^{18}$ S.~Seager,$^{18,19}$ \newauthor
 D.~L.~Buzasi,$^{20}$ L.~Fox-Machado,$^{21}$  A.~Hasanzadeh,$^{22}$  E. Niemczura,$^{10}$ \newauthor P.~ Quitral-Manosalva,$^{1,23}$  M.~J.~P.~F.~G.~Monteiro,$^{1,23}$ I.~Stateva,$^{24}$ \newauthor  P.~De Cat,$^{25}$ 
 A.~Garc\'{\i}a Hern\'andez,$^{26}$ H.~Ghasemi,$^{27}$ G.~Handler,$^{9}$
D. Hey,$^{2,28}$ \newauthor J.~M.~Matthews,$^{29}$ J.~M.~Nemec,$^{30}$ J.~Pascual-Granado,$^{11}$  H.~Safari,$^{22}$ \newauthor  J.~C.~Su\'arez,$^{11,26}$ R.~Szab\'o,$^{5,6}$  A.~Tkachenko$^{7}$ and W.~W.~Weiss$^{31}$ \\
\\ \\
\parbox{\textwidth}{
	\textit{The authors' affiliations are shown in Appendix\,\ref{sec:affiliations}. \\
}}
}
 
\begin{document} 
\maketitle 
\newpage 
\begin{abstract} 
We present the first results from the Transiting Exoplanet Survey Satellite (TESS) on the rotational and pulsational variability of magnetic chemically peculiar A-type stars. We analyse TESS 2-min cadence data from sectors 1 and 2 on a sample of 83 stars. {Five} new rapidly oscillating Ap (roAp) stars are announced. One of these pulsates with periods around 4.7\,min, making it the shortest period roAp star known to date. {Four out of the five new roAp stars are multiperiodic. Three of these, and the singly-periodic one show the presence of rotational mode splitting.} Individual frequencies are provided in all cases. In addition, seven previously known roAp stars are analysed. Additional modes of oscillation are found in some stars, while in others we are able to distinguish the true pulsations from possible aliases present in the ground-based data. We find that the
pulsation amplitude in the TESS filter is typically a factor 6 smaller than that in the $B$ filter which is usually used for ground-based observations. For four roAp stars we set constraints on the inclination angle and magnetic obliquity, through the application of the oblique pulsator model. We also confirm the absence of roAp-type pulsations down to amplitude limits of 6 and 13~$\umu$mag, respectively, in two of the best characterised non-oscillating Ap (noAp) stars. We announce {27} new rotational variables along with their rotation periods, and {provide different rotation periods for seven other stars}. Finally, we discuss how these results challenge state-of-the-art pulsation models for roAp stars.
\end{abstract} 

\begin{keywords} 
stars: oscillations -- stars: variables -- stars: individual -- stars: magnetic fields -- stars: chemically peculiar
\end{keywords} 

\section{Introduction} 
\label{sec:1}

Ap stars are chemically peculiar stars with enhanced abundances of Si, Cr, Sr, or Eu, permeated by relatively strong magnetic fields \citep{preston74}. That makes them test beds for the modelling of the physical processes responsible for element segregation in stars, such as gravitational settling and radiative levitation. They are relatively slow rotators and their magnetic field axis is usually found to be inclined with respect to the rotation axis. Moreover, as a result of element segregation in the presence of magnetic fields, some of the chemical elements are unevenly distributed at the surface \citep[e.g.][]{kochukhov11}. {As a consequence of the magnetic field being inclined to the rotation axis} \citep{stibbs1950}, many of these stars show light, spectral, and magnetic field variations over a period of rotation, and are commonly known as $\alpha^2$ Canum Venaticorum variables ({\it hereafter}, {$\alpha^2$~CVn stars}) {\citep{samus17}}. Still, for some stars no such rotational variations are detected indicating that either the rotation period is too long to be measured within the observational time span, or the star has a very particular configuration, such as a rotation axis pointing towards the observer or a very {small magnetic obliquity and no chemical spots away from the magnetic poles.}

Some of the Ap stars exhibit high radial-order acoustic pulsations. First discovered by \citet{kurtz1978,kurtz1982}, they are known as the rapidly oscillating Ap stars ({\it hereafter}, {roAp stars}). A number of searches for new pulsators of this class have been made since, through both ground-based surveys \citep{martinez1991,joshi16,paunzen15,holdsworth14b}, as well as in space-based data \citep{balona2010kepler,balona2011b,kurtz2011,holdsworth2014,smalley15,bowman18}. At the time of launch of the NASA Transiting Exoplanet Survey Satellite (TESS) mission \citep{TESS2014}, 61 roAp stars had been discovered \citep{smalley15,joshi16}. They are found {on the main sequence}, among the cooler Ap stars, with temperatures ranging from about 6000\,K to 9000\,K and have large scale magnetic fields with strengths of a few kG {\citep[see][for definitions of the different types of magnetic field measurements performed in Ap stars and their typical values]{mathys2017}}. The survey nature of the TESS satellite opens a new window of opportunity for finding additional members of this rare and unique class of pulsating star. In turn, these observations will set strong constraints on the modelling of key aspects of the stellar physics, such as the macroscopic and microscopic processes that define the distribution of elements in the stellar atmospheres and interiors.

Of particular interest for constraining structural models of Ap stars, as well as their magnetic fields, are pulsators exhibiting multiple modes (the multiperiodic roAp stars). {In the absence of a  strong magnetic field, and assuming all modes within a given frequency range were to be observed, the oscillation spectrum of an roAp star would be expected to show a regular pattern of alternating even and odd degree modes, as asymptotically predicted for high radial-order acoustic oscillations \citep{shibahashi79,tassoul1980}. However, in classical pulsators mode selection is usually present \citep[see][for a discussion]{aerts10} and not all modes in that pattern are necessarily visible. In addition,
because of the strong magnetic field, the waves in the outer layers of roAp stars become magnetoacoustic in nature with a consequent change in the mode oscillation frequencies.  Theory predicts that the oscillation spectra of these stars may still exhibit an almost equidistant set of peaks, whose separation is only a few $\mu$Hz larger than that expected in the absence of the magnetic field, but that they may also show a few modes at frequencies that do not follow that equidistant pattern predicted by the asymptotic analysis. These significant anomalies in the mode frequencies occur when the coupling between the acoustic and magnetic waves is strongest \citep{cunha00,cunha01,cunha06,saio04,saio05}. Indeed, some known multiperiodic roAp stars do show such frequency anomalies \citep[e.g.][]{kurtzetal05,huber08,gruberbauer08}.}
By revisiting most known roAp stars, TESS will discover additional pulsation modes, revealing many of these stars to be multiperiodic, in addition to the multiperiodic roAp stars that the satellite will discover for the first time. Likewise, rotationally-split frequency multiplets discovered by TESS in previously known or in new roAp stars will provide stringent constraints on the rotational inclination and magnetic obliquity of these stars.

The excitation mechanism responsible for the oscillations observed in roAp stars is still a matter of debate. Currently the most widely accepted theory proposes that these oscillations are driven by an opacity mechanism in the hydrogen ionization layers in stars where the strong magnetic field is capable of suppressing envelope convection, at least in some region around the magnetic poles \citep{balmforthetal01}. Based on this theory, {\cite{cunha02} determined the the maximum extent of the region of pulsational instability for roAp stars in the HR diagram ({\it hereafter}, {the theoretical instability strip}), derived by assuming that envelope convection in these stars is suppressed at all latitudes \citep[see][for a general discussion of the classical instability strip]{aerts10}.}

A few problems have been identified by confronting the theoretical results with observations. Among these problems is the existence of some roAp stars which pulsate with frequencies higher than those predicted by the models \citep{cunhaetal13,holdsworth2018c}; this is one of the most interesting challenges for the theory. \cite{cunhaetal13} have shown some evidence that these very high frequencies may be excited by an entirely different mechanism, associated with the turbulent pressure, but a thorough test of this possibility still needs to be performed. In fact, a similar mechanism seems to be responsible for the excitation of high radial-order acoustic modes in some $\delta$\,Scuti stars \citep{antoci14}. 

{Another important open problem concerns the region of pulsational instability for the roAp stars in the HR~diagram. Inspection of the effective temperatures known prior to the launch of TESS indicates the existence of roAp stars that are cooler than the  red ({\it i.e.}, cooler) edge of the theoretical instability strip. Likewise, roAp stars have not yet been found close to the blue {(\it i.e.}, hotter) edge of the theoretical instability strip}. Moreover, many stars with properties very similar to the roAp stars have been searched for pulsations without success. Those stars are commonly known as non-oscillating Ap stars ({\it hereafter}, {noAp stars}). TESS allows us, for the first time, to test whether these apparent disagreements between theory and observations are a consequence of an observational bias, resulting from selection effects on the samples observed so far, or from a lower intrinsic amplitude of the pulsational variability in hotter stars.

In this work we present the analysis of TESS data on Ap stars, from {the 27-d data sets for sectors 1 and 2 (see \citealt{TESS2014} for an explanation of the TESS observing strategy),} observed in 2-min cadence. In Sec.\,\ref{analysis} we describe the sample of stars under study, and describe the data analysis techniques employed. In Sec.\,\ref{results} we present the results of our analysis, including the determination of rotation periods and pulsational variability. Sections\,\ref{new} and \ref{old} present detailed TESS results on the newly discovered roAp stars and previously known roAp stars, respectively. Sec.\,\ref{sec:comp} makes a comparison between pulsation amplitudes as seen by TESS and {Johnson $B$} observations. In Sec.\,\ref{no} we discuss the results on two well characterised noAp stars, and in Sec.\,\ref{conclusion} we summarize our main results and conclude. 

\section{Analysis}
\label{analysis}
In this work we have used data from sectors 1 and 2 of the TESS mission. We present the analysis of 83 stars with spectral types from late B to early F, of which 80 have previously been identified as peculiar, or suspected peculiar, with enhanced Sr, Cr, Eu and/or Si. {The 80 chemically peculiar stars are a subset of the nearly 1400 targets proposed for observation with the 2-min cadence of TESS during the nominal mission, in the context of the roAp programme of the TESS Asteroseismic Science Consortium (TASC). The targets proposed for observation included: (1) 56 previously known roAp stars; (2) 13 well characterised noAp stars; (3) 40 stars observed by the KELT survey \citep{pepper18} and found to show evidence for pulsations in the frequency range from 700 to 3000~$\mu$Hz; (4) 1273 additional stars from the Michigan Spectral Catalogues \citep{houk1975,houk1978,houk1982,houk1988,houk1999} with an Ap spectral classification. In particular, the current subset of 80 Ap stars comprises 7 previously known roAp stars, 2 noAp stars, and 71 Ap stars selected from the Michigan Spectral Catalogues. The three additional stars in the sample of 83 stars reported here, namely, TIC\,152808505, TIC\,350146296, and TIC\,407661867, have not been classified as chemically peculiar in the literature and were not part of the targets selected in the context of the roAp programme. Nevertheless, they were observed with the 2-min cadence by TESS and included in our present sample  because hints of high frequency pulsations have been found in their light curves in the context of a systematic search for pulsations in over 5000 stars with $T_{\rm eff} > 6000$\,K observed with the TESS 2-min cadence. Of the 83 stars in our sample, 77 are listed in the General Catalogue of Ap and Am stars \citep{renson09}.} 

\begin{table*}
\centering
\caption{Properties of the 83 stars analysed in this work. Columns show, from left to right: {(1) TIC identification  number; (2) Henry Draper Catalogue identification number; (3) spectral type according to \citet{renson09}, except {where} noted otherwise; (4) TIC visual magnitude; (5) TIC effective temperature; (6) effective temperature from the infrared flux method; (7) error on effective temperature from the infrared flux method; (8) luminosity assuming no extinction; (9) error on luminosity assuming no extinction; (10) luminosity assuming extinction from \citet{gontcharov18}. Entries with "N/A" indicate that data {are} not available for that star.}}
\small
\begin{tabular}{rrcrrrcccc}
\\
\hline
\multicolumn{1}{c}{TIC}	&	\multicolumn{1}{c}{HD}	&	Spectral	&	\multicolumn{1}{c}{$V$}	&	\multicolumn{1}{c}{{\it T}$_{\rm eff}^{\rm TIC}$}	&	\multicolumn{1}{c}{{\it T}$_{\rm eff}^{\rm IRF}$} &	\multicolumn{1}{c}{$\Delta${\it T}$_{\rm eff}^{\rm IRF}$}	&	log({\it L}/L$_\odot$)$^{\rm noAv}$	&	$\Delta$log({\it L}/L$_\odot$)$^{\rm noAv}$	&	log({\it L}/L$_\odot$)$^{\rm Av}$	\\
        &       &        Type   &         &   \multicolumn{1}{c}{{\rm (K)}}                            &     \multicolumn{1}{c}{{\rm (K)}}               & \multicolumn{1}{c}{{\rm (K)}}  &
                                &       &                       \\ 
\hline
12359289	&	225119	&	B9 Si Cr	&	8.210	&	N/A	&	10911	&	189	&	2.46	&	0.07	&	N/A	\\
12393823	&	225264	&	A1 Si Sr	&	8.280	&	9433	&	9374	&	186	&	1.35	&	0.06	&	1.44	\\
12968953	&	217704	&	A5 Sr	&	10.169	&	7972	&	7881	&	162	&	1.33	&	0.09	&	N/A	\\
24693528	&	14944	&	A0 Eu	&	9.706	&	7522	&	7401	&	153	&	2.05	&	0.07	&	N/A	\\
31870361	&	22488	&	A3 Sr Eu Cr	&	7.510	&	7219	&	6911	&	157	&	1.45	&	0.05	&	1.52	\\
32035258	&	24188	&	A0 Si	&	6.260	&	N/A	&	13136	&	220	&	2.17	&	0.05	&	2.25	\\
38586082	&	27463	&	A0 Eu Cr	&	6.340	&	8669	&	8594	&	176	&	1.57	&	0.05	&	1.63	\\
38586127	&	27472	&	F0 Eu Cr Sr	&	9.956	&	7309	&	7391	&	155	&	1.14	&	0.05	&	1.23	\\
41259805	&	43226	&	A0 Sr Eu	&	8.970	&	8034	&	8293	&	172	&	1.08	&	0.05	&	1.23	\\
52368859	&	10081	&	A0 Sr Eu	&	9.620	&	9076	&	8726	&	239	&	1.87	&	0.06	&	N/A	\\
69855370	&	213637	&	F1 Eu Sr	&	9.647	&	6587	&	6433	&	148	&	0.69	&	0.05	&	0.77	\\
89545031	&	223640	&	B9 Si Sr Cr	&	5.170	&	N/A	&	13161	&	207	&	2.21	&	0.06	&	2.26	\\
92705248	&	200623	&	A2 Sr Eu Cr	&	9.070	&	8808	&	8889	&	199	&	1.16	&	0.07	&	1.25	\\
115150623	&	201018	&	A2 Cr Eu Sr	&	8.640	&	9080	&	9891	&	198	&	1.32	&	0.06	&	1.43	\\
116881415	&	3135	&	F3 Si Cr	&	9.600	&	7137	&	6765	&	147	&	1.08	&	0.06	&	1.15	\\
118114352	&	3772	&	A9 Si	&	10.001	&	6789	&	6869	&	150	&	1.01	&	0.05	&	1.07	\\
129636548	&	203585	&	A0 Si	&	5.760	&	N/A	&	10725	&	185	&	1.91	&	0.05	&	1.95	\\
139191168	&	217522	&	A5 Sr Eu Cr	&	7.540	&	6918	&	6888	&	151	&	0.89	&	0.05	&	0.95	\\
141028198	&	35361	&	A2 Cr Eu	&	9.869	&	8192	&	7444	&	157	&	1.74	&	0.06	&	N/A	\\
141610473	&	41613	&	A3 Eu Cr	&	9.672	&	7183	&	6910	&	149	&	1.28	&	0.06	&	N/A	\\
144276313	&	221760	&	A2 Sr Cr Eu	&	4.690	&	8495	&	8498	&	264	&	1.80	&	0.06	&	1.85	\\
152086729	&	224962	&	F0 Sr	&	10.160	&	6768	&	6801	&	147	&	1.38	&	0.06	&	N/A	\\
152808505	&	216641	&	 F3IV/V$^*$	&	8.280	&	6430	&	6640	&	160	&	0.95	&	0.05	&	1.04	\\
159834975	&	203006	&	A2 Cr Eu Sr	&	4.800	&	8790	&	9057	&	218	&	1.48	&	0.06	&	1.53	\\
167695608	&		&	F0p SrEu(Cr)$^{**}$	&	11.513	&	7185	&	7460	&	157	&	1.10	&	0.06	&	N/A	\\
167751145	&	52280	&	A0 Sr Cr Eu	&	9.816	&	7621	&	7772	&	177	&	1.30	&	0.06	&	1.42	\\
182909257	&	6783	&	B8 Si	&	7.960	&	N/A	&	11511	&	197	&	1.86	&	0.06	&	2.00	\\
183802606	&	8700	&	A0 Si Cr Fe	&	9.569	&	8754	&	8982	&	179	&	1.77	&	0.06	&	N/A	\\
183802904	&	8783	&	A2 Sr Eu Cr	&	7.800	&	8648	&	8346	&	181	&	1.78	&	0.06	&	1.89	\\
206461701	&	209364	&	A5 Sr Eu Cr	&	10.028	&	7144	&	6962	&	150	&	1.42	&	0.06	&	N/A	\\
206648435	&	215983	&	A0 Sr Eu Cr	&	9.660	&	8184	&	8395	&	169	&	1.24	&	0.06	&	1.33	\\
207208753	&	20505	&	A2 Cr Sr	&	9.816	&	9310	&	8418	&	169	&	1.45	&	0.06	&	N/A	\\
211404370	&	203932	&	A5 Sr Eu	&	8.812	&	7544	&	7366	&	157	&	0.92	&	0.20	&	N/A	\\
219340705	&	222349	&	F Sr	&	9.217	&	6222	&	6231	&	153	&	0.39	&	0.05	&	0.45	\\
231844926	&	10840	&	B9 Si	&	6.790	&	N/A	&	10471	&	185	&	1.90	&	0.05	&	1.98	\\
232066526	&	11090	&	A Sr	&	10.782	&	8750	&	8587	&	171	&	1.32	&	0.06	&	N/A	\\
234346165	&	16504	&	B8 Si	&	9.050	&	9480	&	10315	&	175	&	2.14	&	0.06	&	N/A	\\
235007556	&	221006	&	A0 Si	&	5.660	&	N/A	&	13863	&	237	&	2.24	&	0.05	&	2.30	\\
237336864	&	218495	&	A2 Eu Sr	&	9.378	&	8283	&	7941	&	162	&	0.92	&	0.05	&	1.04	\\
262613883	&	63728	&	A0 Eu Cr Si	&	9.356	&	9359	&	8847	&	176	&	1.39	&	0.06	&	1.57	\\
262956098	&	3988	&	A0 Cr Eu Sr	&	8.340	&	7691	&	7646	&	166	&	1.67	&	0.06	&	1.82	\\
266905315	&	225234	&	A3 Sr	&	8.872	&	8343	&	7993	&	163	&	1.00	&	0.05	&	1.13	\\
270304671	&	209605	&	F0 Sr Eu	&	9.576	&	8044	&	7943	&	163	&	1.24	&	0.06	&	1.33	\\
271503787	&	2883	&	F4 Sr	&	9.380	&	6359	&	6423	&	149	&	0.64	&	0.06	&	0.75	\\
277688819	&	208217	&	A0 Sr Eu Cr	&	7.200	&	8368	&	8318	&	172	&	1.13	&	0.06	&	1.21	\\
277748932	&	208759	&	A0 Sr Eu Cr	&	9.984	&	8955	&	8308	&	165	&	1.34	&	0.06	&	N/A	\\
278804454	&	212385	&	A3 Sr Eu Cr	&	6.850	&	8672	&	8806	&	175	&	1.50	&	0.06	&	1.56	\\
279091054	&	50861	&	A3 Sr Eu	&	9.746	&	7372	&	7493	&	164	&	1.20	&	0.06	&	1.32	\\
279573219	&	54118	&	A0 Si	&	5.140	&	N/A	&	10848	&	182	&	1.88	&	0.06	&	2.00	\\
280051011	&	18610	&	A2 Cr Eu Sr	&	8.170	&	8628	&	8371	&	178	&	1.50	&	0.05	&	1.64	\\
281668790	&	3980	&	A7 Sr Eu Cr	&	5.720	&	8747	&	7448	&	159	&	1.26	&	0.05	&	1.32	\\
304096024	&	11346	&	A2 Sr Eu Cr	&	9.898	&	8098	&	7339	&	153	&	1.69	&	0.06	&	N/A	\\
306573201	&	66195	&	A0 Sr Eu Cr	&	8.660	&	N/A	&	8892	&	194	&	1.39	&	0.06	&	1.56	\\
\hline
\end{tabular}
\label{properties}
\end{table*}

\begin{table*}
\centering
\contcaption{Properties of the 83 stars analysed in this work.}
\small
\begin{tabular}{rrcrrrcccc}
\\
\hline
\multicolumn{1}{c}{TIC}	&	\multicolumn{1}{c}{HD}	&	Spectral	&	\multicolumn{1}{c}{$V$}	&	\multicolumn{1}{c}{{\it T}$_{\rm eff}^{\rm TIC}$}	&	\multicolumn{1}{c}{{\it T}$_{\rm eff}^{\rm IRF}$} &	\multicolumn{1}{c}{$\Delta${\it T}$_{\rm eff}^{\rm IRF}$}	&	log({\it L}/L$_\odot$)$^{\rm noAv}$	&	$\Delta$log({\it L}/L$_\odot$)$^{\rm noAv}$	&	log({\it L}/L$_\odot$)$^{\rm Av}$	\\
        &       &        Type   &         &   \multicolumn{1}{c}{{\rm (K)}}                            &     \multicolumn{1}{c}{{\rm (K)}}               & \multicolumn{1}{c}{{\rm (K)}}  &
                                &       &                       \\ 
\hline
306893839	&	68561	&	B9 Si	&	8.020	&	N/A	&	11290	&	207	&	2.34	&	0.06	&	N/A	\\
307031171	&	69578	&	 F6III  Sr$^{***}$	&	9.560	&	6720	&	5905	&	148	&	1.18	&	0.06	&	1.35	\\
307288162	&	71006	&	A0 Si	&	9.243	&	N/A	&	12403	&	213	&	2.34	&	0.07	&	N/A	\\
307642246	&	72634	&	A0 Eu Cr Sr	&	7.280	&	8947	&	9348	&	191	&	2.13	&	0.11	&	2.31	\\
308085294	&	74388	&	B8 Si	&	6.990	&	N/A	&	12165	&	223	&	2.48	&	0.06	&	2.66	\\
309148260	&	69862	&	A2 Sr Eu Cr	&	10.112	&	8134	&	7485	&	156	&	1.45	&	0.06	&	N/A	\\
316913639	&	222638	&	A0 Sr Eu Cr	&	8.650	&	9714	&	9961	&	195	&	1.51	&	0.06	&	1.64	\\
327597288	&	206653	&	B9 Si	&	7.210	&	N/A	&	11384	&	205	&	2.00	&	0.06	&	2.13	\\
336731635	&	214985	&	A0 Si	&	11.103	&	9439	&	11631	&	199	&	2.18	&	0.18	&	N/A	\\
348717688	&	19918	&	A5 Sr Eu Cr	&	9.350	&	8074	&	7484	&	159	&	1.10	&	0.05	&	1.22	\\
348898673	&	54399	&	A2 Sr Cr Eu	&	9.725	&	7506	&	7466	&	156	&	1.70	&	0.06	&	N/A	\\
349409844	&	58448	&	B8 Si	&	6.920	&	N/A	&	11536	&	204	&	1.77	&	0.05	&	1.88	\\
350146296	&	63087	&	 A7IV$^{***}$	&	9.408	&	7690	&	7450	&	160	&	0.82	&	0.05	&	N/A	\\
350146577	&	63204	&	B9 Si	&	8.307	&	9737	&	11389	&	200	&	1.86	&	0.05	&	2.03	\\
350272314	&	222925	&	F8 Sr Eu	&	9.020	&	5579	&	5723	&	136	&	1.63	&	0.05	&	N/A	\\
350519062	&	38719	&	A0 Cr Sr Eu	&	7.510	&	8937	&	8967	&	186	&	1.62	&	0.05	&	1.75	\\
358467700	&	65712	&	A0 Si Cr	&	9.340	&	8768	&	8941	&	178	&	1.46	&	0.05	&	N/A	\\
364424408	&	30374	&	A0 Sr Eu Cr	&	10.052	&	7608	&	8420	&	167	&	1.50	&	0.06	&	N/A	\\
372913684	&	65987	&	B9 Si Sr	&	7.620	&	N/A	&	10796	&	186	&	2.33	&	0.06	&	2.49	\\
382512330	&	64369	&	B9 Si	&	8.839	&	9042	&	9701	&	214	&	1.71	&	0.06	&	N/A	\\
389531041	&	206193	&	F5 Sr	&	9.920	&	6513	&	6495	&	152	&	1.16	&	0.06	&	1.25	\\
389922504	&	40277	&	A1 Sr Cr Eu	&	8.350	&	N/A	&	9123	&	188	&	1.22	&	0.05	&	1.36	\\
391927730	&	56981	&	F0 Sr	&	9.589	&	6984	&	6817	&	155	&	0.78	&	0.05	&	0.95	\\
392761412	&	207259	&	A0 Eu Sr Cr	&	8.830	&	8036	&	7901	&	163	&	1.33	&	0.06	&	1.47	\\
394045029	&	211333	&	F6IV Sr$^{***}$	&	8.550	&	6292	&	6349	&	141	&	1.14	&	0.05	&	1.25	\\
394124612	&	218994	&	A3 Sr	&	8.570	&	7451	&	7368	&	154	&	1.42	&	0.06	&	1.52	\\
407661867	&	37584	&	A3V$^{***}$	&	8.330	&	9363	&	8750	&	180	&	1.10	&	0.05	&	N/A	\\
410451752	&	66318	&	A0 Eu Cr Sr	&	9.652	&	9057	&	8849	&	177	&	1.48	&	0.06	&	1.63	\\
431380369	&	20880	&	A3 Sr Eu Cr	&	7.953	&	8242	&	7884	&	167	&	1.40	&	0.06	&	1.51	\\
434103853	&	221531	&	F5 Sr	&	8.340	&	6513	&	6436	&	156	&	0.64	&	0.05	&	0.70	\\
\hline
 \multicolumn{9}{l}{$^*$ Spectral type from \citep{houk1978}} \\
\multicolumn{9}{l}{$^{**}$ Spectral type from \citet{holdsworth14b}}\\
\multicolumn{9}{l}{$^{***}$ Spectral type from \citet{houk1975}}\\
\end{tabular}
%\label{parameters}
\end{table*}

\subsection{Properties of the sample}
\label{sec:prop}
The properties of our sample are given in Table\,\ref{properties}. Two effective temperatures, $T_{\rm eff}$, are provided, one obtained from the TESS Input Catalogue (TIC)\footnote{https://mast.stsci.edu/portal/Mashup/Clients/Mast/Portal.html} the other derived using the Infrared Flux Method \citep[IRFM;][]{1977MNRAS.180..177B}. 

To calculate the effective temperature through the Infrared Flux Method,
the stellar spectral energy distributions (SEDs) were obtained using
literature broad-band photometry: 2MASS \citep{2006AJ....131.1163S},
Tycho $B$ and $V$ \citep{2000A&A...355L..27H}, APASS9 $B$, $V$, $g'$,
$r'$ and $i'$ \citep{2015AAS...22533616H}, USNO-B1 $R$
\citep{2003AJ....125..984M} and WISE \citep{2012yCat.2311....0C}. The
photometry was converted to fluxes, and the best-fitting
\citet{1993KurCD..13.....K} model flux distribution was obtained using a
weighted {Levenberg-Marquardt~\citep{levenberg44,marquardt63} non-linear least-squares fitting procedure, as implemented in \cite{1992nrfa.book.....P}}.
The fitted flux distributions were then numerically integrated to
determine the stellar bolometric fluxes. Interstellar reddening was
assumed to be zero. {The IRFM was then used to determine $T_{\rm eff}$
values and their uncertainties from the three individual 2MASS photometry bands. The final
values given in Table\,\ref{properties} are the weighted mean and uncertainty of the three 2MASS values.}

The luminosities for our sample of stars were computed by taking the Gaia DR2 parallaxes \citep{GDR218,Gaia18}, when available, and the Hipparcos parallaxes \citep{hipparcos07}, otherwise. In both cases we considered the uncertainties quoted in the respective catalogues, but for the Gaia data we inflated the formal errors by 30 per cent, as suggested in the catalogue for stars of magnitude $G<12$. We note also that Gaia DR2 parallaxes do not account for stellar multiplicity, which could, in some cases, influence the luminosities derived. For the bolometric correction we considered the expression of \citet{flower96} {(where we used the corrected version of the coefficients, published in table 1 of  \citet{torres10})}, and assumed an error of 0.13\,mag. This error was estimated by computing the bolometric corrections for six roAp stars for which a detailed bolometric flux computation is available \citep{brunttetal08,brunttetal10,perrautetal11,perrautetal13,perrautetal15,perrautetal16} and comparing them with the values predicted by the \citet{flower96} expression. The root mean square of the difference between the two bolometric correction values was then adopted as the error for all stars. {As the six stars used in this error estimate have effective temperatures lower than $\sim 9100$\,K, hence, do not cover the full temperature range in our sample, we have, in addition, computed bolometric corrections using the calibration proposed by \citet{netopil08}. The latter was derived based on peculiar stars and is valid in the range $7500\, {\rm K} <T_{\rm eff} < 19000\, {\rm K}$. We found a maximum absolute difference of 0.12\,mag between the bolometric corrections from \citet{flower96} and \citet{netopil08} in our sample, which is comparable to the error assumed}. Finally, we considered the $V$ magnitude from the TIC catalogue and assumed an uncertainty of 0.02\,mag, corresponding to the typical spread in results seen in literature. The luminosities and associated errors computed from these data are given in the 8$^{\rm th}$ and 9$^{\rm th}$ columns of Table\,\ref{properties}. No extinction was considered in that calculation. In addition, {we computed the luminosity corrected for extinction} (10$^{\rm th}$ column in the same table), by considering the extinction published by \citet{gontcharov18}. We find that the root mean square of the difference between the logarithmic luminosity values derived with and without accounting for extinction is $0.1$.

\begin{figure*}
\centering
\includegraphics[width=0.95\linewidth,angle=0]{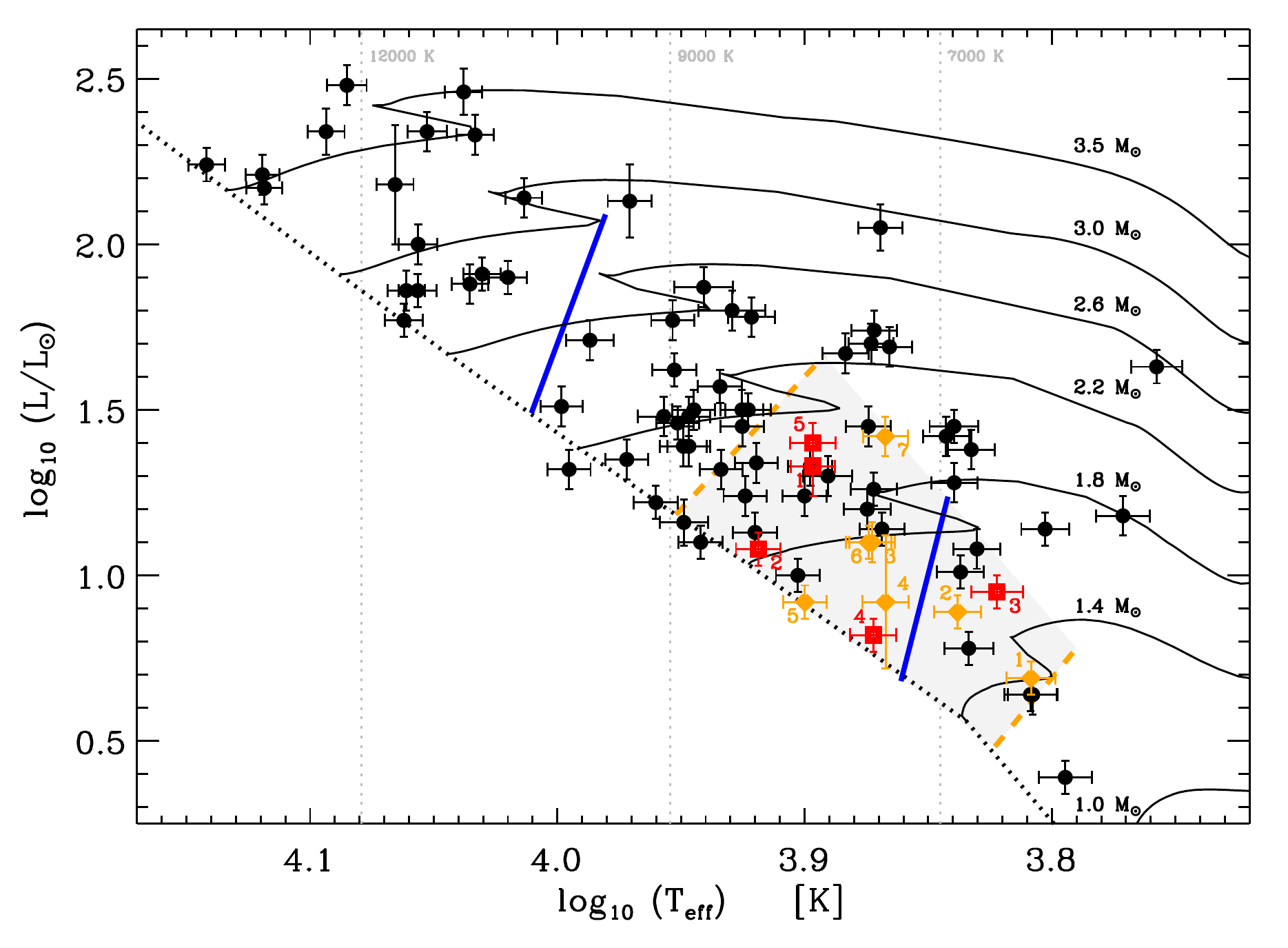}	
\caption{HR diagram showing the position of the 83 stars in our sample. The thick dotted line indicates the zero-age main-sequence and the continuous non-straight lines mark the evolutionary sequences for masses between 1.0\,M$_\odot$ and 3.5\,M$_\odot$ \citep[from][grid B]{marques08}. The thick, blue, continuous straight lines indicate the {theoretical instability strip} from \citet{cunha02} and the orange-dashed lines enclosing the shaded area indicate the region where the 61 roAp stars known prior to the TESS launch were found.  {This region was established by adopting the effective temperatures from \citet{joshi16} for all stars, except TIC\,363716787, for which we adopted the value from the TIC catalogue, as the authors did not provide one, and by computing the roAp stars' luminosities from the data sources considered for the stars in the TESS sample, without considering extinction. } The numbered red squares show the position of the 5 new roAp stars discovered in the TESS data: {1: TIC\,12968953, 2: TIC\,41259805, 3: TIC\,152808505, 4: TIC\,350146296, 5: TIC\,431380369};  the numbered orange circles show the 7 previously known roAp stars observed in sectors 1 and 2: {1: TIC\,69855370, 2: TIC\,139191168, 3: TIC\,167695608, 4: TIC\,211404370, 5: TIC\,237336864, 6: TIC\,348717688, 7: TIC\,394124612}; and the black circles show the remaining stars in the sample.}
\label{HRdiagram}
\end{figure*}

Fig.\,\ref{HRdiagram} shows the position in the HR diagram of all stars in our sample. {Here we have used the effective temperatures and luminosities derived in this work, as listed in columns 6 to 9 of Table\,\ref{properties}.}

\subsection{Data analysis {strategy}} 
\label{dataAna}

In this work we have used the {\sc{PDC\_SAP}} fluxes from the {\sc{fits}} data files retrieved from the Mikulski Archive for Space Telescopes (MAST)\footnote{http://archive.stsci.edu/tess/all\_products.html}. {Ten different teams were involved in the analysis of the data. One of these teams, {\it hereafter}, the reference team, analysed the data on all stars, both in search for rotational and pulsational variability, so that homogeneous results could be provided. Two other teams analysed the full sample in search for rotational variability. The other 7 teams analysed subsamples of the target list for rotational and/or pulsational variability. The distribution of the targets was done in such a way that the rotational variability of each star was inspected by a minimum of three teams and the pulsational variability by a minimum of two. 
The results from the teams other than the reference team were used to: (1) identify apparent inconsistencies; (2) in the case of rotational variability, compute the standard deviation of the periods derived by the different teams.} When inconsistencies were found, the teams were asked to revisit their analysis. The most common inconsistency was the derivation of rotational frequencies corresponding to different harmonic values. In those cases, inspection of the light curve and, when available, of the pulsational multiplets, allowed for the identification of the true rotation period. Rotation periods and pulsation frequencies were only considered after these inconsistencies had been understood and removed and only when at least two teams had confirmed a detection.

For the reference team, the data analysis procedure involved the following steps. The extracted fluxes were converted to magnitudes, with the time stamps being corrected for the zero-point offset. Obvious outliers from the light curves were removed by hand in each case {(less than 1\,per\,cent in most cases)}. Where possible, the data from both sectors were combined to increase the time base, and reduce the noise level in the amplitude spectra.

{For all stars}, the data were first analysed at low frequencies {($0-0.12$\,mHz)} in the search for rotational modulation. Both a visual inspection of the light curve, and the computation of a Fourier transform were conducted. If modulation was identified, the signal and any harmonics were fitted by non-linear least-squares to the light curve to optimise the solution of the rotation frequency. In the cases where a harmonic has a higher amplitude, we used its frequency to determine the period as the frequency precision is higher for a higher amplitude signal. Phase-folded light curves were produced for every determined rotation period to ensure by visual inspection that the correct frequency (i.e. not a harmonic) was chosen.

{As a by-product of this process, we also identified low-frequency variability which we assumed to be either g-mode $\gamma$\,Doradus ({\it hereafter}, {$\gamma$\,Dor}) pulsations, or low overtone p-mode pulsations as seen in $\delta$\,Scuti ({\it hereafter}, {$\delta$\,Sct}) stars. We made a note of these stars, but did not perform an in-depth analysis.}

Once any low-frequency information had been extracted from the light curve, we iteratively pre-whitened the data to remove the rotation signals and low-frequency instrumental artefacts in the range $0-0.12$\,mHz ($0-10$\,d$^{-1}$). This was done on a star-by-star basis; the amplitude limit of the pre-whitening was determined by the noise level in the high-frequency range {where rapid oscillations have been seen in other stars ($0.7-3.6$\,mHz)}. 

Subsequently, an amplitude spectrum was calculated to the Nyquist frequency of the data in the search for high-frequency pulsations. If variability was detected, the mode frequency was fitted with {linear least-squares (where the frequency is fixed and the amplitude and phases are derived to minimise the residuals) and non-linear least-squares (where all three parameters are free)} to optimize the frequency, amplitude and phase of the fit to the data. If the star was found to be multiperiodic, these fits were performed simultaneously on all frequencies. In the cases where variability was not detected, we took the limit of the non-detection as either four times the error of the highest noise peak in the data, or the top of the Fourier `grass' {(the approximate background peaks of the amplitude spectrum which resemble mown grass)}, whichever was found to be greater to be conservative. 

As mentioned above, a number of additional teams analysed the sample of stars under study (typically three per star for the pulsational variability, and typically five per star for the rotational variability). Different analysis tools were used by these teams, including: Period04 \citep{period04}, the {CLEANest \citep{foster95}} and {Phase Dispersion Minimization (PDM) algorithms \citep{stellingwerf78}} as implemented in the Peranso light curve and period analysis software~\citep{peranso}\footnote{http://www.cbabelgium.com/peranso/index.htm}, the harmonic-function fitting program LCfit \citep{sodor12}, and the SigSpec code \citep{reegen07}. The spread in the rotation periods derived by the different teams, as measured by the standard deviations listed in Table\,\ref{rot}, likely results from the different tools applied in the data analyses, but also from differences in the pre-processing of the light curve, including the filtering of outliers and bad data sections, and in the normalization of the light curves. {In this respect, we note that the reference team normalised each stellar light curve to zero by removing the mean magnitude of the entire sector. Some of the other teams normalised each light curve by the maximum or minimum of the light curve, and others still, by the median of the raw electron counts.}

\section{Results}
\label{results}
A summary of the rotational and pulsational variability results obtained for the 83 stars in our sample is presented in Table\,\ref{rot}. In the column ``Variability" we identify whether a star is found to be a variable and, when that is the case, the type(s) of variability present. When the star was not previously known to exhibit a given type of variability, the variability type is prefixed by the word ``new". Likewise, if the star was known to exhibit rotational variability, but we find evidence that the rotation period published in the literature is incorrect, we write ``new rot per" in that column. {For new $\delta$\,Sct and/or $\gamma$\,Dor variables we further use the term ``susp." to indicate these are suspected variables of these types. This is to remind the reader that no detailed analysis of the low-frequency variability was performed in this work, as mentioned in Sec.\,\ref{analysis}, {thus we cannot be certain that the variability is only due to pulsation.}}

{As the TESS pixels are large ($20.25$\,arcsec) there is a relatively high possibility of contamination from nearby sources. This is summarised by the contamination ratio for each star as provided by the TIC. For the 83 stars we study here, the contamination ratio is below 0.1 for all but one star (TIC\,410451752). This gives us confidence that the variability we report is for the target star in question, and not associated with a contaminating source. In addition to checking the contamination ratio, we searched the literature for discussion on stellar binarity or multiplicity. For most stars reported as new rotational or roAp pulsational variables, there is no mention in the literature of multiplicity. The case of TIC\,410451752, as well as cases where multiplicity may be a source of confusion are {discussed}, on a star by star basis, in Secs.~\ref{cont}, \ref{sec:opm} and \ref{cont2}.}

\subsection{Rotational variability}
\label{res:rot}
\begin{figure*}
\centering
\includegraphics[width=0.49\textwidth]{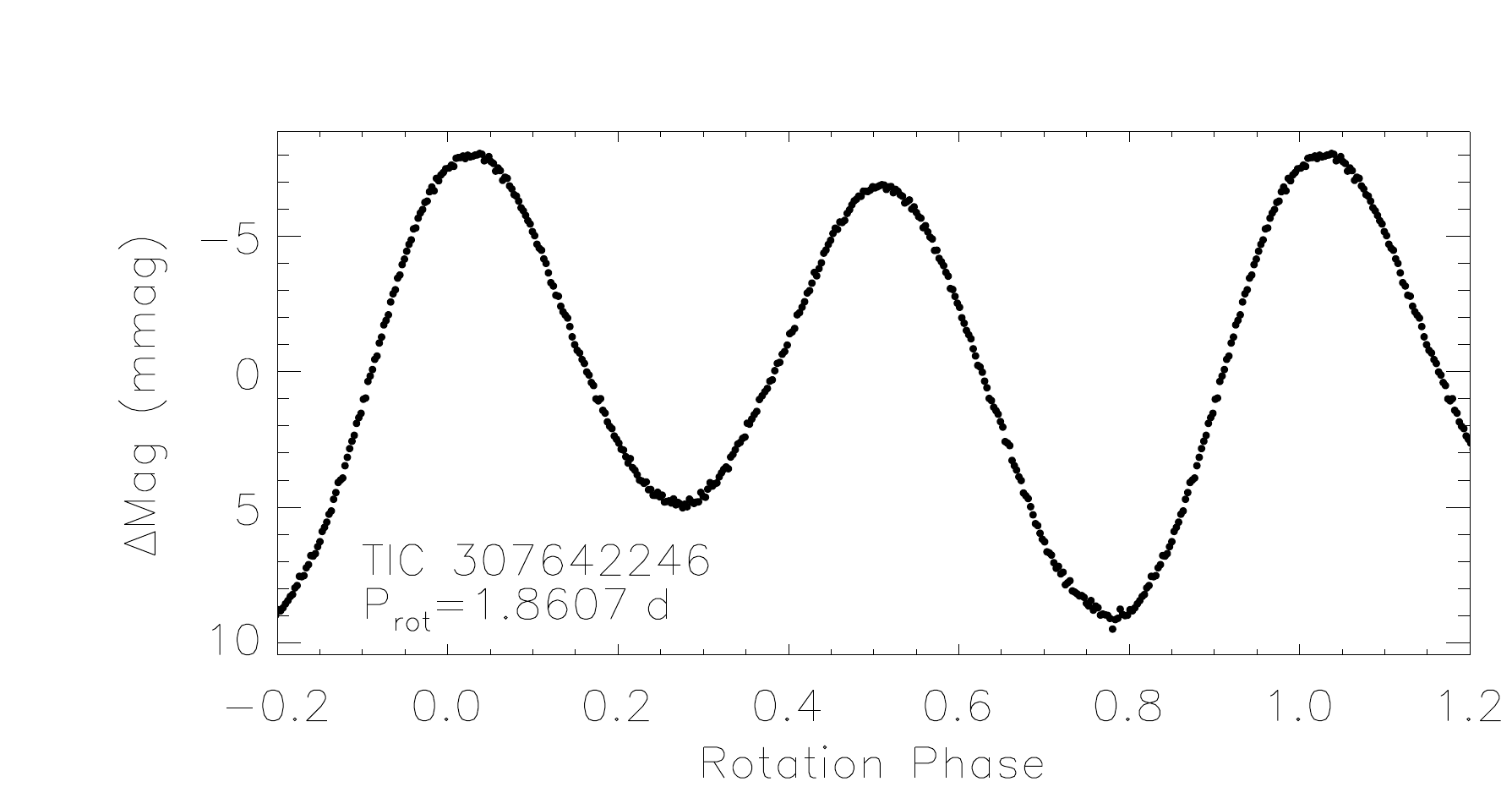}\hfill
\includegraphics[width=0.49\textwidth]{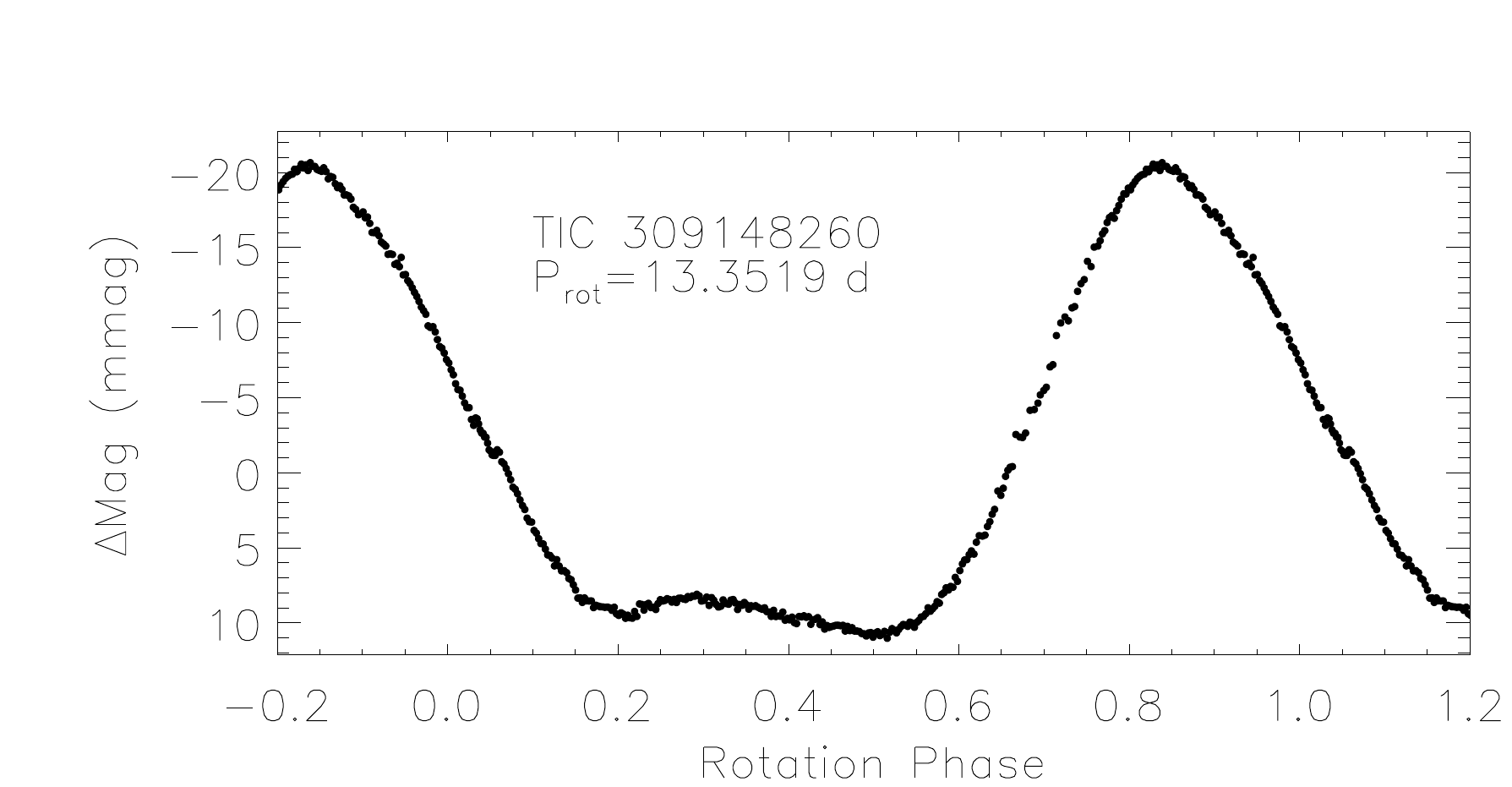}\\
\includegraphics[width=0.49\textwidth]{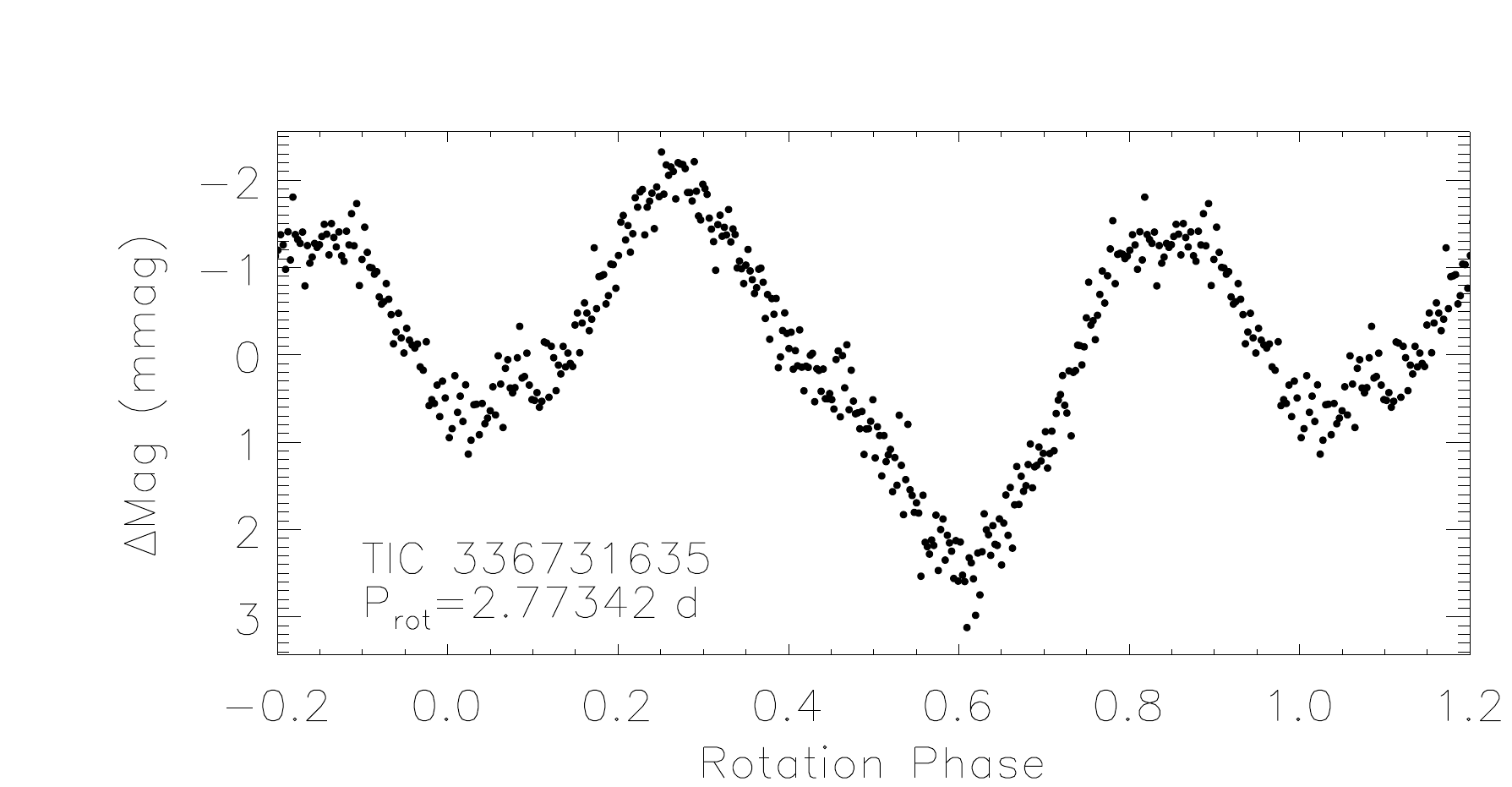}\hfill
\includegraphics[width=0.49\textwidth]{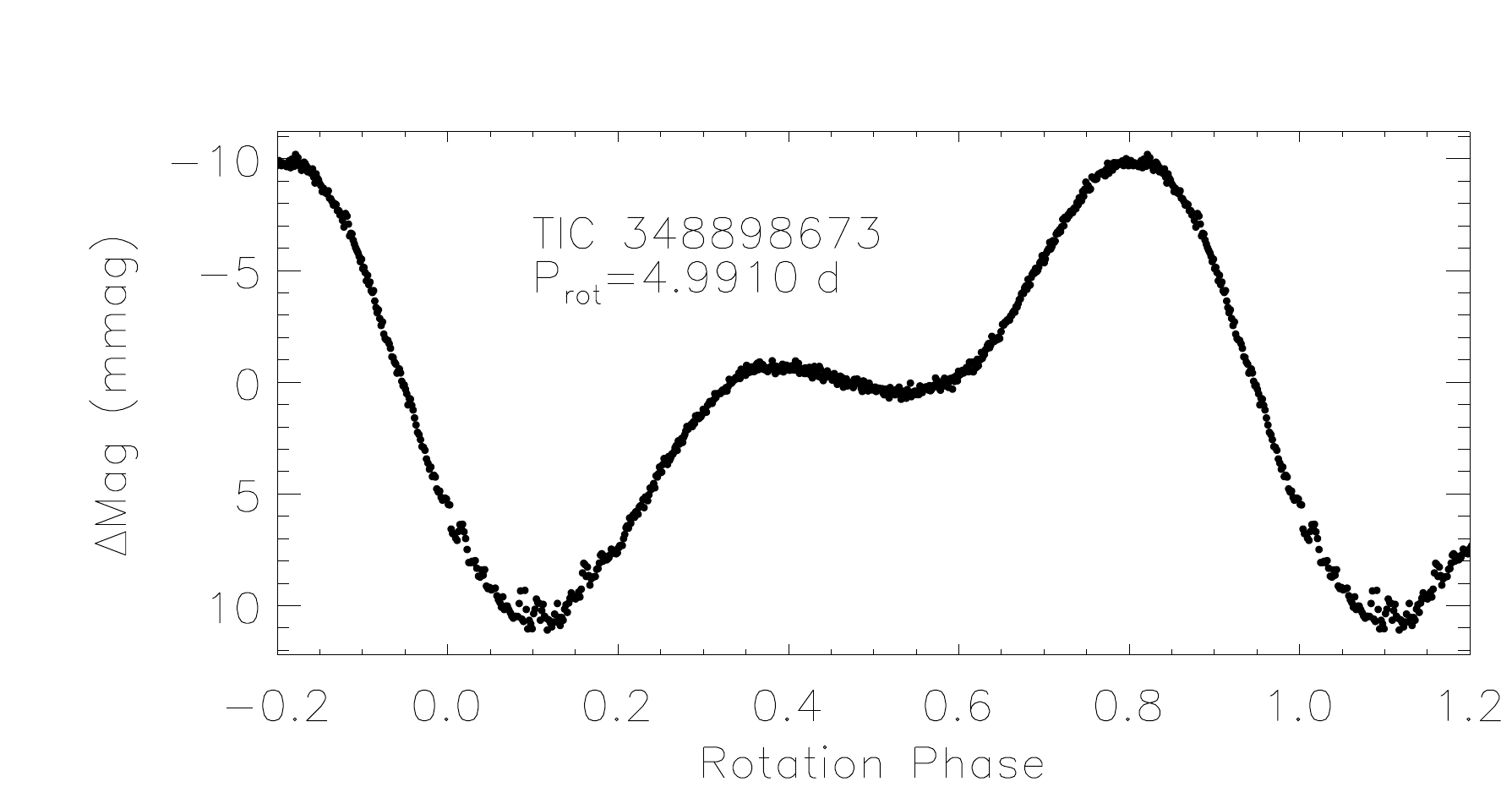}\\
\includegraphics[width=0.49\textwidth]{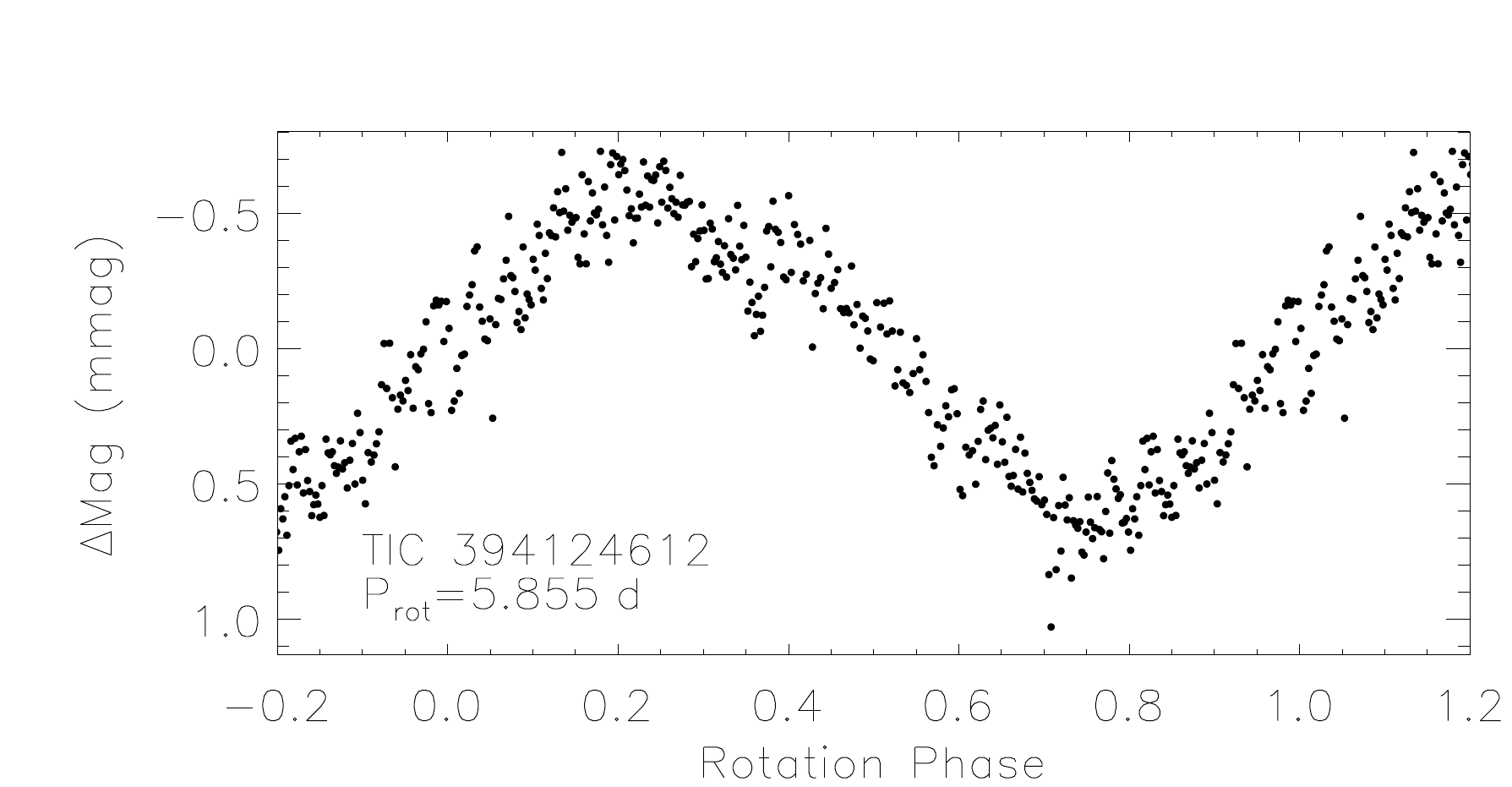}\hfill\hfill

\caption{The phase-folded light curves for the five stars where we provide a new determination of the rotation period. Top left: TIC\,307642246, top right: TIC\,309148260, middle left: TIC\,336731635, middle right:TIC\,348898673, bottom left: TIC\,394124612. The data have been binned 50:1 for clarity.}
\label{fig:rotwrong}
\end{figure*}

We have identified {27} new rotational variables among the 83 targets and found {five} known rotational variables to have previously misidentified rotation periods. Of these, in three cases the published values correspond to a harmonic of the true rotation frequency.
The phase-folded light curves for the {five} stars mentioned above are shown in Fig.\,\ref{fig:rotwrong}. The double-wave nature of the light curves from TIC\,307642246, TIC\,336731635 and TIC\,348898673 explains why these stars were previously identified to have half of the true rotation period. {For TIC\,309148260 and TIC\,394124612} the reasons why the published values are so different from the ones determined now are unclear.

{In addition to the rotation periods derived by the reference team and associated formal uncertainties, we provide in Table\,\ref{rot} the standard deviations on the rotation periods derived by all teams analysing each star. These correspond to the spread in results arising from the application of different analysis techniques. For comparison, the literature values for the rotation periods are also provided, when available.} 

{Besides the five stars mentioned above, for which our rotation periods are strikingly different from the published results, there are a few other cases in which we found rotation periods differing significantly from those in the literature. To identify the most extreme cases, we added in quadrature the formal error and standard deviation of our rotation periods, and adopted that quantity, $\sigma_{\rm rot}$, as a conservative measurement of the uncertainty on our results. We found that for another two stars the absolute difference between the literature value, taking into account the published uncertainties, and the value we obtained is larger than 10 $\sigma_{\rm rot}$. For those stars, namely TIC 12359289 and TIC 372913684, we established that the rotation period derived from TESS data is significantly different from the previously published values. Moreover, while this work was under review an independent study of magnetic O, B and A stars \citep{david2019} has reported rotation periods derived from the same TESS data used in this work for twelve stars in our sample. While for nine stars the agreement is clear (with eight rotation periods agreeing within the formal errors and one within $\sigma_{\rm rot}$), for the remaining three stars the rotation periods reported by \cite{david2019} are somewhat different. In particular, for TIC\,89545031 the period in that study, $P=3.7349\pm 0.0005$~d, differs by 2.6 $\sigma_{\rm rot}$ from the period derived in our work; for TIC\,237336864 the same authors find $P=4.183\pm 0.006$~d, which differs from the period we derived by 4.8 $\sigma_{\rm rot}$ ; and for TIC\,279573219 their period, of $3.2759 \pm 0.0002$~d differs from ours by 2.6 $\sigma_{\rm rot}$. In fact, these three stars have some strong low-frequency instrumental systematics, which are either not removed by the TESS science pipeline, or introduced by the TESS reduction pipeline. It is, thus, more likely that different approaches to the analysis of their light curves lead to differences in the derived periods. While the differences found are not as significant as in the cases discussed earlier in this section, they do call for caution when analysing TESS PDC SAP fluxes when strong low-frequency instrumental noise is present. }

{Finally, we note, for future comparison, that for TIC\,118114352 and TIC\,407661867 rotation periods were derived, but considered insecure, hence were not included in the table. We provide provisional values here for future reference.} For TIC\,118114352, the amplitude spectrum shows a peak consistent with a rotation period of 1.388\,d. However, there is a second peak close to it, corresponding to a period of 1.093\,d. If one is the rotation signal, then the other may be due to a pulsation mode or contamination from another star. The second star is TIC\,407661867. For this star there is a peak in the amplitude spectrum corresponding to a period of 0.562\,d but with other similar amplitude peaks in the vicinity and the period is short for an Ap star. The field for this star is crowded (despite the target star being the brightest), so the source of the variability is also questionable. 

\begin{table*}
\centering
\caption{Variability properties of the 83 stars analysed in this work. {Columns show, from left to right: (1) TIC identification number; (2) Henry Draper Catalogue identification number:} (3) {variability type; here the word `new' is used to indicate stars in which the property under consideration was first discovered or measured in this work. For new $\delta$~Sct and $\gamma$~Dor variables we use ``susp." to indicate the stars are suspected to be variables of the indicated type}; (4) published {rotation} period, if available; (5) rotation period determined in this work; (6) formal uncertainty in the rotation period; (7) standard deviation computed from the set of rotation periods derived by the different teams (see text for details); (8) {pulsation amplitude limit} for stars in which pulsational variability has not been detected; (9) references for known rotational variables; {``new rot var'' means new rotational variable -- all of these appear to be $\alpha^2$~CVn stars;  ``new rot per'' means new rotation period.}}
\small
\begin{tabular}{rrcrrrrrr}
\\
\hline
\multicolumn{1}{c}{TIC}	&	\multicolumn{1}{c}{HD}	&	\multicolumn{1}{c}{Variability type}	&	\multicolumn{1}{c}{$P_{\rm rot}{\rm (lit)}$}	&	\multicolumn{1}{c}{$P_{\rm rot}$}	&	\multicolumn{1}{c}{$\Delta$$P_{\rm rot}$}	&	\multicolumn{1}{c}{SDV} & 	\multicolumn{1}{c}{Pulsation limit} 	&	\multicolumn{1}{c}{Ref}	\\
    &       &                       &    \multicolumn{1}{c}{(d)}            &    \multicolumn{1}{c}{(d)}            &      \multicolumn{1}{c}{(d)}                  &  \multicolumn{1}{c}{(d)}  &   \multicolumn{1}{c}{($\umu$mag)}      &        \\           
\hline
12359289	&	225119	&	rot var	    &	2.9	&	3.06395	&	0.00041	&	0.0030	&	19	&	1,2	\\
12393823	&	225264	&	new rot var	&		&	1.42353	&	0.00023	&	0.00016	&	20	&		\\
12968953	&	217704	&	new roAp	&		&		&		&		&		&		\\
24693528	&	14944	&	susp. new $\delta$~Sct / $\gamma$~Dor 	&		&		&		&		&		&		\\
31870361	&	22488	&	susp. new $\delta$~Sct / $\gamma$~Dor 	&		&		&		&		&		&		\\
32035258	&	24188	&	$\alpha^2$~CVn 	    &	2.2301 &	2.23047	&	0.00004	&	0.0013	&	7	&	3,17	\\
38586082	&	27463	&	$\alpha^2$~CVn; susp. new $\delta$~Sct	&	2.833 $\pm$ 0.010	&	2.8349	&	0.0001	&	0.0030	&		& 18		\\
38586127	&	27472	&		        &		&		&		&		&	30	&		\\
41259805	&	43226	&	new roAp; new rot var	&		&	1.71441	&	0.00011	&	0.00038	&		&		\\
52368859	&	10081	&	$\alpha^2$~CVn	        &	1.57032 $\pm$ 0.00003	&	1.57056	&	0.00006	&	0.0023	&	40	&	5	\\
69855370	&	213637	&	roAp	    &		&		&		&		&		&		\\
89545031	&	223640	&	$\alpha^2$~CVn	        &	3.73524	&	3.72251	&	0.00097	&	0.0047	&	30	&	4	\\
92705248	&	200623	&	$\alpha^2$~CVn	&	2.18 $\pm$ 0.01	&	2.1577	&	0.0002	&	0.0021	&	30	&	4,9	\\
115150623	&	201018	&	new rot var	&		&	1.50048	&	0.00005	&	0.00050	&	24	&		\\
116881415	&	3135	&	susp. new $\delta$~Sct / $\gamma$~Dor 	&		&		&		&		&		&		\\
118114352	&	3772	&	susp. rot variable$^{+}$	&		&		&		&		&	30	&		\\
129636548	&	203585	&	new rot var	&		&	3.11016	&	0.00056	&	0.0052	&	8	&		\\
139191168	&	217522	&	roAp	    &		&		&		&		&		&		\\
141028198	&	35361	&	new rot var	&		&	6.3035	&	0.0009	&	0.013	&	30	&		\\
141610473	&	41613	&	new rot var	&		&	4.0954	&	0.0004	&	0.0029	&	23	&		\\
144276313	&	221760	&	$\alpha^2$~CVn; susp. new $\delta$~Sct / $\gamma$~Dor 	&	12.5	&		&		&		&		&	4,2	\\
152086729	&	224962	&	susp. new $\delta$~Sct / $\gamma$~Dor 	&		&		&		&		&		&		\\
152808505	&	216641	&	new roAp star	&		&	1.876660$^*$	& 0.00895	&		&		&		\\
159834975	&	203006	&	$\alpha^2$~CVn	        &	2.12204	&	2.12230	&	0.00009	&	0.0025	&	10	&	17	\\
167695608	&		    &	roAp	    &		&		&		&		&		&		\\
167751145	&	52280	&		        &		&		&		&		&	42	&		\\
182909257	&	6783	&	new rot var	&		&	3.14108	&	0.00082	&	0.0044	&	18	&		\\
183802606	&	8700	&	new rot var	&		&	2.27015	&	0.00017	&	0.0022	&	25	&		\\
183802904	&	8783	&	$\alpha^2$~CVn	        &	19.396 $\pm$ 0.05	&	19.408	&	0.017	&	0.04	&	14	&	6	\\
206461701	&	209364	&		        &		&		&		&		&	36	&		\\
206648435	&	215983	&	new rot var	&		&	5.1094	&	0.0022	&	0.018	&	42	&		\\
207208753	&	20505	&	$\alpha^2$~CVn	        &	2.04401 $\pm$ 0.00005	&	2.04334	&	0.00019	&	0.0012	&	40	&	6	\\
211404370	&	203932	&	roAp; new rot var	    &		&	6.442	&	0.012	&	0.083	&		&		\\
219340705	&	222349	&	new rot var	&		&	5.129	&	0.036	&	0.0064	&	16	&		\\
231844926	&	10840	&	$\alpha^2$~CVn	        &	2.09784	&	2.0971	&	0.0001	&	0.0012	&	8	&	17	\\
232066526	&	11090	&	new rot var	&		&	2.91982	&	0.00016	&	0.0082	&	46	&		\\
234346165	&	16504	&	new rot var	&		&	3.3040	&	0.0003	&	0.0091	&	19	&		\\
235007556	&	221006	&	$\alpha^2$~CVn	        &	2.3148 $\pm$ 0.0004	&	2.31206	&	0.00036	&	0.0022	&	18	&	4,11	\\
237336864	&	218495	&	roAp; new rot var	&		&	4.2006	&	0.0001	&	0.0037	&		&		\\
262613883	&	63728	&	$\alpha^2$~CVn	        &	1.83993 $\pm$ 0.00004	&	1.84015	&	0.00017	&	0.00075	&	33	&	5	\\
262956098	&	3988	&		        &		&		&		&		&	15	&		\\
266905315	&	225234	&		        &		&		&		&		&	24	&		\\
270304671	&	209605	&	$\alpha^2$~CVn	        &	7.8132 $\pm$ 0.0007	&	7.8896	&	0.0050	&	0.061	&	38	&	6	\\
271503787	&	2883	&	new rot var	&		&	6.056	&	0.013	&	0.024	&	18	&		\\
277688819	&	208217	&	$\alpha^2$~CVn; noAp	&	8.44475 $\pm$ 0.00011	&	8.3200	&	0.0084	&	0.057	&	13	&	4,7	\\
277748932	&	208759	&	new rot var	        &		&	4.4501	&	0.0019	&	0.0036	&	43	&		\\
278804454	&	212385	&	$\alpha^2$~CVn	        &	2.48 $\pm$ 0.04	&	2.5062	&	0.0002	&	0.0027	&	14	&	4,12	\\
279091054	&	50861	&		        &		&		&		&		&	30	&		\\
279573219	&	54118	&	$\alpha^2$~CVn 	    &	3.27533	&	3.2724	&	0.0010	&	0.00089	&	8	&	17	\\
280051011	&	18610	&		        &		&		&		&		&	15	&		\\
281668790	&	3980	&	$\alpha^2$~CVn; noAp	&	3.9516 $\pm$ 0.0003	&	3.9517	&	0.0001	&	0.00054	&	6	&	4,13	\\
304096024	&	11346	&	new rot var	&		&	7.116	&	0.006	&	0.097	&	38	&		\\
306573201	&	66195	&	new rot var	&		&	4.88938	&	0.00063	&	0.0057	&	17	&		\\
\hline
\end{tabular}
\label{rot}
\end{table*}

\begin{table*}
\centering
\contcaption{Variability properties of the 83 stars analysed in this work.}
\small
\begin{tabular}{rrcrrrrrr}
\\
\hline
\multicolumn{1}{c}{TIC}	&	\multicolumn{1}{c}{HD}	&	\multicolumn{1}{c}{Variability type}	&	\multicolumn{1}{c}{$P_{\rm rot}{\rm (lit)}$}	&	\multicolumn{1}{c}{$P_{\rm rot}$}	&	\multicolumn{1}{c}{$\Delta$$P_{\rm rot}$}	&	\multicolumn{1}{c}{SDV} & 	\multicolumn{1}{c}{Pulsation limit} 	&	\multicolumn{1}{c}{Ref}	\\
    &       &                       &    \multicolumn{1}{c}{(d)}            &    \multicolumn{1}{c}{(d)}            &      \multicolumn{1}{c}{(d)}                  &  \multicolumn{1}{c}{(d)}  &   \multicolumn{1}{c}{($\umu$mag)}      &        \\           
\hline306893839	&	68561	&	$\alpha^2$~CVn 	&	4.2334	&	4.23415	&	0.00016	&	0.0017	&	18	&	3,17	\\
307031171	&	69578	&		&		&		&		&		&	19	&		\\
307288162	&	71006	&	new rot var	&		&	1.52073	&	0.00026	&	0.000051	&	40	&		\\
307642246	&	72634	&	$\alpha^2$~CVn; new rot per	&	0.93062 $\pm$ 0.00001	&	1.8607	&	0.0002	&	0.00083	&	12	&	6	\\
308085294	&	74388	&	new rot var	&		&	4.3063	&	0.0019	&	0.0051	&	11	&		\\
309148260	&	69862	&	$\alpha^2$~CVn; new rot per	&	0.518885 $\pm$ 0.000003	&	13.3519	&	0.0107	&	0.25	&	48	&	6	\\
316913639	&	222638	&	new rot var	&		&	2.34691	&	0.00026	&	0.0021	&	18	&		\\
327597288	&	206653	&	$\alpha^2$~CVn	&	1.788 $\pm$ 0.005	&	1.786898	&	0.000058	&	0.00094	&	13	&	4,14	\\
336731635	&	214985	&	$\alpha^2$~CVn; new rot per	&	1.3851 $\pm$ 0.0001	&	2.77342	&	0.00219	&	0.0019	&	70	&	8	\\
348717688	&	19918	&	roAp	&		&		&		&		&		&		\\
348898673	&	54399	&	$\alpha^2$~CVn; new rot per 	&	2.50142 $\pm$ 0.00008	&	4.9910	&	0.0011	&	0.0060	&	25	&	6	\\
349409844	&	58448	&	$\alpha^2$~CVn 	&	0.831 $\pm$ 0.003	&	0.83088	&	0.00005	&	0.00017	&	10	&	4,15	\\
350146296	&	63087	&	new roAp star	&		&	2.66121	&	0.00029	&	0.0012	&		&		\\
350146577	&	63204	&	$\alpha^2$~CVn	&	1.83817 $\pm$ 0.00002	&	1.83764	&	0.00004	&	0.00036	&	20	&	5	\\
350272314	&	222925	&		&		&		&		&		&	20	&		\\
350519062	&	38719	&	rot var	&	4.02107	&	4.0237	&	0.0004	&	0.0034	&	9	&	10	\\
358467700	&	65712	&	$\alpha^2$~CVn	&	1.9457	&	1.94639	&	0.00054	&	0.00047	&	55	&	4	\\
364424408	&	30374	&	$\alpha^2$~CVn	&	1.55631 $\pm$ 0.00003	&	1.55682	&	0.00014	&	0.00059	&	50	&	6	\\
372913684	&	65987	&	$\alpha^2$~CVn	&	1.44962 $\pm$ 0.00018	&	1.45610	&	0.00015	&	0.00033	&	15	&	4,16	\\
382512330	&	64369	&	$\alpha^2$~CVn	&	0.89113	$\pm$ 0.00001 &	0.8912	&	0.0001	&	0.00012	&	25	&	5	\\
389531041	&	206193	&	roAp candidate$^{++}$; new rot var	&		&	6.030	&	0.022	&	0.041	&		&		\\
389922504	&	40277	&	new rot var	&		&	0.849585	&	0.000008	&	0.00035	&	14	&		\\
391927730	&	56981	&	new rot var	&		&	3.7843	&	0.0018	&	0.0021	&	22	&		\\
392761412	&	207259	&	$\alpha^2$~CVn 	&	2.16 $\pm$ 0.01	&	2.1557	&	0.0002	&	0.0020	&	23	&	4,9	\\
394045029	&	211333	&		&		&		&		&		&	17	&		\\
394124612	&	218994	&	roAp; new rot per 	&	1.09058	&	5.855	&	0.008	&	0.031	&		&	10	\\
407661867	&	37584	&	susp. rot variable$^+$, roAp candidate	&		&		&		&		&		&		\\
410451752	&	66318	&	new rot var	&		&	0.77688	&	0.00052	&	0.00021	&	57	&		\\
431380369	&	20880	&	new roAp; new rot var	&		&	5.2434	&	0.0026	&	0.042	&		&		\\
434103853	&	221531	&	new rot var	&		&	3.2584	&	0.0066	&	0.0023	&	15	&		\\
\hline
\multicolumn{9}{l}{$^{+}$: The star shows two low frequency peaks, but it is unclear whether either of them is a rotation signal {(see text for details)}.}\\
\multicolumn{9}{l}{$^{++}$: The formal significance of the pulsation detection is 4.9 sigma, thus marginal.}\\
\multicolumn{9}{l}{$^{*}$: Rotation period inferred from the analysis of pulsations (see text for details).}\\
\multicolumn{9}{l}{Ref: 1: \cite{renson09}; 2: \cite{catalano98b}; 3: \cite{paunzen98}; 4: \cite{samus17}; 5: \cite{bernhard15a};}\\
\multicolumn{9}{l}{6: \cite{hummerich16}; 7: \cite{manfroid1997} ; 8: \cite{bernhard15b}; 9: \cite{renson01};}\\
\multicolumn{9}{l}{10: \cite{TESSVC18}; 11: \cite{manfroid85}; 12: \cite{renson78}; 13: \cite{maitzen1980}; 14: \cite{Hensberge77};}\\
\multicolumn{9}{l}{15: \cite{manfroid1983}; 16: \cite{north84}; 17: \cite{1997ESASP1200.....E}; 18: \cite{manfroid81}.}
\end{tabular}
\end{table*}

\subsubsection{{Contamination and multiplicity}}
\label{cont}
%{In an attempt to rule out any false positive claims of variability, we checked the literature for discussion on stellar binarity or multiplicity. For most stars, there is no mention in the literature of multiplicity and we discuss here, on a star by star basis, those where binary contamination may be a factor. Furthermore, as the TESS pixels are large ($20.25$\,arcsec) there is a relatively high possibility of contamination from nearby sources. This is summarised by the contamination ratio for each star as provided by the TIC. For the 83 stars we study here, the contamination ratio is below 0.1 for all but one star (TIC\,410451752). This gives us confidence that the variability we report is for the target star in question, and not associated with a contaminating source.}

{We found TIC\,410451752 to be a new rotationally variable star with a period of $0.77688\pm0.00052$\,d, {which we measured from the rotation frequency and its harmonic}. {This star has a contamination ratio of 0.8}.  To ensure we attributed this signal to the correct star, we investigated the surrounding bright objects which may have fallen in the photometric aperture. We found that one, TIC\,410451777 (HD\,66295) is classified as B8/9p\,Si by \citet{hartoog1976}. The rotation period is known to be 2.45\,d \citep{netopil2017} which is significantly different from what we find for TIC\,410451752. The other stars in the region have no rotation period reported in the literature. We are thus confident we attributed the rotation signal to the correct star, {although we cannot say for certain}.}

{Of the other stars for which we find new {rotational variability}, there are 6 stars with discussion in the literature pertaining to multiplicity. {TIC\,152808505 and TIC\,394124612 are also pulsational variables, hence we discuss their cases in dedicated sections (Sec.~\ref{sec:opm} and Sec.~\ref{cont2}, respectively) below.}

TIC\,12393823: this star is a known spectroscopic binary with a period of {5.400945\,$\pm$\,0.000040\,d} and an eccentricity of $e=0.267\pm0.008$ \citep{gonzalez2009}. This period is not associated with the rotation period we derive here {($P=1.42353\pm 0.00023$\,d)}. There is no mention of the companion to TIC\,12393823 in the literature, indicating a less luminous object. We therefore assume that the derived rotation period is for the primary component of TIC\,12393823, i.e. the A1\,SiSr star.}

{TIC\,115150623: this star is reported in the literature as an sdB+F binary \citep{kilkenny2015}. However, this is cast into doubt when considering the Gaia parallax of $4.03\pm0.08$\,mas. In the absence of extinction, this provides an absolute magnitude of about $+1.7$ which corresponds to a mid-A star. Since the composite spectrum of an sdB+F binary shows both signatures of the hot star and cool star, a combined apparent magnitude would be between $V=3$ and $V=5$. Finally, TIC\,115150623 was classified as Ap\,CrEuSr by \citet{houk1982}, which is consistent with the rotation period {of $P=1.50048\pm 0.00005$\,d} we found.}

{TIC\,129636548: the multiplicity of this star has been reported many times. The Washington Double Star Catalogue \citep{mason2001} provides information on three components in the system, with two being of similar magnitude (AB; $V=6.24$ and $V=6.88$), and a fainter third (C; $V=10.30$). {The AB pair {has} been observed to have a variable separation (0.8\,arcsec in 1879 and 0.3\,arcsec in 2014), with the AC separation being about 80\,arcsec over the same period. Given the pixel size of TESS, and the photometric aperture, it is expected that all three components will be contributing to the recorded flux.} \citet{malkov2012} provide an orbital period of 464\,{yr} for the two bright components. There is only one spectral type provided for the entire system: A0\,Si. Therefore, we cannot say which of the three stars the determined rotation period of 3.11\,d is intrinsic to, but assume it is either the A or B component; {given the classification and magnitude difference, component C is likely a G star, which is not expected  to have the stable spots that  we observed.}}

{TIC\,434103853: \citet{north1991} first noted this star to be an SB1 system. \citet{makarov2005} noted that the star is an astrometric binary, for which \citet{pourbaix2000} gave a period of 1416\,d and an eccentricity of 0.165. There is no discussion of the secondary in the literature, but we estimate here a lower limit on the mass of 0.6\,M$_\odot$ {from the results of \citet{pourbaix2000}}. Therefore, depending on the inclination, it is possible that the rotation signal we derive here (3.25\,d) belongs to either the primary or the secondary component.}

\subsection{Pulsational variability}
\label{puls}

 {Five new roAp stars have been identified in our sample, in addition to the seven previously known roAp stars in TESS sectors 1 and 2. Of the five new roAp stars, three were discovered among the list of 80 Ap stars submitted for observation with the TESS 2-min cadence in these sectors. The other two roAp stars, TIC\,152808505 and TIC\,350146296, and the roAp candidate TIC\,407661867, were discovered also among the stars observed in 2-min cadence by TESS, but are A stars not classified as peculiar in the literature. Considering that only the sample of stars previously classified as peculiar was thoroughly searched for roAp-type pulsations in this work, we deduce an incidence of the roAp phenomena of about 4~per~cent among the Ap stars (corresponding to 3 in 73 stars). }
 
 The roAp stars discovered in TESS data are shown as red squares in Fig.\,\ref{HRdiagram}. Their global parameters are within those of the 61 roAp stars known prior to the launch of TESS, bracketed by the orange-dashed lines seen in the same figure. One of them is located outside the theoretical instability strip, marked by the {straight blue} lines. Moreover, none is close to the theoretical blue edge.
 
 Among the stars found to exhibit pulsational variability, there are also two classified as ``roAp candidate" and a few classified as {suspected} $\gamma$\,Dor and/or $\delta$\,Sct stars. We have made a note of these in {the third column of} Table\,\ref{rot}, but leave the full exploitation of the data on those pulsators to forthcoming publications. In particular, for the roAp candidate TIC\,407661867, we found signals of pulsations in the range 0.7 to 0.74\,mHz (periods around 23\,min) which could be $\delta$\,Sct in nature, but could also be roAp. However, the star has not been classified as chemically peculiar; its spectral type is A3\,V in \cite{houk1975}. \cite{McDonald12} obtained $T_{\rm eff}$ = 8430\,K for this star by comparing model atmospheres to spectral energy distributions (SEDs) derived through the combination of data from different sources, and a {luminosity of $\log (L/{\rm L}_\odot)=1.23$, based on the Hipparcos parallax. In this work we find a temperature of $T_{\rm eff} = 8750 \pm 180$\,K and a luminosity based on Gaia data of $\log (L/{\rm L}_\odot)=1.10\pm 0.05$}. The high effective temperature of this star, along with either of the above luminosities, makes it relatively unevolved. For a star with such properties, the expected roAp frequencies are significantly higher than the ones observed \citep[cf. figure\,4, in][]{cunha02}, thus putting into question the roAp origin of the pulsations. As the star is in the TESS continuous viewing zone, we will leave the confirmation, or otherwise, of this roAp candidate to a later stage. 
 
{In the case of the roAp candidate TIC\,389531041 (HD\,206193), we detected a pulsation mode with a frequency of $1.21012\pm 0.00005$\,mHz, in the typical range of the roAp stars. While the signal at this frequency was reported by three independent teams, its significance is only of $4.9\,\sigma$, making the identification of this star as roAp insecure. Its peculiarity is also unclear. It was classified by \cite{renson91} as F5\,Sr, but with a ``peculiarity probability note" of ``doubtful nature". The effective temperature derived here, $T_{\rm eff}= 6495\pm152$\,K, places the star outside the theoretical instability strip, but there are other known roAp stars with similar effective temperatures.  TIC\,389531041 was observed during sector 1, and will not be revisited in TESS's primary mission. Thus, unfortunately, TESS data will not allow us to test the roAp nature of this star during the nominal mission.}

 Finally, for 63 stars in our sample, no pulsational variability was found. For those stars, the 8$^{\rm th}$ column in Table\,\ref{rot} provides the detection limit as defined in Sec.\,\ref{dataAna}. We note, however, that these limits to pulsational variability are in the TESS photometric band, where pulsation amplitudes in roAp stars are significantly smaller than in the $B$ filter most often used in ground-based campaigns. Further discussion on the amplitude comparison between the TESS and $B$ filters is presented in Sec.\,\ref{sec:comp}, based on the analysis of previously known roAp stars. 
 
 In the following sections, we provide details of the analysis and results for 14 stars in our sample found to be of particular interest. These consist of the {five} new roAp stars (Sec.\,\ref{new}), seven previously known roAp stars (Sec.\,\ref{old}), and two well characterised noAp stars (Sec.\,\ref{no}). 

\section{{The} new TESS roAp stars}
\label{new}

\subsection{TIC\,12968953}

TIC\,12968953 (HD\,217704) was classified as A7 by \citet{philip72}. Later, \citet{renson91} \citep[see also,][]{renson09} identified the peculiar nature of this star, classifying it as A5\,Sr. The star was among the targets of the Cape Survey for rapid oscillations in Ap stars \citep[see e.g.][]{martinez1993,martinez1991}, but no evidence was found for high-overtone pulsations at that time \citep{martinez1994}. Its effective temperature of $7880\pm160$\,K, derived in this work, places it on the hot side of the temperature range for the known roAp stars.

\begin{figure}
\centering
\includegraphics[width=0.9\columnwidth]{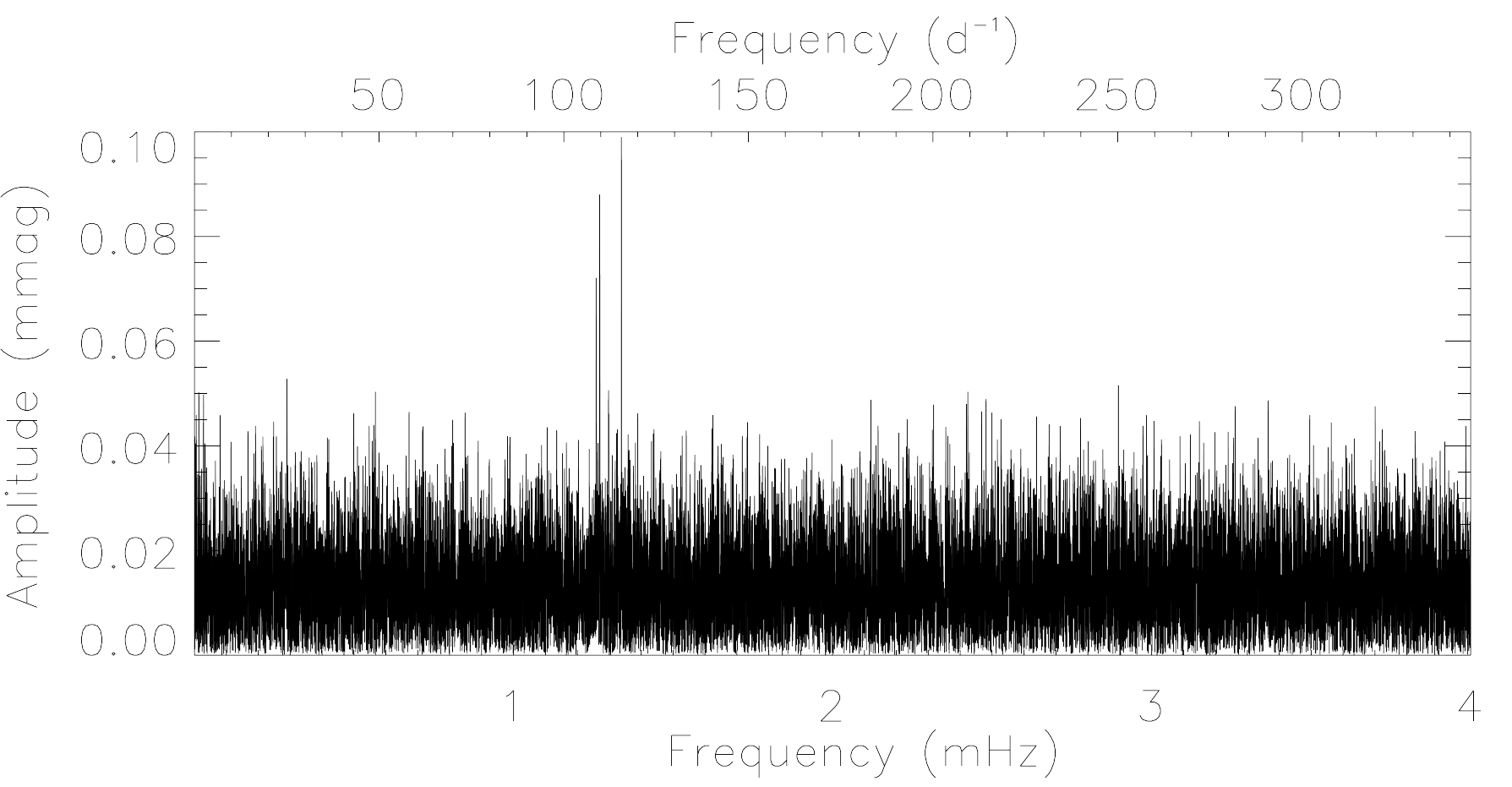}
\includegraphics[width=0.9\columnwidth]{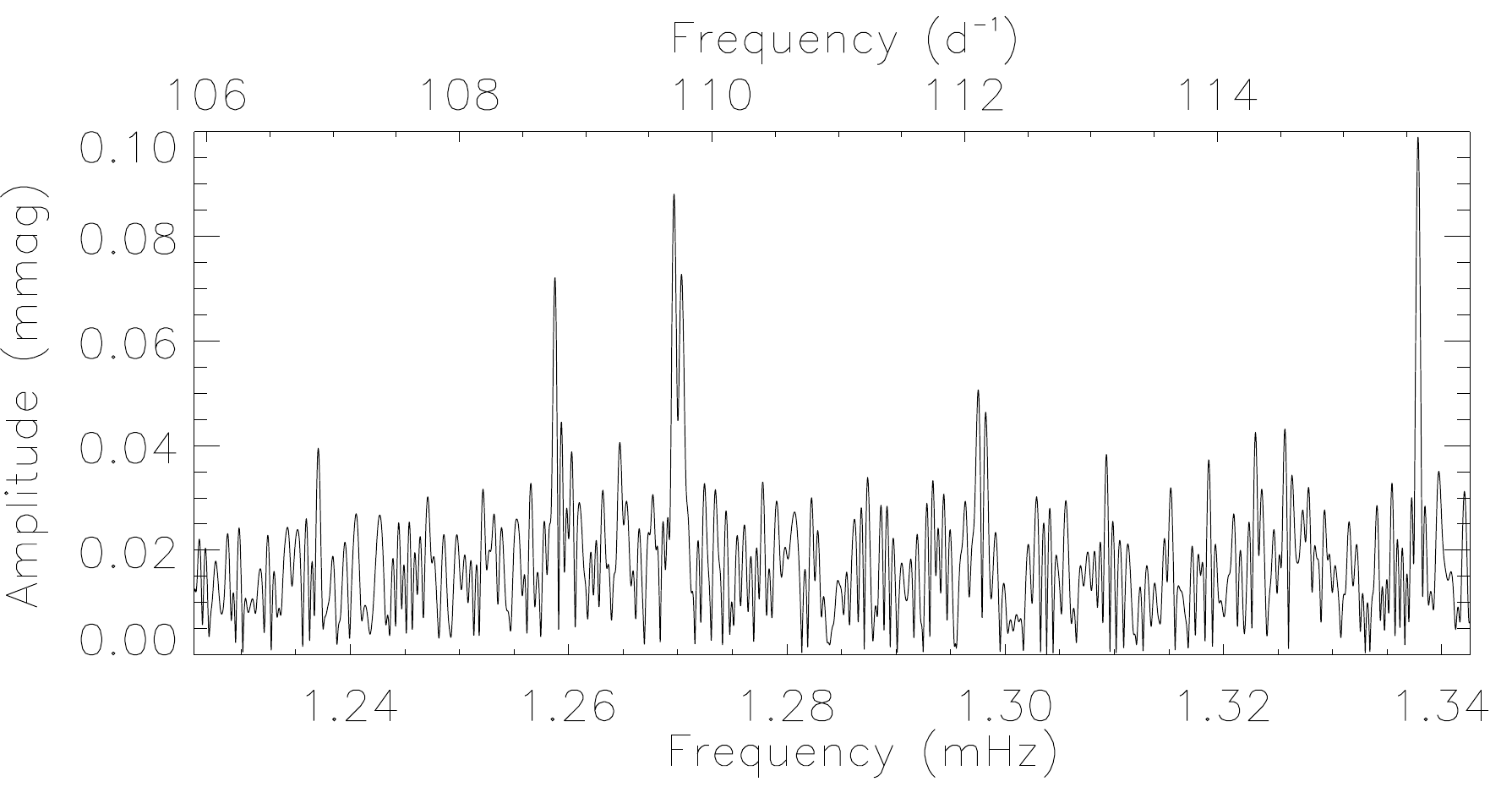}
\includegraphics[width=0.9\columnwidth]{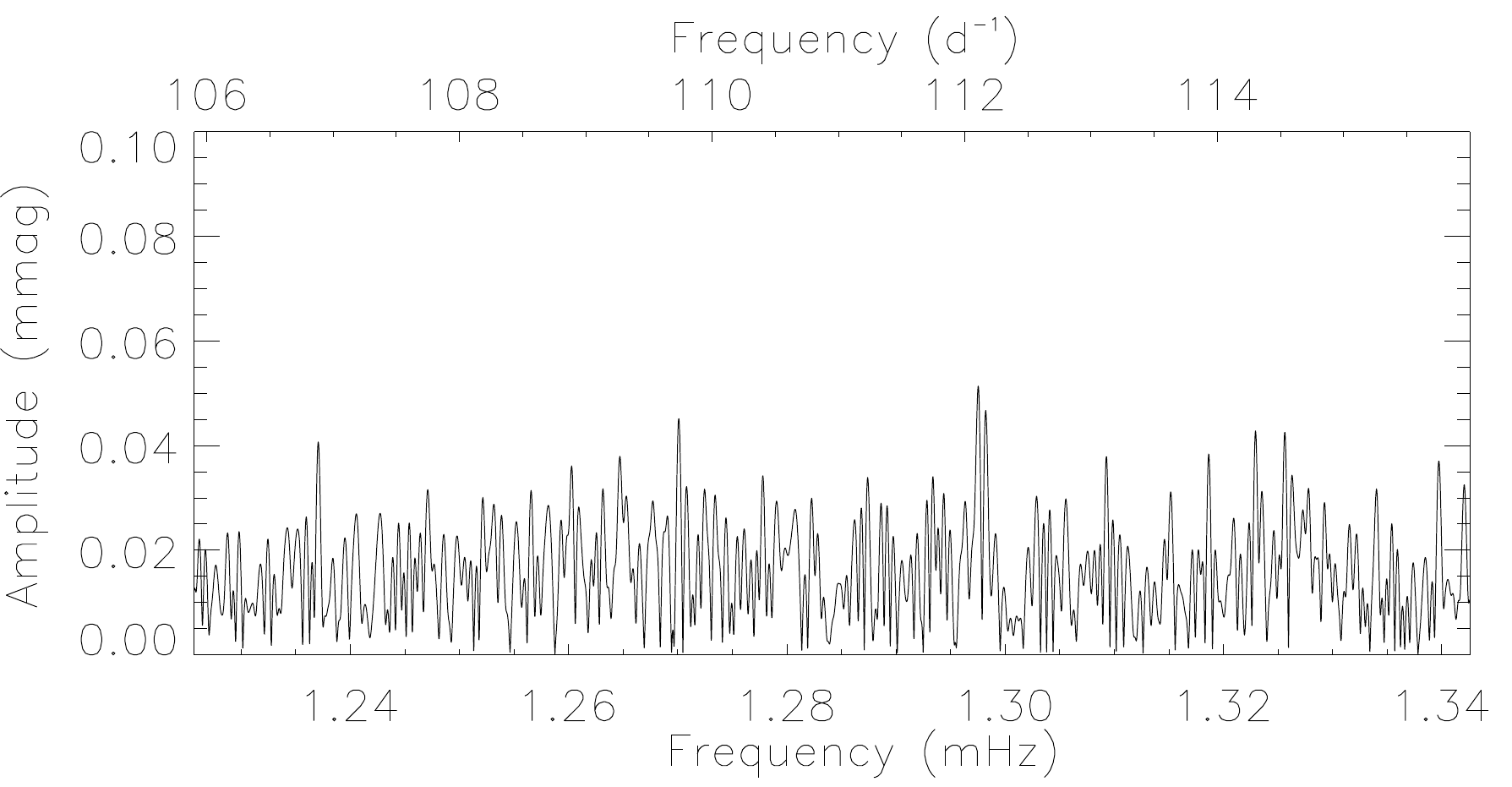}
\caption{Top: full amplitude spectrum of TIC\,12968953 to almost the Nyquist frequency. Clearly evident are the pulsations around 1.3\,mHz. Middle: zoom of the pulsation mode frequencies. Bottom: the amplitude spectrum of the residuals after removing the frequencies shown in Table\,\ref{tab:12968953}, showing no remaining significant signals.}
\label{fig:12968953_ft}
\end{figure}

We analysed TESS sector 2 data for this star and found no indication of rotational modulation of the light curve. This null detection could be a result of a long rotation period, or a rotational inclination or magnetic obliquity near to zero. However, there is clear evidence of four pulsation modes in this star, thus making it a new roAp star. There is no evidence of multiplets in the amplitude spectrum, consistent with a lack of rotational modulation {of the light curve}, nor is there amplitude modulation of the modes. In Fig.\,\ref{fig:12968953_ft} we show the full amplitude spectrum, a detailed view of the mode frequencies, and the amplitude spectrum after removing the mode frequencies displayed in Table\,\ref{tab:12968953}. {There is no clear separation pattern between the mode frequencies which could represent the large or small frequency separations. Unfortunately, the amplitudes in this star are low, so it is unlikely that ground-based $B$ observations will provide more information on the pulsations in TIC\,12968953.}

\begin{table}
\centering
\caption{Details of the pulsation modes found in TIC\,12968953. The zero-point for phases is BJD\,2458367.81699.}
\label{tab:12968953}
\begin{tabular}{lccrr}
\hline
ID & Frequency & Amplitude & \multicolumn{1}{c}{Phase} & \multicolumn{1}{c}{S/N}\\
    & (mHz)  		 & (mmag) & \multicolumn{1}{c}{(rad)}&\\
   &  &  \multicolumn{1}{c}{$\pm 0.013$} & &\\
\hline
$\nu_1$ & $1.25875\pm0.00004$ &	$0.071$ &	$1.45\pm0.17$ & 5.9\\
$\nu_2$ & $1.26961\pm0.00003$ &	$0.083$ &	$-0.02\pm0.15$& 6.4\\
$\nu_3$ & $1.27047\pm0.00005$ &	$0.064$ &	$2.50\pm0.21$ & 4.9\\
$\nu_4$ & $1.33782\pm0.00003$ &	$0.099$ &	$-0.47\pm0.12$ & 8.3\\
\hline
\end{tabular}
\end{table}

%Two modes, $\nu_1$ and $\nu_2$, are separated by 10.86\,$\umu$Hz which may represent the small frequency separation, $\delta\nu$. However, with just two modes, this is conjecture. 

\subsection{TIC\,41259805}

\begin{figure}
\centering
\includegraphics[width=0.9\columnwidth]{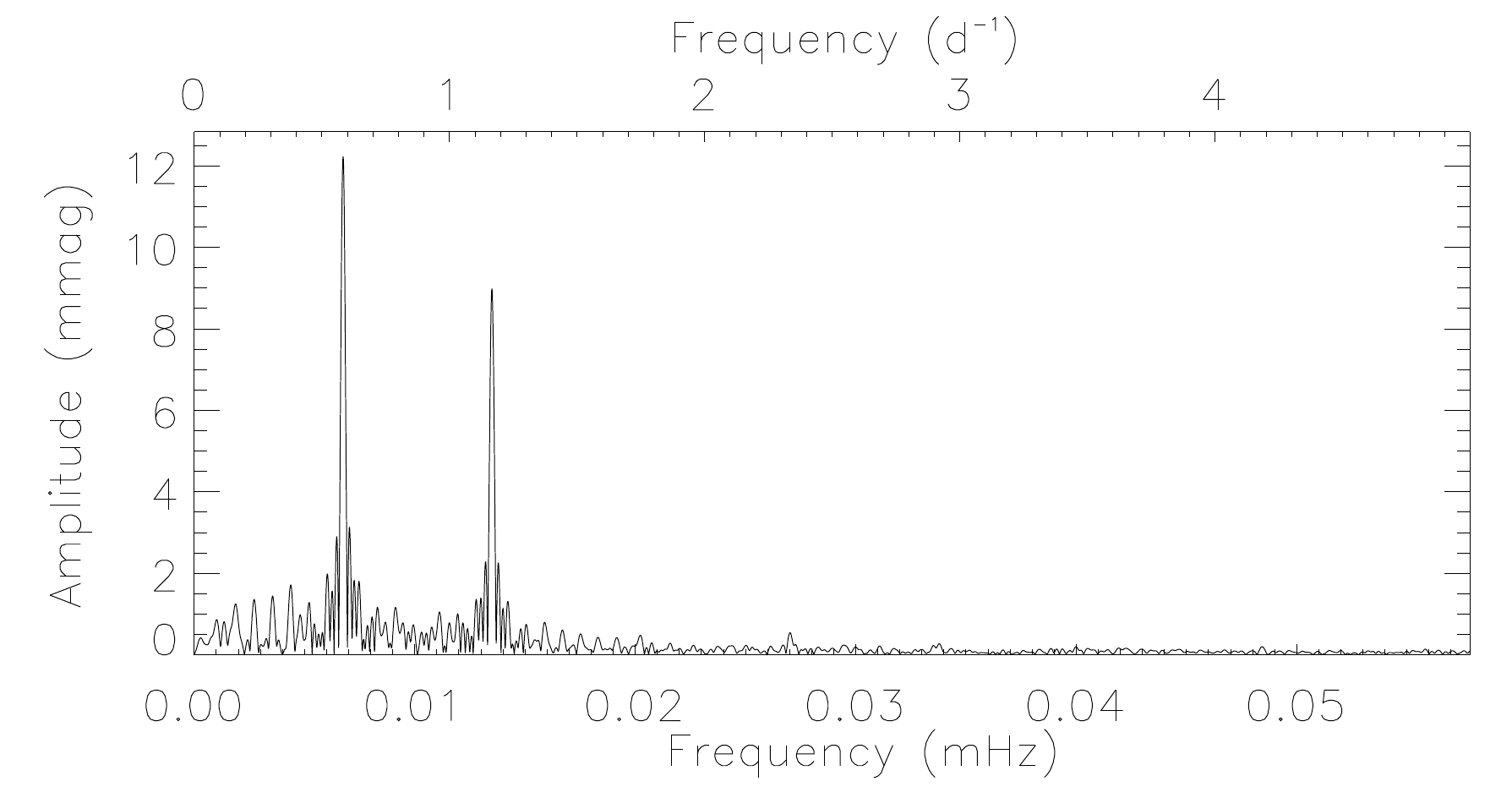}
\includegraphics[width=0.9\columnwidth]{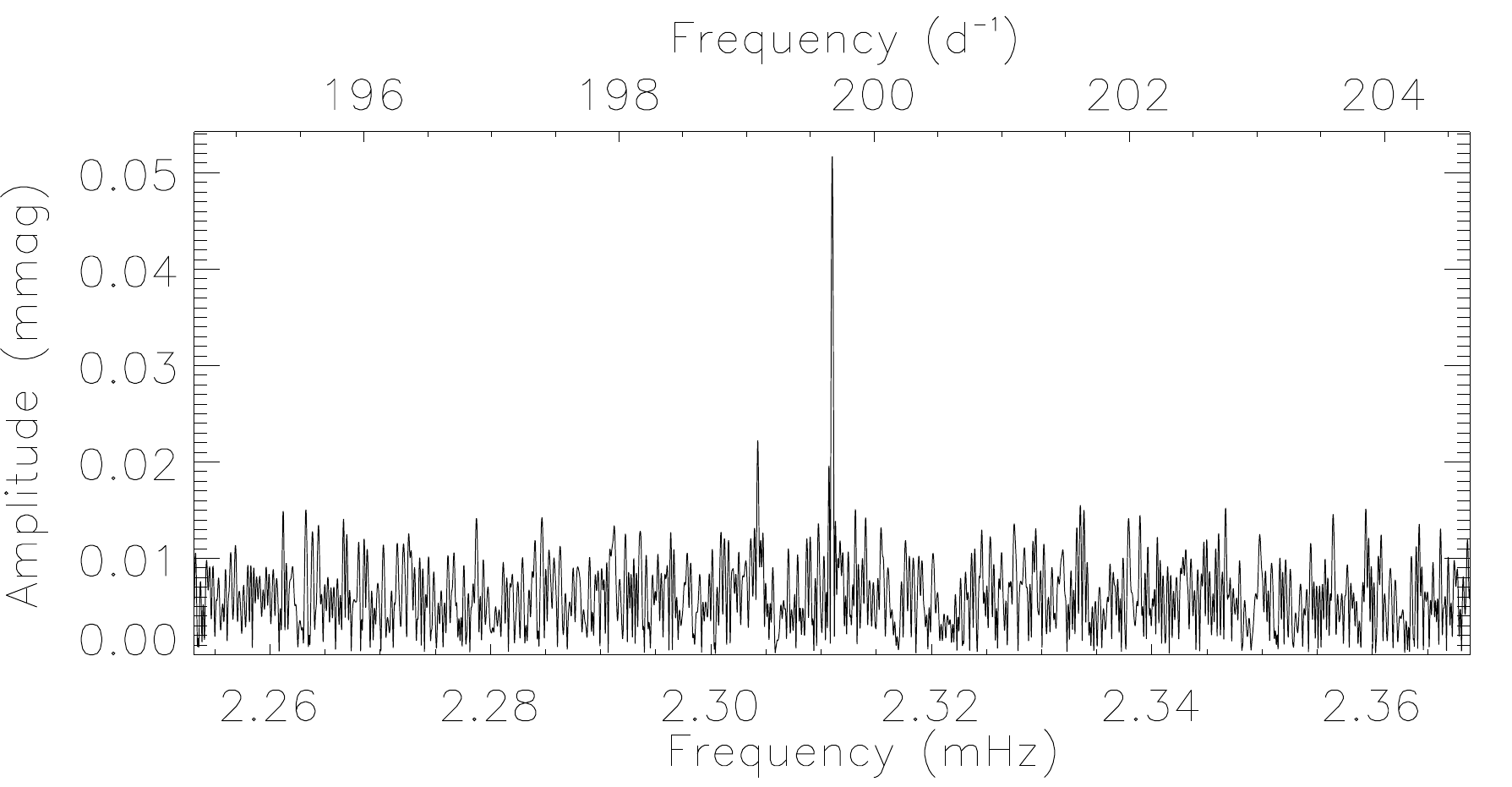}
\includegraphics[width=0.9\columnwidth]{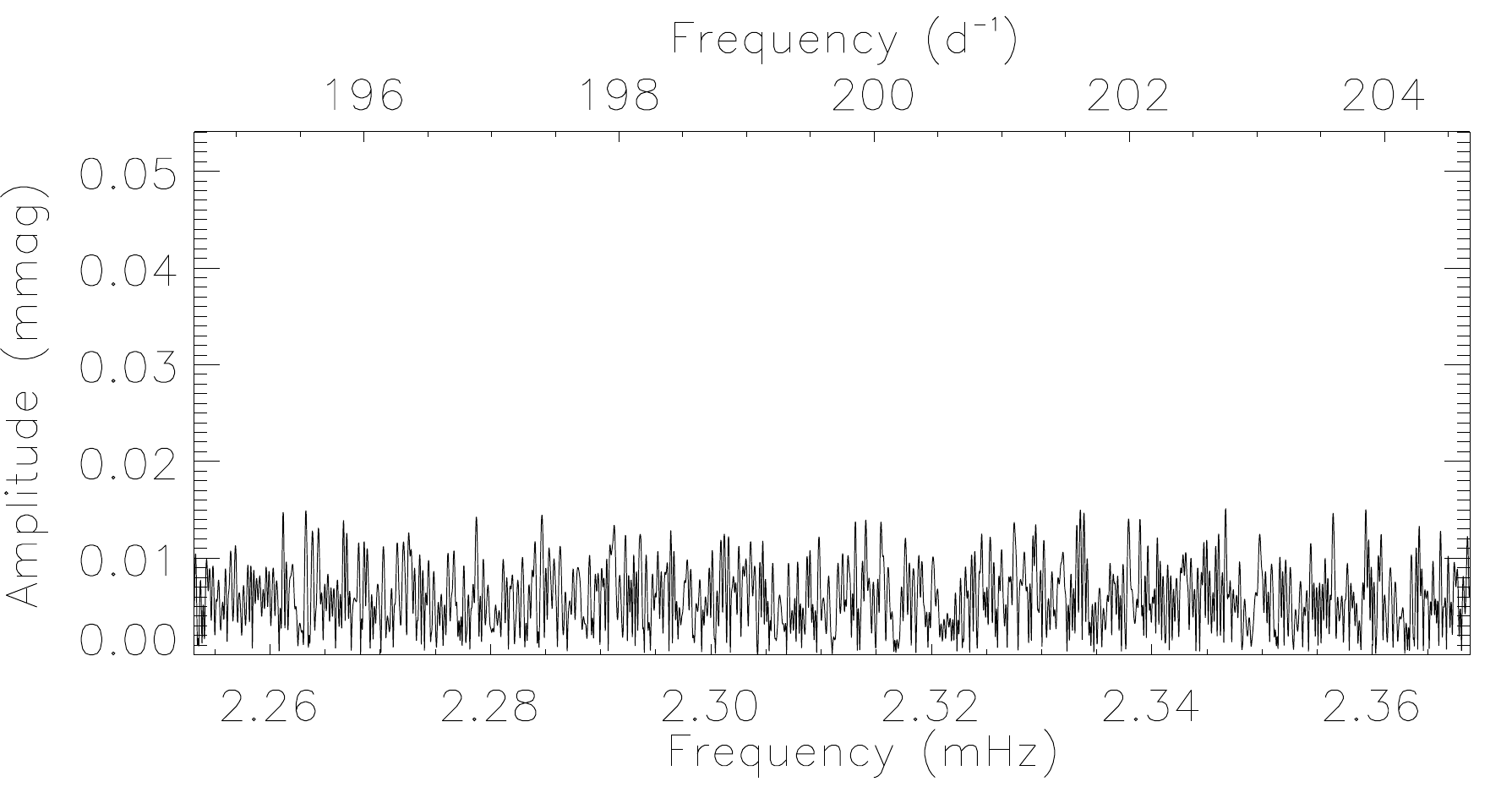}
\caption{Top: amplitude spectrum at low frequency showing the rotation signal in TIC\,41259805. The rotation frequency is the highest amplitude peak. Middle: zoom of the pulsation frequencies found in this star. Bottom: the amplitude spectrum of the residuals after removing the frequencies shown in Table\,\ref{tab:41259805} showing no remaining significant signals.}
\label{fig:41259805_ft}
\end{figure}

\begin{table}
\centering
\caption{Details of the pulsation frequencies found in TIC\,41259805. The zero-point for phases is BJD\,2458353.40492.}
\label{tab:41259805}
\begin{tabular}{lccrr}
\hline
ID & Frequency & Amplitude & \multicolumn{1}{c}{Phase} & \multicolumn{1}{c}{S/N}\\
    & (mHz)  		 & (mmag) & \multicolumn{1}{c}{(rad)}\\
       &  &  \multicolumn{1}{c}{$\pm 0.005$} & &\\
\hline
$\nu-\nu_{\rm rot}$ & $2.30425\pm0.00002$ &	$0.022$ & $-1.06\pm0.22$ & 4.4\\
$\nu$ & $2.31100\pm0.00001$ & $0.052$ & $-1.84\pm0.09$ & 10.4\\
\hline
\end{tabular}
\end{table}

TIC\,41259805 (HD\,43226) was classified Ap Sr(Eu) by \citet{houk1975} and A0 SrEu by \citet{renson91}.  Its fundamental parameters were obtained by \citet{McDonald12} by comparing model atmospheres to spectral energy distributions inferred from different data sources. They determined $T_{\rm eff}=8054$\,K, which is similar to the TIC effective temperature, and $L/{\rm L}_\odot = 13.74$, which is between the two values presented in Table\,\ref{properties}. The effective temperature determined in this work, $T_{\rm eff}=8290\pm170$\,K, places this star among the hottest roAp stars known. 

TIC\,41259805 was observed during both sectors 1 and 2, and will continue to be observed for the rest of cycle 1 apart from sector 9. There is a clear rotation signal in this star, as shown by the amplitude spectrum in the top panel of Fig.\,\ref{fig:41259805_ft}, corresponding to a rotation period of $1.71441\pm0.00011$\,d. {This is the shortest measured rotation period for an roAp star.} The double-wave nature of the light curve, as indicated by the  presence of the strong second harmonic in the amplitude spectrum, is an indication that we see both magnetic poles of this star during the rotation cycle (under the assumption that the spots are located at the magnetic poles).

There are two clear pulsation frequencies (Table\,\ref{tab:41259805}) in the amplitude spectrum of TIC\,41259805, as shown in the middle panel of Fig.\,\ref{fig:41259805_ft}. These peaks are split by the rotation frequency of the star, implying a triplet with a missing sidelobe at $\nu+\nu_{\rm rot}$. We fitted the assumed triplet with linear least-squares by forcing the sidelobe frequencies to be split by exactly the rotation frequency. {The high-frequency sidelobe is present, but not statistically significant at a S/N of 1.8.}

\subsection{TIC\,152808505}
\label{sec:opm}

TIC\,152808505 (HD\,216641) is one of the three stars in our list not identified as peculiar in the literature. It was classified F3\,IV/V by \citet{houk1978}. Through a Bayesian method, using parallaxes and multiband photometry, \citet{bailerJones11} derived two values for its effective temperature, from two different models: $T_{\rm eff}=6710\pm230$\,K and $T_{\rm eff}=6670\pm225$\,K. These values agree with the effective temperature of $6640\pm160$\,K derived in this work. This effective temperature places this star among the coolest roAp stars known and outside of the theoretical instability strip.

{Several sources in the literature cite TIC\,152808505 as a multiple system \citep[e.g.][]{turon1993,mason2001,fabricius2002}. The two components are separated by about 0.3\,arcsec and have Hipparcos magnitudes of 8.92 and 9.41. Therefore, the temperature measurements discussed above {have been} calculated using the combined flux of both stars.}

\begin{figure}
\centering
\includegraphics[width=0.9\columnwidth]{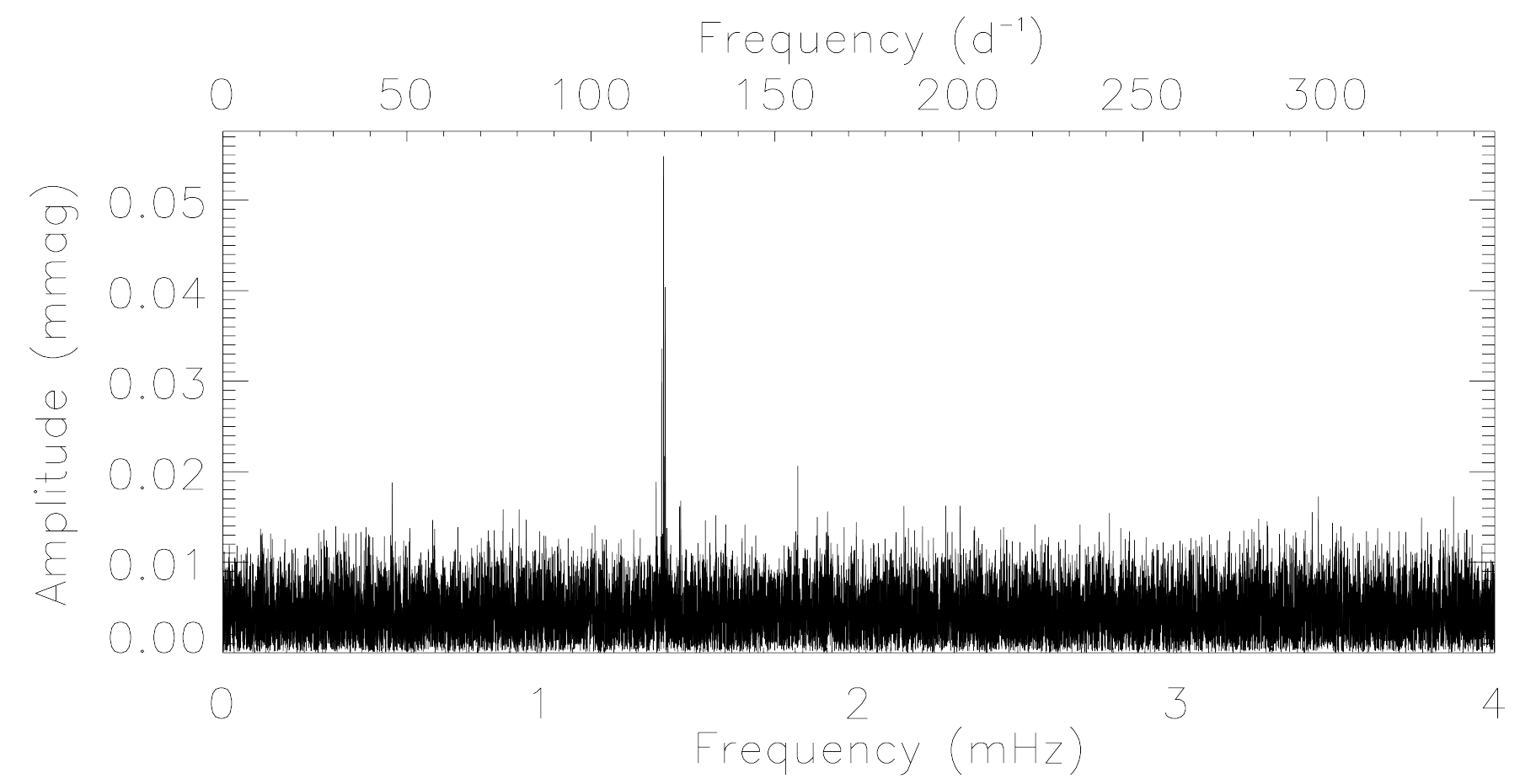}
\includegraphics[width=0.9\columnwidth]{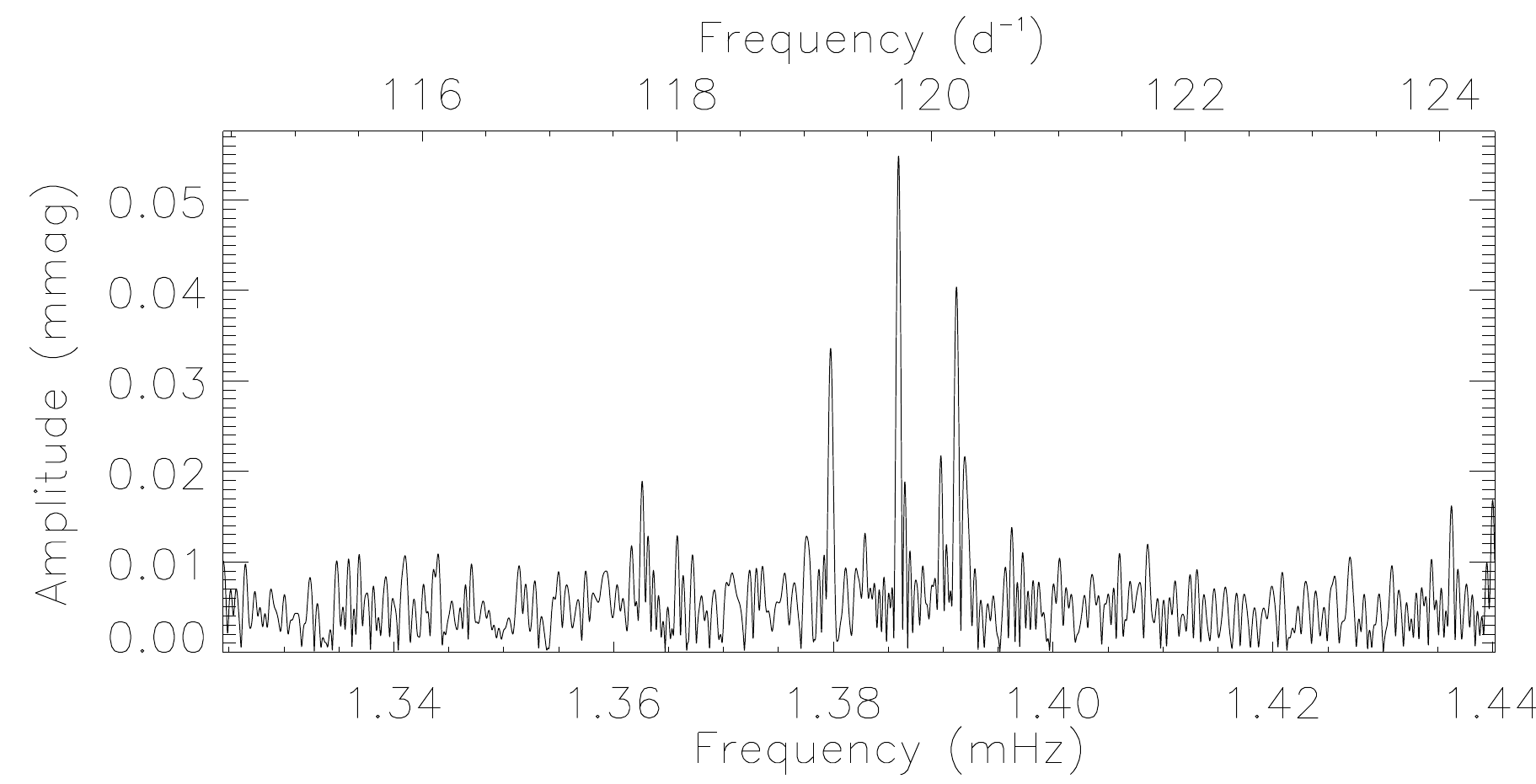}
\includegraphics[width=0.9\columnwidth]{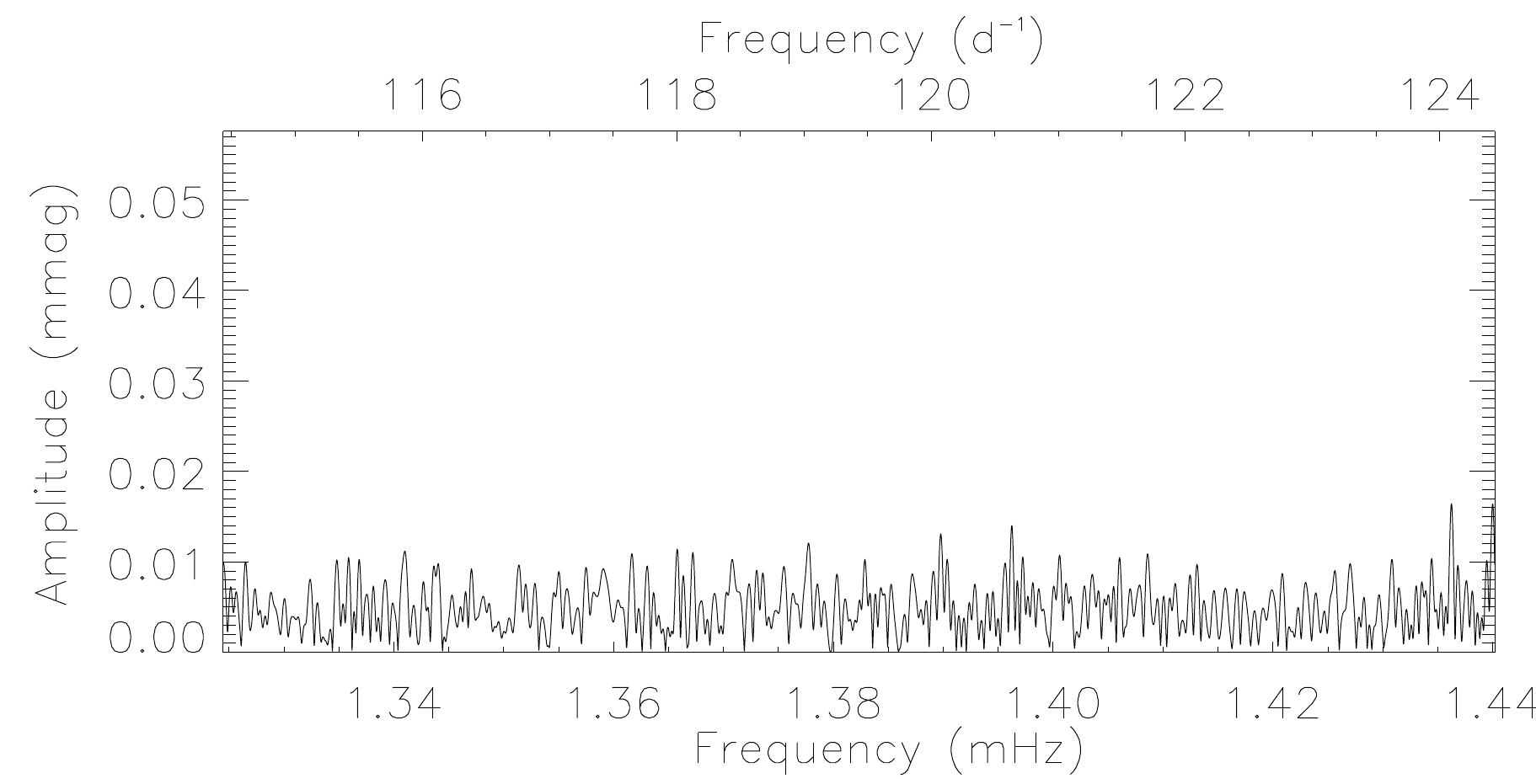}
\caption{Top: amplitude spectrum of TIC\,152808505 to almost the Nyquist frequency. Middle: zoom of the pulsation modes found in this star. Bottom: the amplitude spectrum of the residuals after removing the frequencies shown in Table\,\ref{tab:152808505_OPM} showing no remaining significant signals.}
\label{fig:152808505_ft}
\end{figure}

Observed in TESS sector 1, the data for TIC\,152808505 do not show low-frequency light variations that could be attributed to rotation. There are, however, clear signals of pulsation in this star (Fig.\,\ref{fig:152808505_ft}). We detected 5 significant peaks in the data, as are shown in Table\,\ref{tab:152808505_OPM}. There are three frequencies that are split by the same frequency which we interpret as a rotationally split triplet. This allows us to infer a rotation period of $1.8766\pm0.0090$\,d. Therefore, we find three independent modes in TIC\,152808505. {We assume all three modes are in a single star of this binary pair, however we cannot say which. We also note that the mode amplitudes will be diluted due to the flux from the other component.}

\begin{table}
\centering
\caption{Linear least-squares fit to the pulsations and force fitted sidelobes in TIC\,152808505. The zero-point for the fit is BJD\,2458340.30085, and has been chosen as such to force the sidelobe phases to be equal. 
}
\label{tab:152808505_OPM}
\begin{tabular}{llcrr}
\hline
ID & \multicolumn{1}{c}{Frequency} & Amplitude & \multicolumn{1}{c}{Phase} & \multicolumn{1}{c}{S/N}\\
    & \multicolumn{1}{c}{(mHz)}  		 & (mmag) & \multicolumn{1}{c}{(rad)}&\\
           &  &  \multicolumn{1}{c}{$\pm 0.004$} & &\\

\hline
$\nu_1$ & $1.36260\pm0.00005$ & $0.019$ & $2.16\pm0.21$ & 4.8\\
$\nu_2-\nu_{\rm rot}$ & $1.37976$ & $0.033$ & $1.09\pm0.12$ &8.3 \\
$\nu_2$ & $1.38593\pm0.00002$ & $0.055$ & $0.35\pm0.07$ & 13.8\\
$\nu_2+\nu_{\rm rot}$ & $1.39210$ & $0.020$ & $1.09\pm0.19$& 5.0 \\
$\nu_3$ & $1.39119\pm0.00002$ & $0.041$ & $-2.84\pm0.10$ & 10.3\\
 
\hline
\end{tabular}
\end{table}

{As discussed in Sec.\,\ref{sec:1}, the rotation axis in Ap stars is often inclined with respect to the magnetic axis. In the simple oblique pulsator model the pulsation axis is assumed to be coincident with the magnetic axis \citep{kurtz1982,ss85a,ss85b,dg85}, whereas in the ``improved'' oblique pulsator model the pulsation axis generally lies in the plane defined by the rotation and magnetic axes \citep{bigot02,bigot2011}, and is not necessarily coincident with the magnetic axis.} Oblique pulsation results in a changing view of the pulsation axis as the star rotates. Such a configuration allows constraints to be placed on the geometry of the star. These constraints are derived through the analysis of sidelobes to the pulsation mode which are separated by the rotation frequency of the star. For a pure dipole mode, one expects a triplet, and for a pure quadrupole mode, a quintuplet (i.e. a $2\ell+1$ multiplet).

The triplet here shows unequal sidelobe amplitudes which is a signature of the effect of Coriolis force on the pulsations \citep[e.g.][]{bigot02}. 
Under the assumption that the triplet represents a dipole mode, we forced the sidelobes to be equally separated from the pulsation frequency by the assumed rotation frequency, then fitted the triplet (after removing the two other modes from the data) by linear least-squares to test the oblique pulsator model. By choosing the zero-point in time such that the phases of the first sidelobes are equal, we are able to show that the mode is slightly distorted as the central peak in the triplet has a different phase. The results of this test are shown in Table\,\ref{tab:152808505_OPM}.

Furthermore, we are able to provide constraints on the geometry of the pulsation through the relation of \citet{kurtz1990b}:
\begin{equation}
\label{eq:OPM_trip}
    \tan i\tan\beta = {\frac{A_{+1}^{(1)}+A_{-1}^{(1)}}{A_0^{(1)}}}, 
\end{equation}
where $A_{\pm1}^{(1)}$ are the dipole sidelobe amplitudes, $A_0$ is the amplitude of the central peak, $i$ is the inclination angle and $\beta$ is the angle of obliquity. This holds true under the assumption of a pure, non-distorted, dipole pulsation ($\ell=1$, $m=0$), with sidelobes generated from rotation alone.

\begin{figure}
\centering
\includegraphics[width=1.\columnwidth]{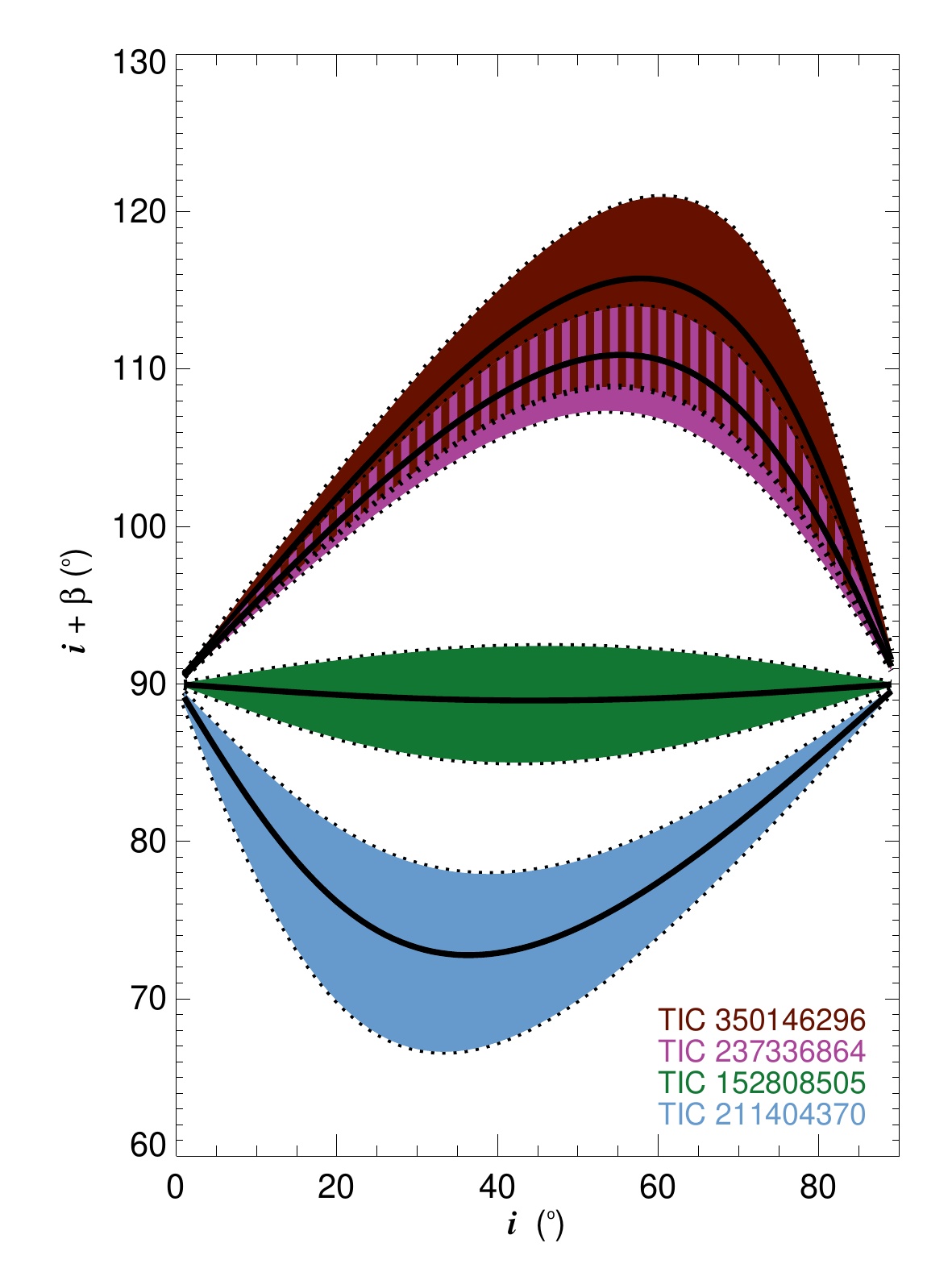}
\caption{{Display of the possible $i+\beta$ combinations, for the four stars where the analysis of the multiplets allows us to set constraints on the stellar pulsation geometry. The full lines and coloured regions limited by dashed lines correspond, respectively, to the values and uncertainties of $i+\beta$ for each star. In stars for which $i+\beta \leq 90^{\circ}$ only one pulsation pole is observed. Different colours show the results for different stars.
%(in black and white printing, the order in which a star is listed in the bottom-right-hand-side corner corresponds to the order in which the $i+\beta$ region allowed for that star appears in the figure). 
The brown and purple stripes indicate regions of overlap of possible solution for TIC\,350146296 and TIC\,237336864.}}
\label{ibeta}
\end{figure}

For TIC\,152808505, we measure $\tan i\tan\beta=0.96\pm0.12$. Although we cannot separate $i$ or $\beta$, we show the values which satisfy this relation in Fig.\,\ref{ibeta}. The results show that $i+\beta\simeq90\degr$ implying we only see one pulsation pole.

\subsection{TIC\,350146296}
\label{sec:350146296}

TIC\,350146296 (HD\,63087) is another star in our sample for which references to chemical peculiarity do not exist in the literature. The only reference for this star is \citet{houk1975} who classified it as A7\,IV. The effective temperature derived in this work, $T_{\rm eff}=7450\pm160$\,K places it well within the theoretical instability strip. The analysis presented here is based on sector 1 and 2 data, however this star is in TESS's continuous viewing zone and will have a complete 13-sector data set at the end of TESS cycle 1.

\begin{figure}
\centering
\includegraphics[width=0.9\columnwidth]{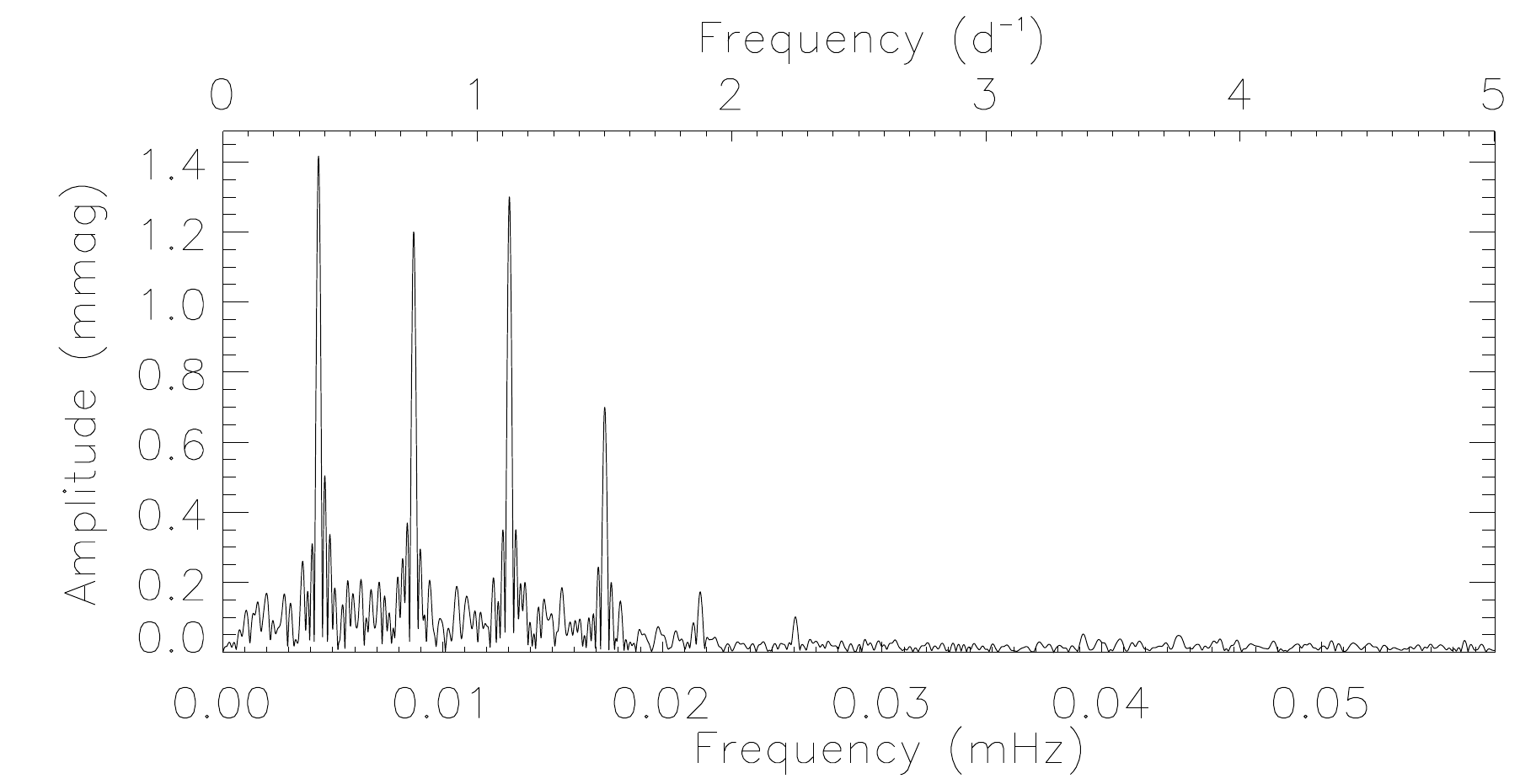}
\includegraphics[width=0.9\columnwidth]{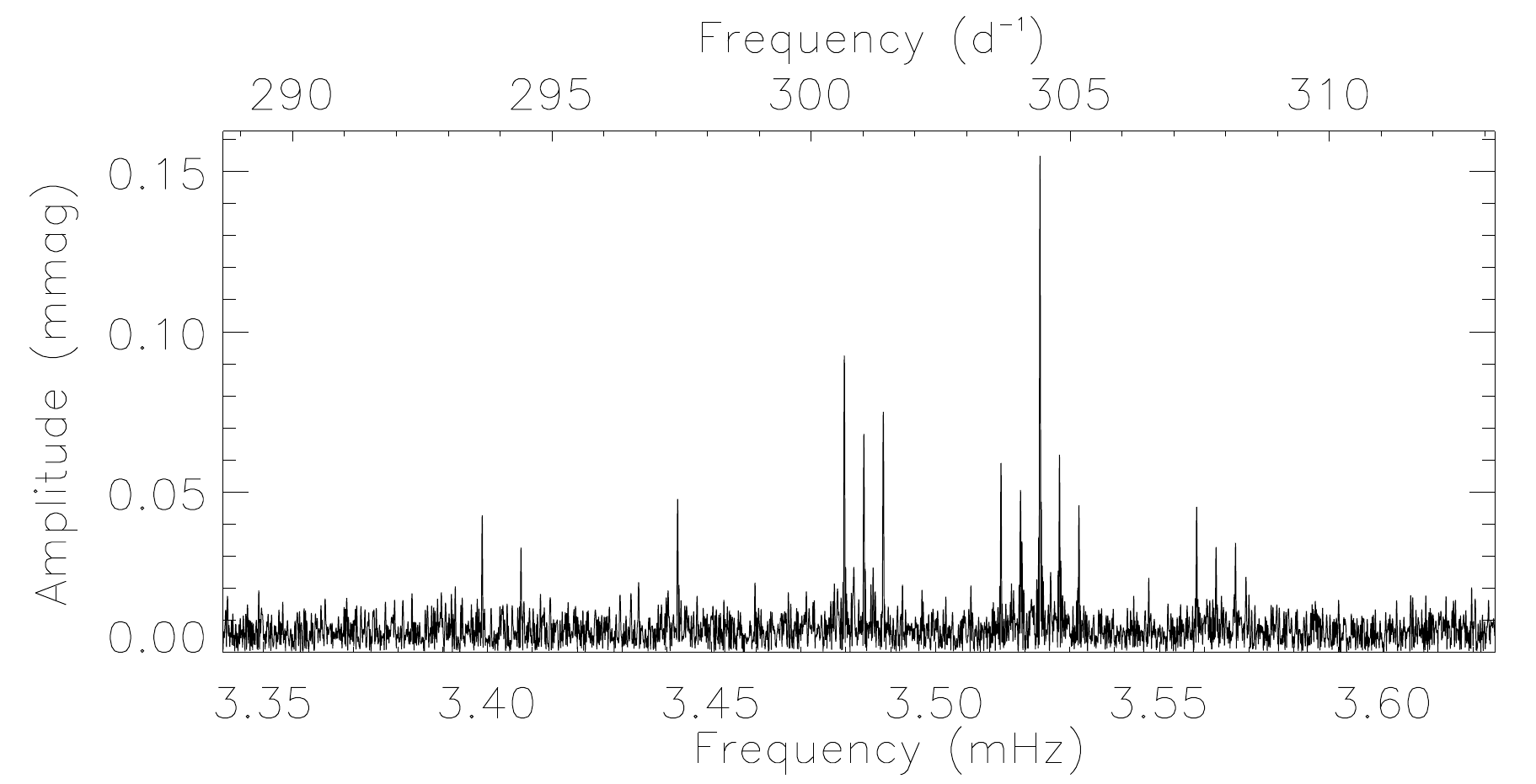}
\includegraphics[width=0.9\columnwidth]{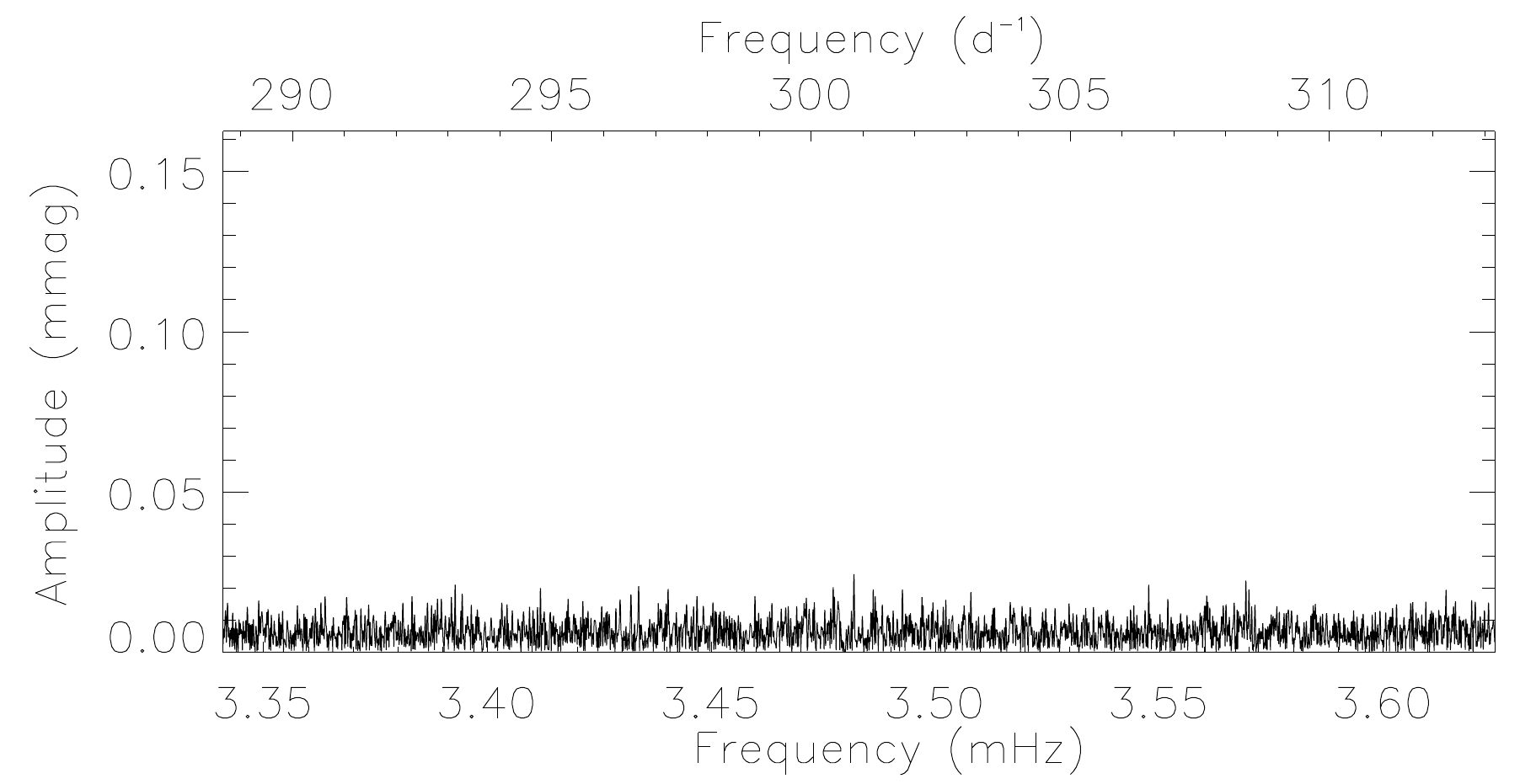}
\caption{Top: amplitude spectrum at low frequency showing the rotation signal of TIC\,350146296 and its harmonics. Middle: zoom of the pulsation modes found in this star. Bottom: the amplitude spectrum of the residuals after removing the frequencies shown in Table\,\ref{tab:350146296} showing no remaining significant signals.}
\label{fig:350146296_ft}
\end{figure}

\begin{table*}
\centering
\caption{Details of a linear least squares fit to the pulsation modes found in TIC\,350146296. The zero-point for the phases, BJD\,2458353.55189, was chosen to force the sidelobes of $\nu_3$ to be equal. The frequencies of all sidelobes have been forced to be split from the central component by integer values of the rotation frequency.}
\label{tab:350146296}
\begin{tabular}{llccr}
\hline
ID & \multicolumn{1}{c}{Frequency} & Amplitude & \multicolumn{1}{c}{Phase} & \multicolumn{1}{c}{S/N}\\
    & \multicolumn{1}{c}{(mHz)}  		 & (mmag) & \multicolumn{1}{c}{(rad)} &\\
                   &  &  \multicolumn{1}{c}{$\pm 0.005$} & &\\
\hline
$\nu_1-\nu_{\rm rot}$ & $3.398781\pm0.000021$&	0.044	& $2.35\pm0.12$ &	8.2\\
$\nu_1+\nu_{\rm rot}$ & $3.407455\pm0.000021$&	0.036	& $2.27\pm0.14$ &	6.9\\
$\nu_2$ & $3.442436\pm0.000012$&	0.051	& $-2.10\pm0.10$ &	9.5\\
$\nu_3-\nu_{\rm rot}$ & $3.479660$	         &  0.095	& $2.79\pm0.06$ &	18.0\\
$\nu_3$ & $3.484009\pm0.000009$&	0.068	& $2.97\pm0.08$ &	12.8\\
$\nu_3+\nu_{\rm rot}$ & $3.488359$	         &  0.074	& $2.79\pm0.07$ &	13.9\\
$\nu_4-2\nu_{\rm rot}$ & $3.514631$	         &  0.060	& $1.51\pm0.09$ &	11.4\\
$\nu_4-1\nu_{\rm rot}$ & $3.518980$	         &  0.053	& $1.94\pm0.10$ &	9.9\\
$\nu_4$ & $3.523329\pm0.000004$&	0.159	& $1.44\pm0.03$ &	30.0\\
$\nu_4+1\nu_{\rm rot}$ & $3.527678$	         &  0.062	& $1.39\pm0.08$ &	11.8\\
$\nu_4+2\nu_{\rm rot}$ & $3.532027$	         &  0.045	& $1.51\pm0.12$ &	8.5\\
$\nu_5-\nu_{\rm rot}$ & $3.558301$	         &  0.043	& $2.56\pm0.12$ &	8.2\\
$\nu_5$ & $3.562650\pm0.000020$&	0.029	& $2.72\pm0.18$ &	5.5\\
$\nu_5+\nu_{\rm rot}$ & $3.566999$           &	0.032	& $2.41\pm0.16$ &	6.1\\
\hline
\end{tabular}
\end{table*}

There is clear rotational modulation in this star as demonstrated by the high amplitude {harmonic series} in the top panel of Fig.\,\ref{fig:350146296_ft}. We determine the rotation period to be $2.66121\pm0.00029$\,d. The pulsation spectrum is rich. In order of increasing frequency we find: 
{a doublet split by twice the rotation frequency, a singlet, a triplet, a quintuplet and another triplet, all split by the rotation frequency}. The details of the pulsations are shown in Table\,\ref{tab:350146296}. Further to the {rich amplitude spectrum}, TIC\,350146296 hosts the highest frequency pulsations of any known roAp star (3.562\,mHz; $P = 4.68$\,min), significantly surpassing those found in HD\,134214 (2.947\,mHz, $P = 5.66$\,min; \citealt{kreidl1985}, \citealt{gruberbauer2011}).

 {The pulsations observed in TIC\,350146296 may be above the star's acoustic cut-off frequency, $\nu_{\rm ac}$. The roAp stars with frequencies above the acoustic cut-off frequency are a challenge to theory \citep{cunhaetal13}. This is because, while the strong magnetic field provides a natural way to keep the wave energy of such high frequency pulsations within the star \citep{sousa08}, the opacity mechanism is unable to excite them, as discussed in Sec.\,\ref{sec:1}. {We estimate a mass of {$1.55\pm 0.09$\,M$_\odot$} for TIC 350146296, from the mass-luminosity relation published by \cite{eker15}.} Based on that mass and on the effective temperature and luminosity listed in Table\,\ref{properties}, we compute the cut-off frequency by scaling from the solar value using $\nu_{\rm ac}\propto g/\sqrt{T_{\rm eff}}$, where $g$ is the surface gravity \citep[e.g.][]{chaplin13}. Taking a solar value of $\nu_{{\rm ac}_\odot}$ = 5.106\,mHz \citep{jimenez06}, we find  $\nu_{\rm ac}$= 2.91 $\pm 0.79$\,mHz, for TIC\, 350146296. Unfortunately, the large uncertainties in the star's global parameters prevent us from concluding whether the star is indeed pulsating above the cut-off frequency.}

Since the frequencies in this star are so high, there is a non-negligible suppression of the pulsation amplitude due to the length of the exposures. Calculated using the expression \citep{murphy14}
\begin{equation}
\frac{A}{A_0} = {\rm sinc}\frac{\pi T_{\rm exp}}{P_{\rm puls}},
\end{equation}
where $T_{\rm exp}$ and $P_{\rm puls}$ are the exposure time and pulsation period, we find that the intrinsic amplitude ($A_0$) of the central component of the quintuplet is 0.217\,mmag in the TESS filter. Given the non-optimum red filter of TESS, it is possible that this pulsation is detectable from the ground with $B$ observations (cf. Sec.\,\ref{sec:comp}).

With such a rich pulsation spectrum, we are also able to determine the large frequency separation, $\Delta\nu$, for this star, defined as the difference in frequency of modes of the same degree and consecutive radial orders. The frequency difference between most of the modes is 39.3\,$\umu$Hz. This spacing may correspond to the large frequency separation, if the modes are of the same degree, or to half of it, if they are of alternating even and odd degrees. To establish which of these is the most likely option, we consider again the mass estimated above and the global parameters given in Table\,\ref{properties}, and scale from the Sun, to derive an estimate of the large frequency separation for TIC\,350146296. Using the scaling relation $\Delta\nu \propto\sqrt{\left<\rho\right>}$, where $\left<\rho\right>$ is the stellar mean density {\citep[e.g.][]{chaplin13}}, and adopting  135\,$\umu$Hz as the solar large frequency separation \citep[e.g.][]{stello09}, we estimate the large frequency separation for this star to be $87\pm16\,\umu$Hz. This value clearly points towards the modes seen in TIC\,350146296 as being alternating even and odd degrees. 

This remains the case even if extinction is accounted for in the computation of the luminosity. Although no extinction value is provided for this star, we saw from Sec.\,\ref{sec:prop} that the root mean square of the difference between the logarithmic luminosity values derived with and without accounting for extinction is 0.1. If we were to assume that the logarithmic luminosity of TIC\,350146296 is 0.1 larger than the value considered before, we would find a large frequency separation of 73.5\,$\umu$Hz, still pointing towards the same conclusion. {Interestingly, there is one spacing, between $\nu_2$ and $\nu_3$, which is slightly larger at 41.6\,$\umu$Hz. This may indicate that  $\nu_2$ and $\nu_4$ correspond to modes of different degrees, a possibility that would also naturally explain the difference seen in their multiplet structures.} 
All of the splittings are shown in Table\,\ref{tab:350146296_split}.

\begin{table}
\centering
\caption{Details of the splittings between pulsation modes in TIC\,350146296.}
\label{tab:350146296_split}
\begin{tabular}{lc}
\hline
IDs & Splitting \\
    & ($\umu$Hz) \\
\hline
$\nu_2-\nu_1$ & $39.318\pm0.016$\\
$\nu_3-\nu_2$ & $41.573\pm0.014$\\
$\nu_4-\nu_3$ & $39.320\pm0.009$\\
$\nu_5-\nu_4$ & $39.321\pm0.020$\\
\hline
\end{tabular}
\end{table}

{Now that we have justified the identification of the modes as alternating even and odd degree modes, we suggest the following degree identifications for the multiplets: $\nu_1$ dipole; $\nu_2$ radial, or quadrupole; $\nu_3$ dipole; $\nu_4$ quadrupole and $\nu_5$ dipole.}
From this, we are able to provide constraints on the geometry of the star by applying the oblique pulsator model to the various multiplets we detect, as described in Section~\ref{sec:opm}. For a pure, non-distorted mode, we expect the phases of each peak in the multiplet to be the same. We find that, in the cases of the dipole triplets, there is good agreement between the phases of all of the peaks in a multiplet. 
{This is also the case for the quadrupole mode ($\nu_4$) where there is good agreement between all but the $\nu_4-\nu_{\rm rot}$ phase}.

{Returning to the oblique pulsator model and the geometry of the star, as discussed above, we can use the relative amplitudes of the sidelobes and the pulsation peak to see the relationship between $i$ and $\beta$. Applying equation\,(\ref{eq:OPM_trip}) to the two triplets, $\nu_3$ and $\nu_5$, in TIC\,350146296, we find that $\tan i\tan\beta=2.50\pm0.22$ and $2.58\pm0.53$, respectively.} {We have also identified the doublet as the rotational sidelobes of an undetected $\nu_1$, hence a dipole triplet for these frequencies. The undetected frequency $\nu_1$ could plausibly be lost in the noise, it is possible that the dipole modes have different geometries, as was suggested for KIC\,10195926 \citep{kurtz2011}. This result will become clear with more sectors of data for this star, which is in the continuous viewing zone.}

Now considering the quintuplet ($\nu_4$), {the relationship between the amplitudes of the sidelobes of a non-distorted quadrupole mode ($\ell=2$, $m=0$) is given by} \citep{kurtz1990b}:
\begin{equation}
\label{eq:OPM_quad}
    \tan i\tan\beta = 4\frac{A_{-2}^{(2)}+A_{+2}^{(2)}}{A_{-1}^{(2)}+A_{+1}^{(2)}}
\end{equation}
where $A_{\pm1}^{(2)}$ and $A_{\pm2}^{(2)}$ are the amplitudes of the first and second sidelobes of the mode. With this relation we find $\tan i\tan\beta = 3.91\pm0.36$, {which differs from the relation derived from the two dipole modes by 3.3$\sigma$ and 2.1$\sigma$, respectively. This difference may be a result of the magnetic field perturbing the different degree modes differently, a result of the broad TESS filter, a combination of these, or something else entirely. This star warrants a detailed study reserved for a later publication when more data are available.}

Using the combined results of the dipole triplets, {by taking their average $\tan i\tan\beta$ and summing the corresponding errors in quadrature}, we calculate the allowed values of $i$ and $\beta$ which satisfy equation\,(\ref{eq:OPM_trip}) and show them in Fig.\,\ref{ibeta}. As we have no information on either $i$ or $\beta$, they cannot be disentangled. 

\subsection{TIC\,431380369}

TIC\,431380369 (HD\,20880) was classified as Ap\,Sr(EuCr) by \citet{houk1975}. \citet{martinez1993} measured Str\"omgren and H$\beta$ indices for this star: $V = 7.957; b-y = 0.097; m_1 = 0.212; c_1 = 1.005; \beta = 2.870$, from which the parameters $\delta m_1 = -0.009$ and $\delta c_1 = 0.095$ can be derived; neither of these indices are indicative of an Ap star. {The H$\beta$ index indicates an equivalent spectral type near mid-A \citep[see][for calibration relations for the Str\"omgren and H$\beta$ indices for F and A stars, respectively]{crawford1975, crawford1979,moon85}.} The relatively high temperature of $7880\pm170$\,K derived in this work may explain why the Str\"omgren indices are essentially normal.

Observed in sector 2, the TESS data for TIC\,431380369 show this star to be a rotationally modulated variable with a period of $5.2434\pm0.0026$\,d. The light curve is significantly affected by instrumental artefacts during the last 5 days of sector 2, which may lead one to conclude that the period is double to what we present, but the pulsation analysis discussed below provides us with confidence in our determination.

\begin{figure}
\centering
\includegraphics[width=0.9\columnwidth]{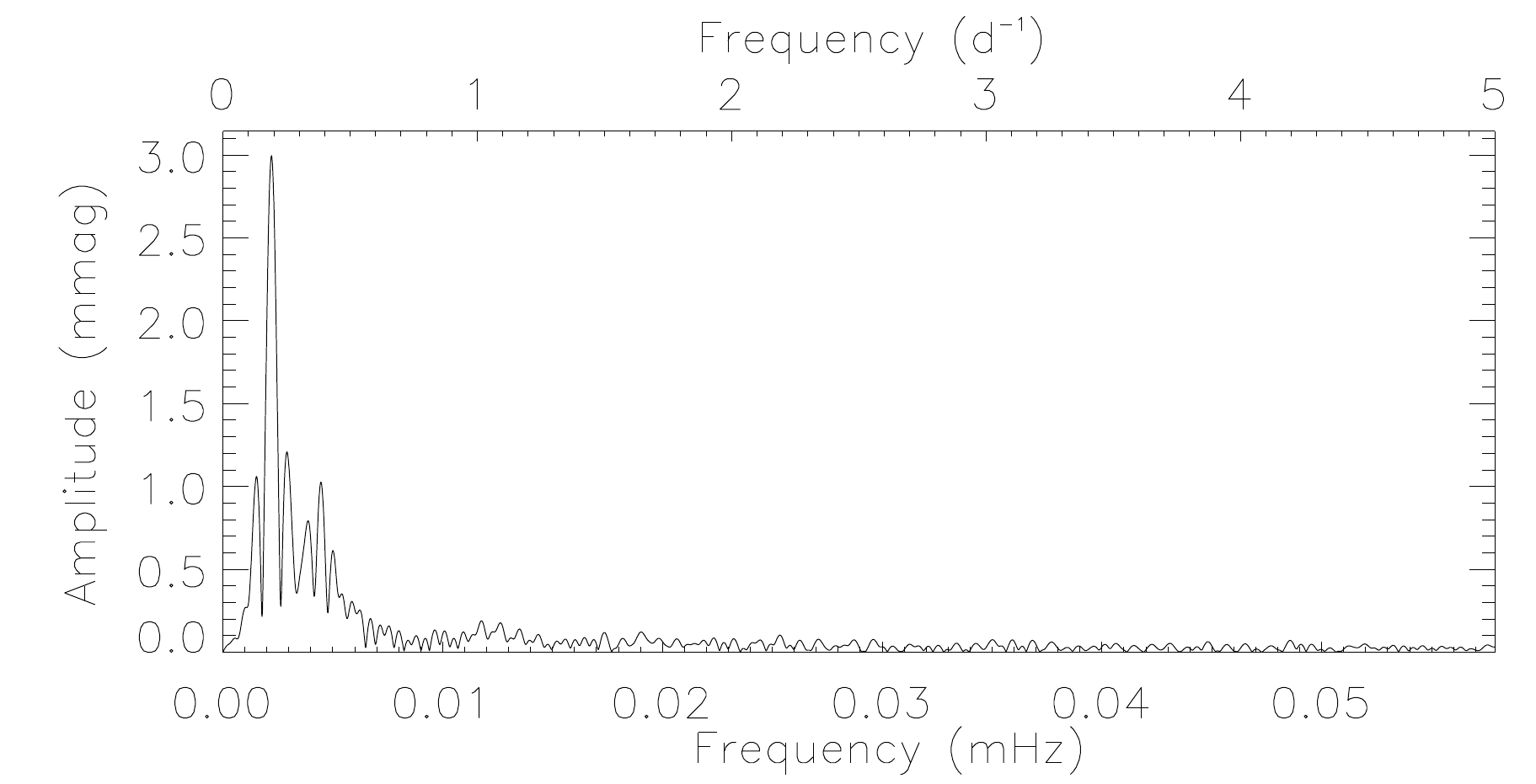}
\includegraphics[width=0.9\columnwidth]{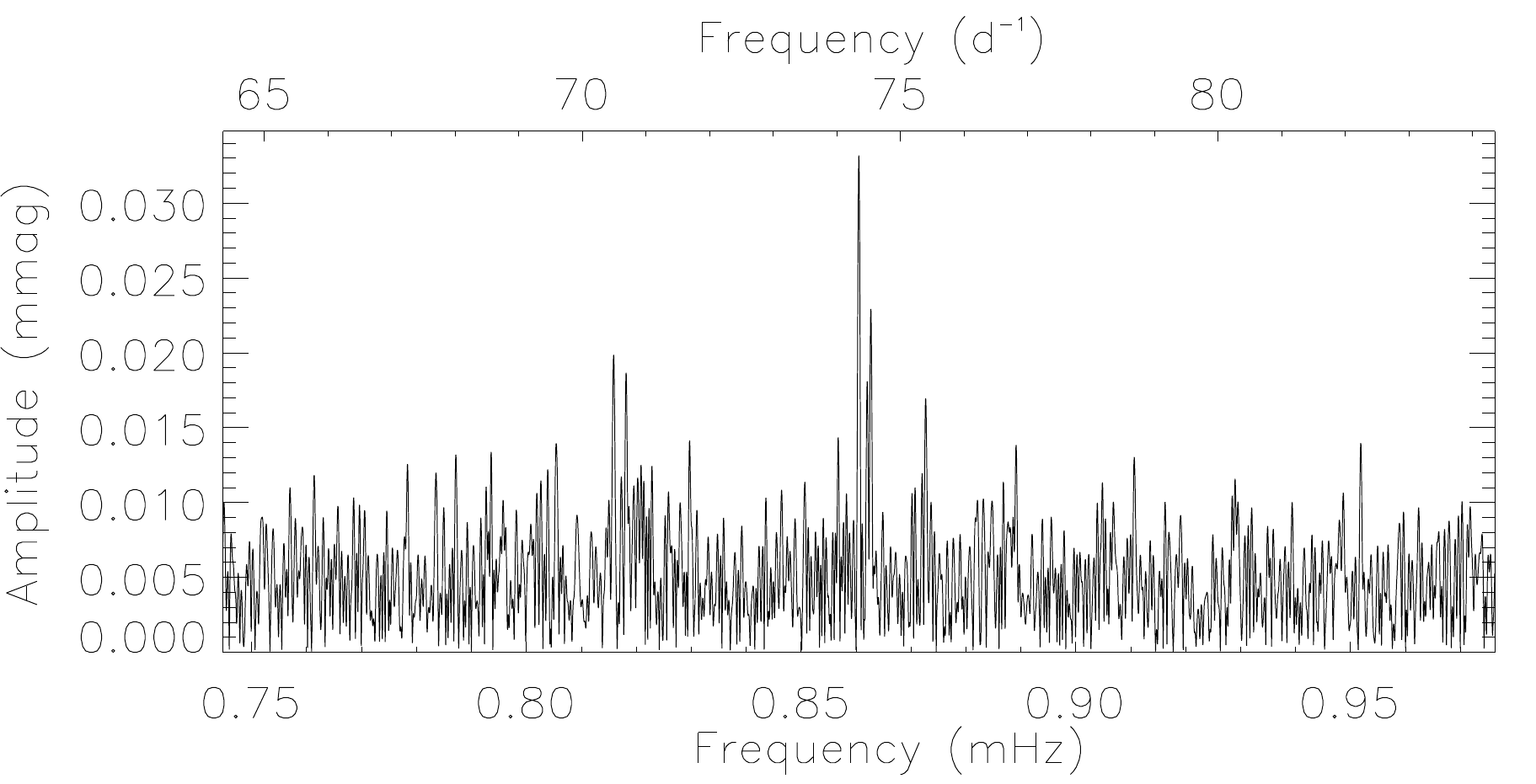}
\includegraphics[width=0.9\columnwidth]{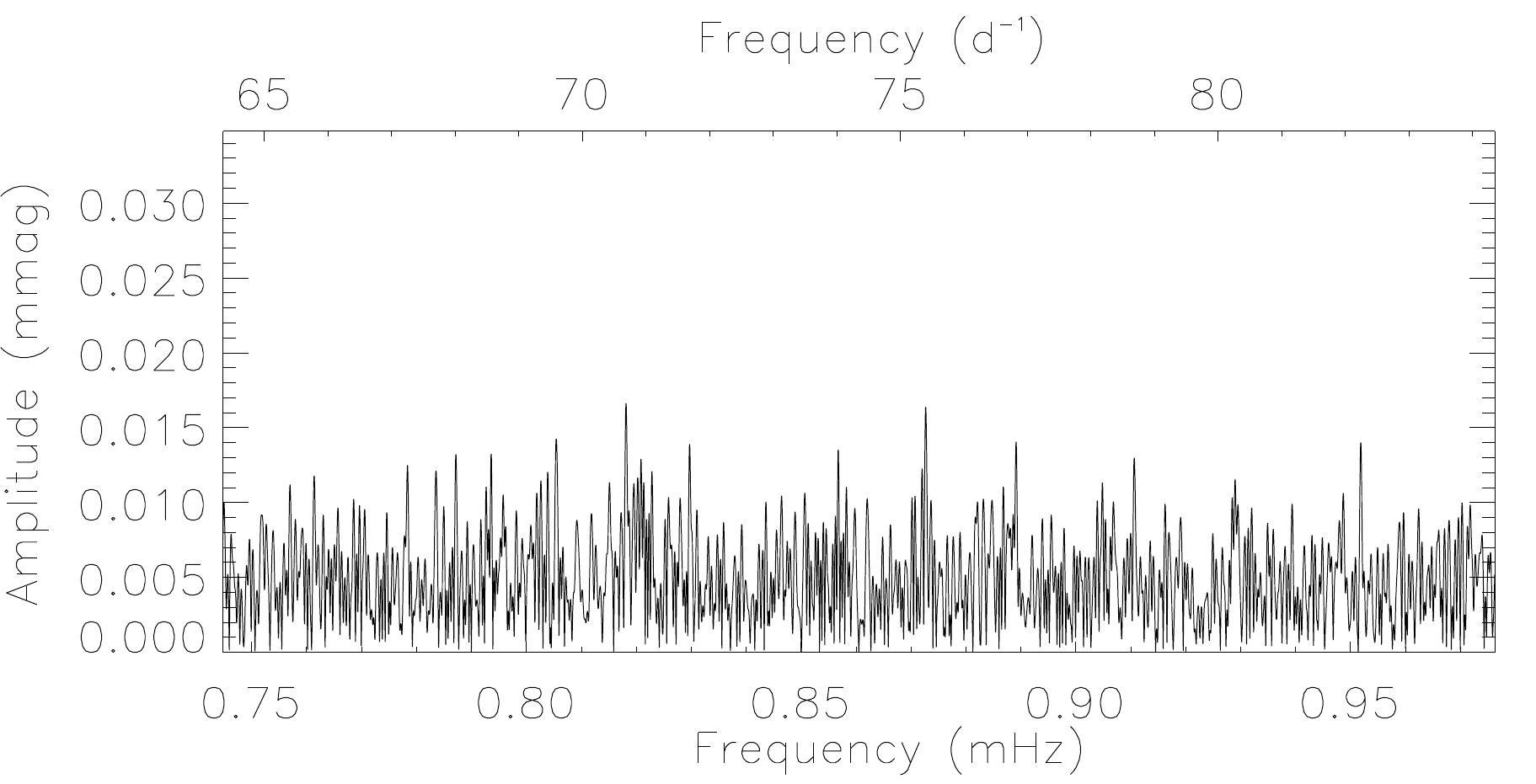}
\caption{Top: amplitude spectrum at low frequency showing the rotation signal of TIC\,431380369. Middle: zoom of the pulsation modes found in this star. Bottom: the amplitude spectrum of the residuals after removing the significant frequencies given in Table\,\ref{tab:431380369_lobes} showing no remaining significant signals.}
\label{fig:431380369_ft}
\end{figure}

There are three frequencies that can be extracted with confidence in this star, as shown in Fig.\,\ref{fig:431380369_ft} and Table\,\ref{tab:431380369_lobes}. Analysis of these peaks show that two are independent modes, with the third being a rotational sidelobe. The separation of the two modes, $\sim45\,\umu$Hz, could be the large separation. In fact, an analysis similar to that described in Sec.\,\ref{sec:350146296} gives for this star $\Delta\nu = 44 \pm 8 \,\umu$Hz, when extinction is neglected.

Given that one of the modes shows a rotational sidelobe ($\nu_2$), we attempt to extract further sidelobes that are not clearly seen in Fig.\,\ref{fig:431380369_ft}. To do this, we force fitted sidelobes at $\pm\nu_{\rm rot}$ to the two modes with linear least-squares. We find that only the positive sidelobes are significant at the 4~$\sigma$ level, as shown in Table\,\ref{tab:431380369_lobes}. 

\begin{table}
\centering
\caption{Linear least-squares fit to the pulsations and force-fitted sidelobes in TIC\,431380369. The zero-point for phases is BJD\,2458370.481478, and has been chosen such that the phases of the sidelobes of the highest amplitude mode are equal.}
\label{tab:431380369_lobes}
\begin{tabular}{llcrr}
\hline
ID & \multicolumn{1}{c}{Frequency} & Amplitude & \multicolumn{1}{c}{Phase} & \multicolumn{1}{c}{S/N}\\
    & \multicolumn{1}{c}{(mHz)}  		 & (mmag) & \multicolumn{1}{c}{(rad)} &\\
                   &  &  \multicolumn{1}{c}{$\pm 0.004$} & &\\
\hline
$\nu_1-\nu_{\rm rot}$ & $0.81366$ & $0.003$ & $0.13\pm1.34$ & 0.8\\
$\nu_1$               & $0.81587\pm0.00004$ & $0.018$ & $-1.18\pm0.21$ & 4.5 \\
$\nu_1+\nu_{\rm rot}$ & $0.81808$ & $0.017$ & $-0.77\pm0.24$ & 4.3\\
$\nu_2-\nu_{\rm rot}$ & $0.85829$ & $0.011$ & $-3.10\pm0.36$ & 2.8 \\
$\nu_2$ & $0.86050\pm0.00003$ & $0.034$ & $-1.24\pm0.12$ & 8.5 \\
$\nu_2+\nu_{\rm rot}$ & $0.86271$ & $0.024$ & $-3.10\pm0.16$ &6.0\\
\hline
\end{tabular}
\end{table}

TIC\,431380369 will be observed in two more sectors. However, these are well spaced in time which will result in a complex window function, potentially complicating the extraction of the sidelobes and any further modes.

\section{{The} previously known roAp stars}
\label{old}

\subsection{TIC\,69855370}
\label{hd213637}

TIC\,69855370 (HD\,213637) was classified as A\,(pEuSrCr) by \citet{houk1988}. \citet{martinez1993} measured Str\"omgren and H$\beta$ indices for this star: $V = 9.611; b-y = 0.298; m_1 = 0.206; c_1 = 0.411; \beta = 2.670$, from which the parameters $\delta m_1 = -0.035$ and $\delta c_1 = -0.031$ can be derived; both of these indices are indicative of an Ap star. The H$\beta$ index indicates an equivalent spectral type near mid-F, so this is one of the coolest roAp stars. This is confirmed by the effective temperature derived in this work, $T_{\rm eff} = 6430 \pm 150$\,K. This value is consistent with the effective temperature determined by \citet{kochukhov2003}, based on the analysis of high resolution spectra from the {Ultraviolet and Visual Echelle Spectrograph (UVES)} on the ESO Very Large Telescope (VLT), from which the author derived $T_{\rm eff} = 6400 \pm 100$\,K from the H$_\alpha$ and H$_\beta$ lines, and $\log g = 3.6 \pm 0.2$ from the ionisation equilibrium for Fe\,{\sc{i}} and Fe\,{\sc{ii}} lines. The author points out that stratification contributes to the uncertainty in the surface gravity determination. In any case, this star, along with a few others such as HD\,101065 (Przybylski's star), sets the lower temperature boundary for the Ap phenomenon, and hence tests our understanding of these stars. 

\citet{mathys2003} first measured the mean magnetic field modulus of TIC\,69855370 to be 5.2\,kG. \citet{kochukhov2003} also derived a mean magnetic field modulus of $B_s = 5.5\pm0.1$\,kG from Zeeman splitting in the Fe\,{\sc{ii}} 6146.26-\AA\ line, and \citet{elkin2015} found $B_s = 5.56 \pm 0.15$\,kG from the partially resolved components of the Fe\,{\sc{ii}} 6146.26-\AA\ line. Clearly, TIC\,69855370 is a strongly magnetic Ap star.

Rapid pulsations were discovered in TIC\,69855370 by \citet{martinez1998} who found two pulsation mode frequencies, $\nu_1 = 1.41089 \pm 0.00011$\,mHz and $\nu_2 = 1.45235 \pm 0.00006$\,mHz\footnote{Note that we have changed the labeling of the modes presented by \citet{martinez1998} to be in increasing frequency order, rather than in decreasing amplitude order, to fit the convention we use in this work.}, although with some uncertainty because of the daily aliases in their data set.  \citet{elkin2015} obtained 2.1\,h of high time resolution, high spectral resolution observations of TIC\,69855370 with the UVES spectrograph on the ESO VLT and reported on the pulsational radial velocity variations at different atmospheric heights. Their 2-h observations were not long enough to resolve the two frequencies found by \citet{martinez1998}.

TIC\,69855370 was, and will only be, observed during TESS sector 2. We detect no indication of rotation in this star, suggesting one of three options: a long rotation period, an unfavourable alignment (with $i$ or $\beta$ close to zero), or the absence of significant spots. The latter would be surprising, given the strong magnetic field. \citet{kochukhov2003} measured a {projected rotational velocity $v_{\rm e}\sin i = 3.5\pm0.5$\,km\,s$^{-1}$}, and suggested a rotation period of 25~d. With this in mind, we favour an unfortunate geometry for the explanation of the lack of a rotation signal.

\begin{figure}
\centering
\includegraphics[width=0.9\columnwidth]{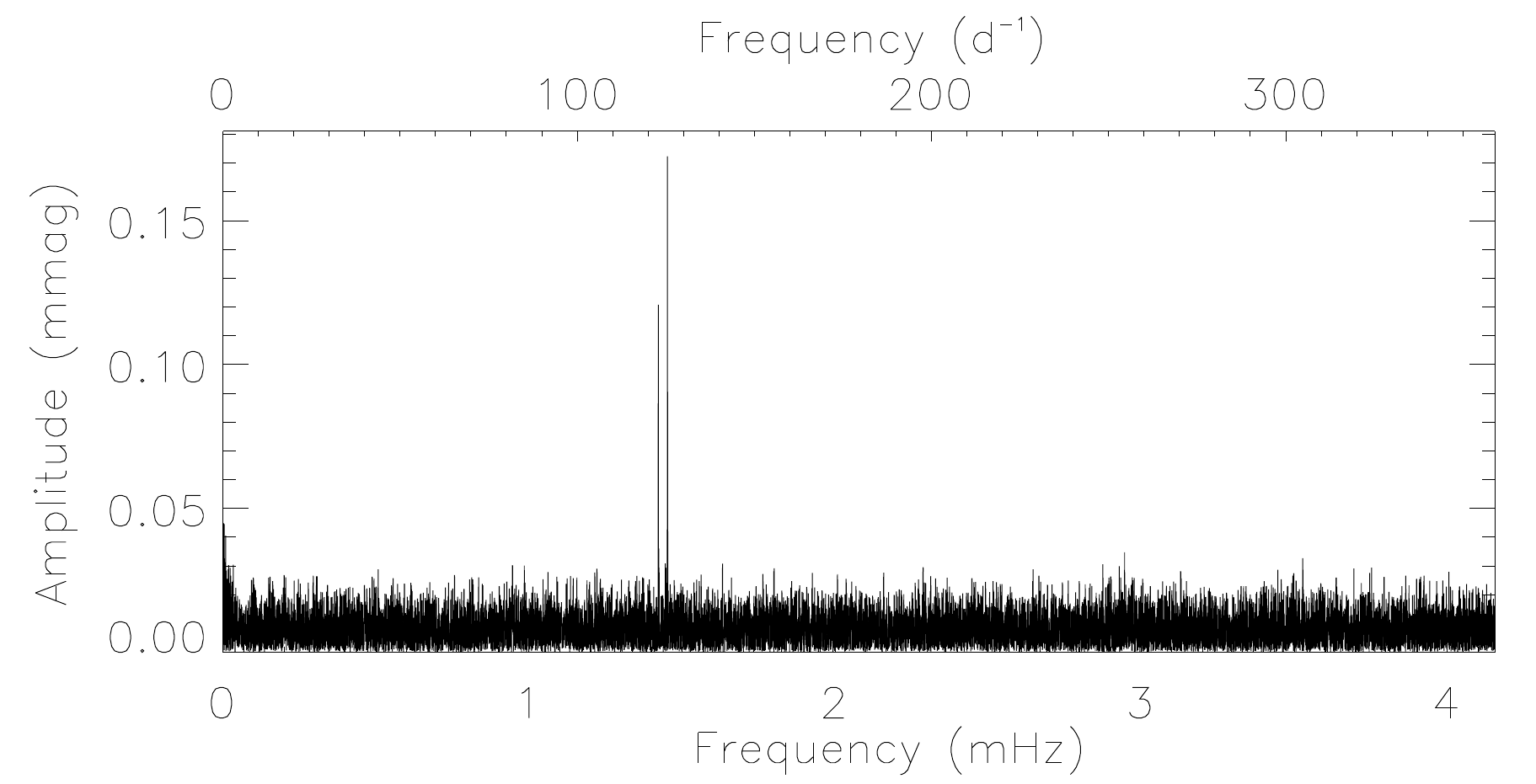}
\includegraphics[width=0.9\columnwidth]{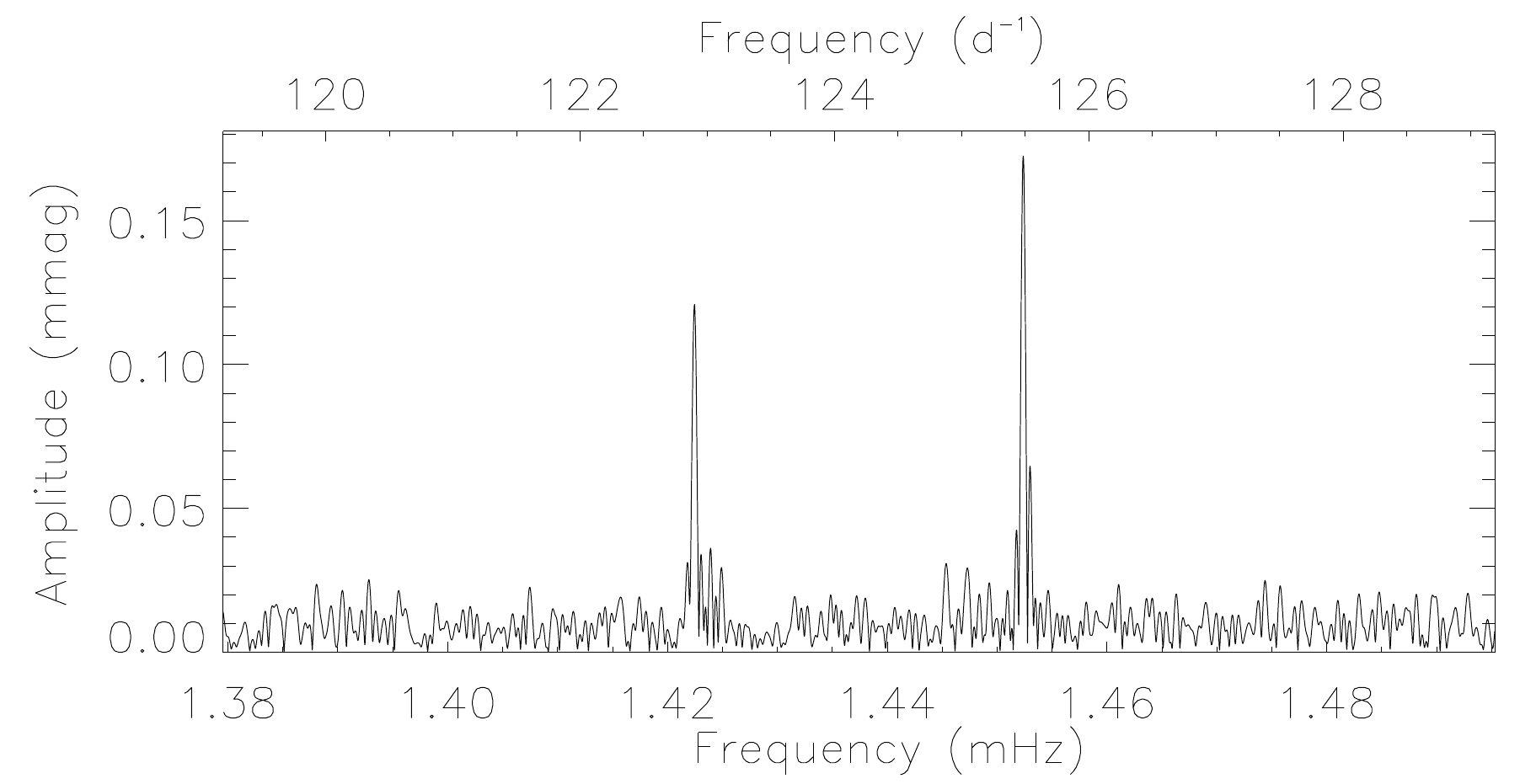}
\includegraphics[width=0.9\columnwidth]{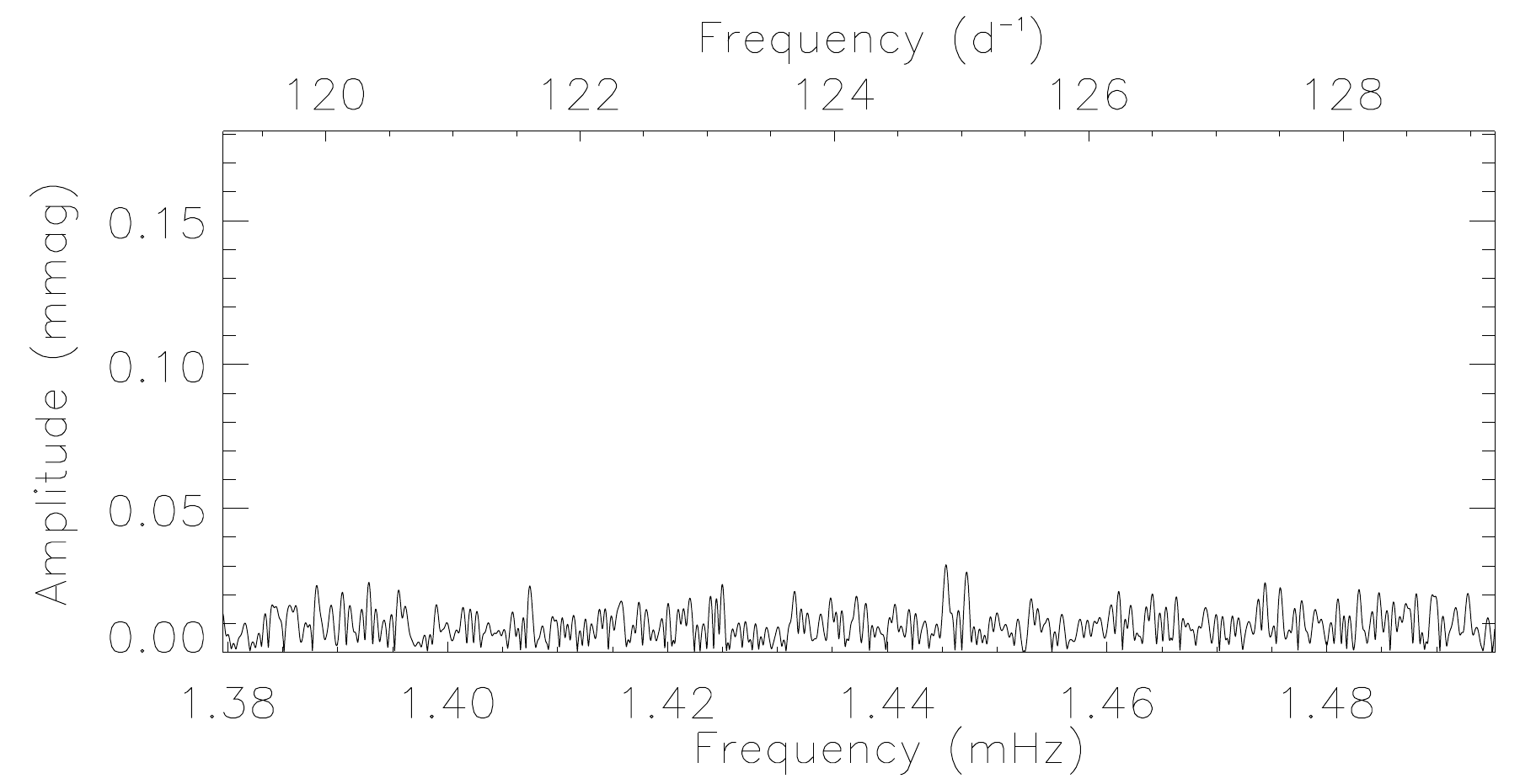}
\caption{Top: full amplitude spectrum of TIC\,69855370 to almost the Nyquist frequency. Middle: zoom of the pulsation modes found in this star. Bottom: the amplitude spectrum of the residuals after removing the frequencies given in Table\,\ref{tab:69855370} showing no significant frequencies remaining.}
\label{fig:69855370_ft}
\end{figure}

Clearly present in the TESS data are the two known pulsation mode frequencies (Fig.\,\ref{fig:69855370_ft}). We fitted these frequencies with non-linear least-squares and show the results in Table\,\ref{tab:69855370}. $\nu_2$ is in agreement with \citet{martinez1998}; $\nu_1$ differs from their value by 11.57\,$\umu$Hz (1\,d$^{-1}$); as they were concerned about daily aliases in their data set, this resolves the issue; the frequencies in Table\,\ref{tab:69855370} do not suffer from aliasing, thanks to the high duty cycle of TESS. 

\begin{table}
\centering
\caption{Details of the pulsation modes found in TIC\,69855370. The zero-point for phases is BJD\,2458367.81716.}
\label{tab:69855370}
\begin{tabular}{lccrr}
\hline
ID & Frequency & Amplitude & \multicolumn{1}{c}{Phase} & \multicolumn{1}{c}{S/N} \\
    & (mHz)  		 & (mmag) & \multicolumn{1}{c}{(rad)} & \\
                       &  &  \multicolumn{1}{c}{$\pm 0.007$} & &\\
\hline
$\nu_1$ & $1.422447\pm0.000014$ & $0.120$ & $1.96\pm0.06$ &17.1\\
$\nu_2$ & $1.452371\pm0.000010$ & $0.172$ & $1.03\pm0.04$ & 24.6\\

\hline
\end{tabular}
\end{table}

We therefore derive the separation between the two pulsation frequencies of $\nu_2 - \nu_1 = 29.924 \pm 0.017$\,$\umu$Hz. This is plausibly the large separation, or half of that. Following the scaling applied in Sec.\,\ref{sec:350146296}, we find for this star $\Delta\nu = 68 \pm 13 \umu$Hz, when extinction is neglected, which is consistent with the observed separation corresponding to half of the large frequency separation, so with {the degree of the modes being of alternating parity}.

\subsection{TIC\,139191168}
\label{hd217522}

TIC\,139191168 (HD\,217522) was classified as an Ap\,(Si)Cr star by \citet{houk1978} with the remark `may be Eu rather than Si'. The Eu suggests a cooler star than Si. \citet{martinez1993} measured Str\"omgren and H$\beta$ indices for this star: $V = 7.525; b-y = 0.289; m_1 = 0.227; c_1 = 0.484; \beta = 2.691$, from which the parameters $\delta m_1 = -0.056$ and $\delta c_1 = -0.015$ can be derived; both of these indices are indicative of an Ap star. The H$\beta$ index indicates an equivalent spectral type near mid-F, so this is among the coolest roAp stars. {Previously, \citet{gelbmann1998} obtained $T_{\rm eff}=6750$\,K as part of the detailed abundance analysis of this star. \citet{kochukhov2002}  derived $T_{\rm eff}=6850$\,K from the Balmer line profiles.} We derive an effective temperature of $6890 \pm 150$\,K, which is consistent with those.

The mean longitudinal magnetic field of TIC\,139191168 was measured to be $-400 \pm 120$\,G by \citet{mathyshubrig1997}. Later, \citet{hubrig2002} measured a mean quadratic field strength of $2000\pm400$\,G which is consistent with the mean field modulus upper limit of 1.5\,kG provided by \citet{ryabchikova2008}.

\citet{kurtz1983} discovered pulsations with periods around 13.72\,min in this star with observations made on 17 nights in 1982 through a Johnson $B$ filter. \citet{kreidl1991} conducted a multi-site observing campaign on TIC\,139191168 in 1989 from four southern hemisphere observatories. They concluded that the pulsation mode amplitudes have short growth and decay times for the pulsations with frequencies near 1.2\,mHz. Indeed, the highest peak near 1.2\,mHz has a greater amplitude in 1989 than in 1982. They also discovered new pulsations with frequencies near 2\,mHz that were completely absent from the 1982 observations, supporting the argument for rapid growth and decay for the pulsations in this star.

%They also discovered new pulsations with frequencies near 2\,mHz that were completely absent from the 1982 observations. They argued that this also supports rapid growth and decay for the pulsations in this star. The highest peak near 1.2\,mHz has a greater amplitude in 1989 than in 1982, and the peaks near 2.0\,mHz are present in 1989 and absent in 1982. 

\citet{medupe2015} obtained further $B$ observations of TIC\,139191168 in 2008 which also showed the presence of the 2.0\,mHz peaks. Importantly, they also obtained high time resolution, high spectral resolution spectra with the ESO VLT UVES spectrograph. Their study of individual lines shows the presence of both the 1.2\,mHz and 2.0\,mHz peaks with amplitudes that were very sensitive to the atmospheric height of the line formation layer for individual elements and ions, largely of rare earth elements. The H$_\alpha$ measurements barely showed the presence of the 2.0\,mHz peak, whereas, for example, that frequency had a higher amplitude than the 1.2\,mHz peak for the Nd\,{\sc{ii}} 5293-\AA\ line.

Previous observations of TIC\,139191168 showed no signs of rotational light variability due to surface spots \citep{vanheerden2012,medupe2015}, {nor} is there spectral line variability observed in this star. {The longitudinal field measurements reported in the catalogue by \citet{bagnulo15}, based on {data collected with the FOcal Reducer and low dispersion Spectrograph 1 (FORS1)}, indicate variability on the time scale of a few years. The rotational broadening corresponding to $v \sin i=2.5$\,km\,s$^{-1}$ was derived by \citet{ryabchikova2008}.} A `mild constraint' on the $v\sin i$ of 3\,km\,s$^{-1}$ was given by \citet{medupe2015}. Consistent with these ground-based observations, we are unable to detect a signal of rotation in the TESS sector 1 data.

\begin{figure}
\centering
\includegraphics[width=0.9\columnwidth]{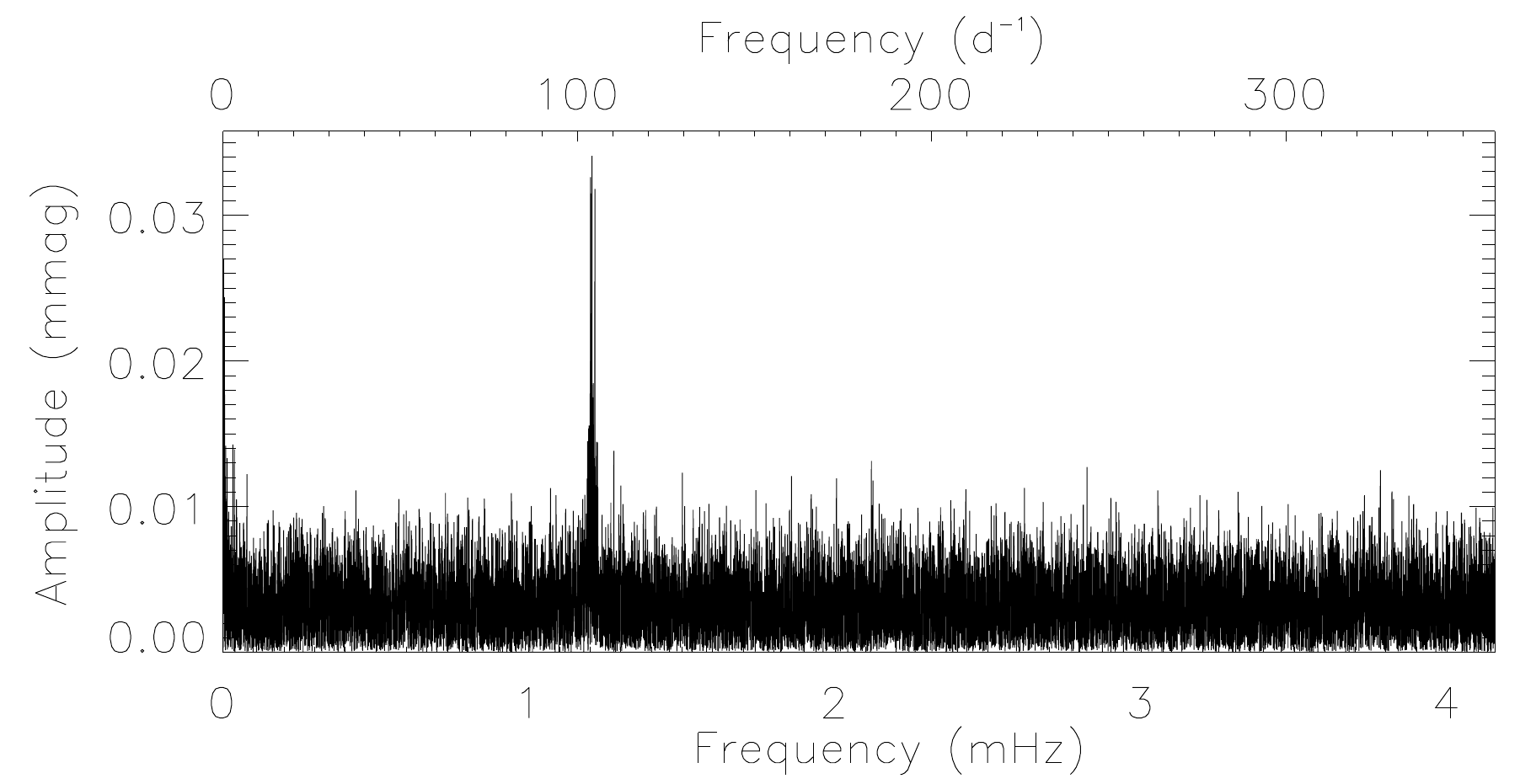}
\includegraphics[width=0.9\columnwidth]{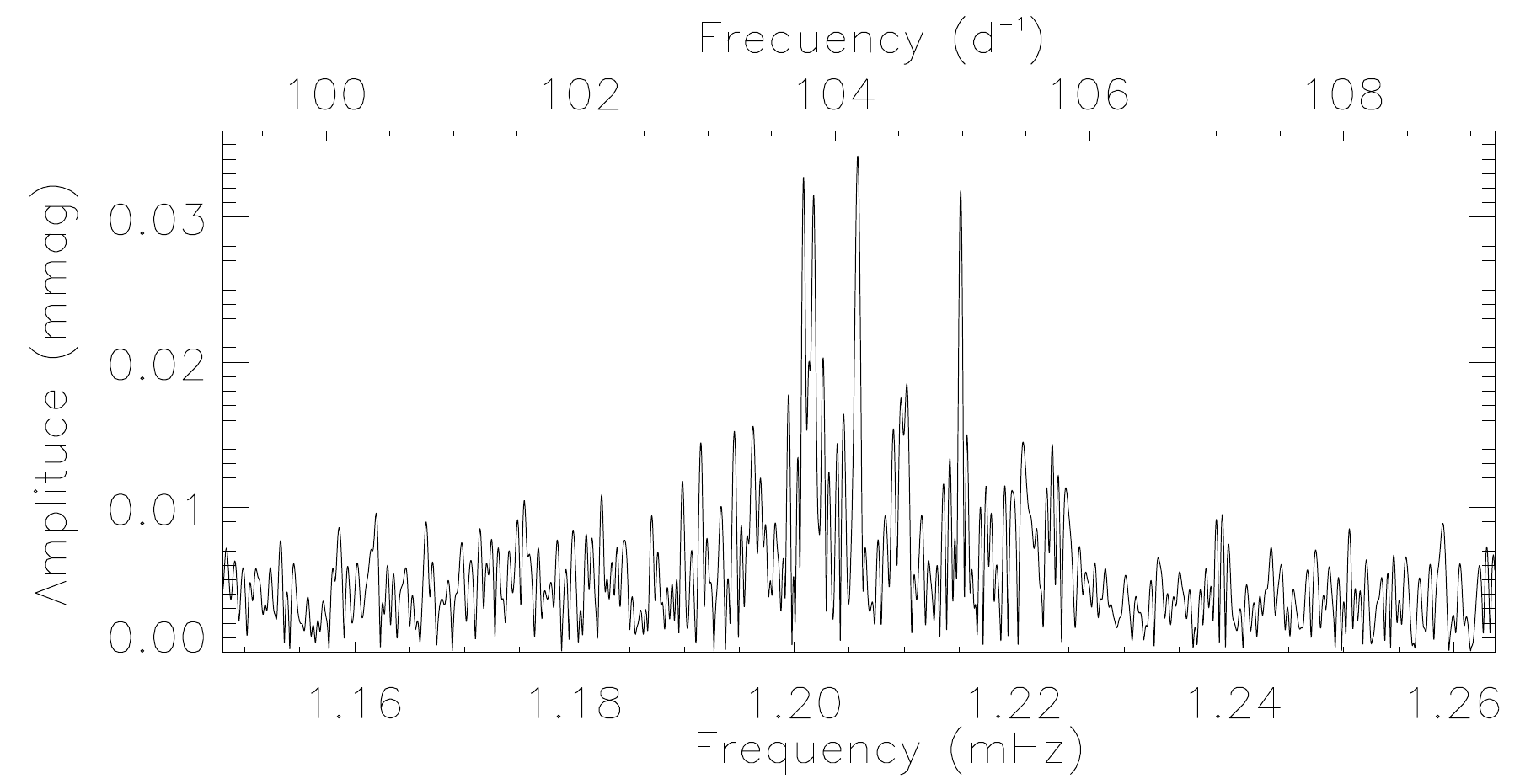}
\includegraphics[width=0.9\columnwidth]{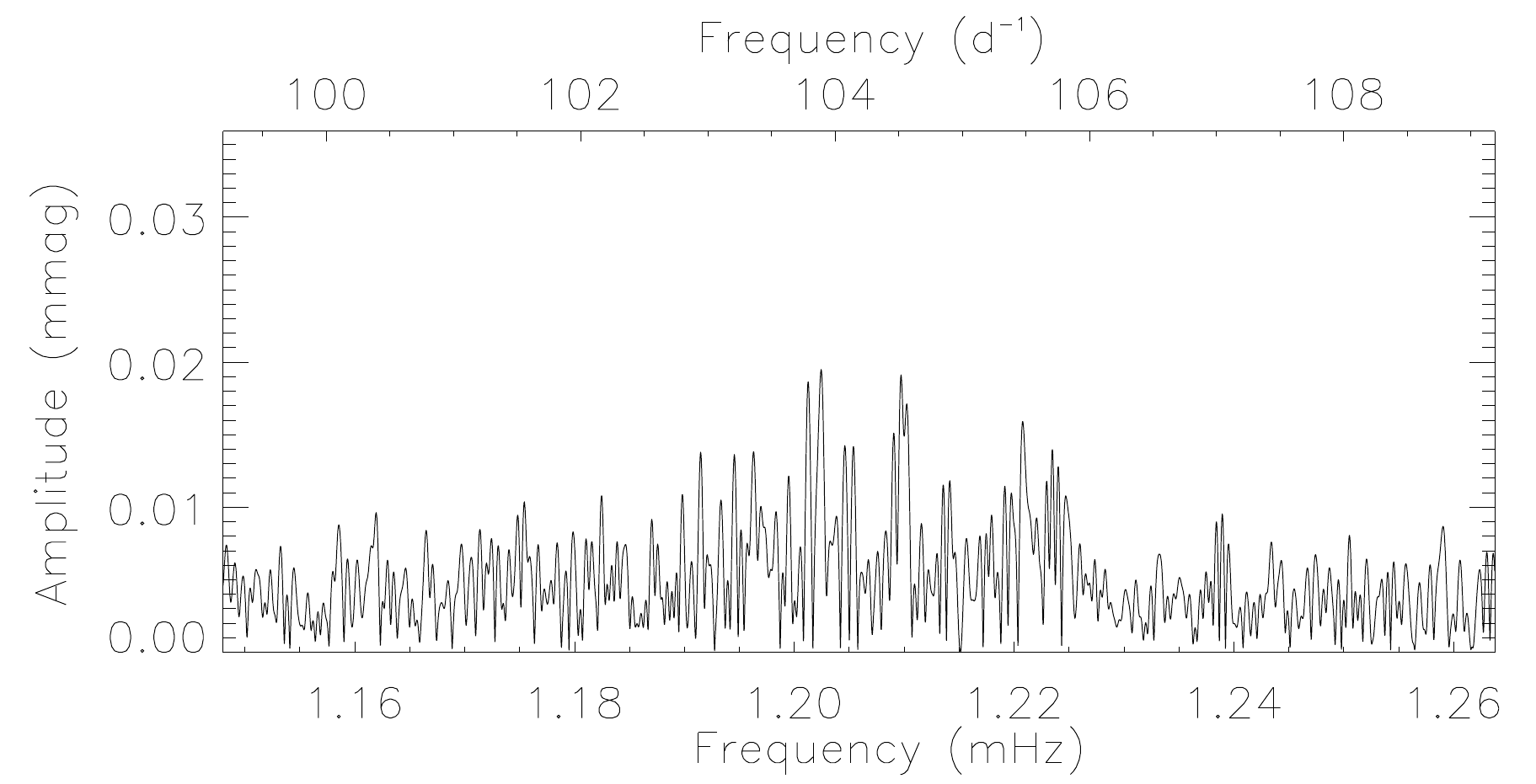}
\caption{Top: full amplitude spectrum of TIC\,139191168 to almost the Nyquist frequency. Middle: zoom of the pulsation modes found in this star. Bottom: the amplitude spectrum of the residuals after removing the frequencies shown in Table\,\ref{tab:139191168}. There is still excess power in the spectrum, indicating further, unresolved modes or frequency modulation in this star.}
\label{fig:139191168_ft}
\end{figure}

The TESS data do clearly show the pulsational variability in TIC\,139191168, though only around 1.2\,mHz (Fig.\,\ref{fig:139191168_ft}). There is no signal at 2\,mHz where pulsations have previously been found. {This is not surprising, given that the spectroscopic studies show that the 2-mHz frequencies are more visible higher in the atmosphere, and the TESS red filter samples more deeply in the atmosphere because of the lower opacity in the red.}

As with the ground-based data, the TESS data show significant evidence of frequency and/or amplitude modulation, or short mode lifetimes. After the removal of four frequencies in TIC\,139191168 (Table\,\ref{tab:139191168}) there is still excess power in the amplitude spectrum (as shown by the bottom panel of Fig.\,\ref{fig:139191168_ft}). The two higher frequency peaks listed in Table\,\ref{tab:139191168} are separated by $\sim9\,\umu$Hz, which could be the small frequency  separation, {{\it i.e.} the separation between the almost degenerate frequencies of modes of radial orders differing by one and degrees differing by two \cite[e.g.][]{cunha07}}, although since earlier studies of this star have suggested growth and decay of mode amplitude, that could also be consistent here.

\begin{table}
\centering
\caption{Details of the pulsation modes found in TIC\,139191168. The zero-point for phases is BJD\,2458339.24058.}
\label{tab:139191168}
\begin{tabular}{lccrr}
\hline
ID & Frequency & Amplitude & \multicolumn{1}{c}{Phase} & \multicolumn{1}{c}{S/N}\\
    & (mHz)  		 & (mmag) & \multicolumn{1}{c}{(rad)} & \\
                           &  &  \multicolumn{1}{c}{$\pm 0.003$} & &\\
\hline
$\nu_1$ & $1.200877\pm0.000023$ & $0.031$ & $-1.73\pm0.10$ & 10.3 \\
$\nu_2$ & $1.201677\pm0.000026$ & $0.029$ & $-2.11\pm0.11$ & 9.7 \\
$\nu_3$ & $1.205792\pm0.000021$ & $0.032$ & $0.67\pm0.09$ & 10.7 \\
$\nu_4$ & $1.215135\pm0.000021$ & $0.031$ & $-1.59\pm0.10$ & 10.3 \\
\hline
\end{tabular}
\end{table}

Although we do not detect the pulsation at 2.0\,mHz (probably a result of the red filter), we agree with \citet{medupe2015} that the importance of TIC\,139191168 lies in the large frequency difference between the 1.2\,mHz and 2.0\,mHz pulsations, since there must be a large number of non-excited modes between those. Unfortunately, we will not be able to solve this problem with the TESS data. {No further observations of this star are planned.}

\subsection{TIC\,167695608}

TIC\,167695608 (TYC 8912-1407-1) was found to be an roAp star by \citet{holdsworth14b} through a survey of A stars in the SuperWASP archive. In the broad-band WASP data, the pulsation was detected at a frequency of $1.53$\,mHz with an amplitude of 0.79\,mmag. The star was spectroscopically classified in the same work as F0p\,SrEu(Cr). As this was not previously known to be a chemically peculiar star, no Str\"omgren nor H$\beta$ indices exist in the literature. The star is relatively faint among the known roAp stars ($V=11.513$), and is a recent addition to the class which explains the lack of previous study in the literature. The temperature derived in this work, $T_{\rm eff} = 7460 \pm 160$\,K (Table\,\ref{properties}), is consistent with that derived from Balmer line fitting of {low-resolution spectra \citep{holdsworth14b}}. Of the known roAp stars, TIC\,167695608 will have the most complete data set from TESS -- it will be observed in 12 of the 13 sectors, being missed only in sector 5. Therefore, the analysis presented here will be greatly improved by the end of TESS's cycle 1.

\begin{figure}
\centering
\includegraphics[width=0.9\columnwidth]{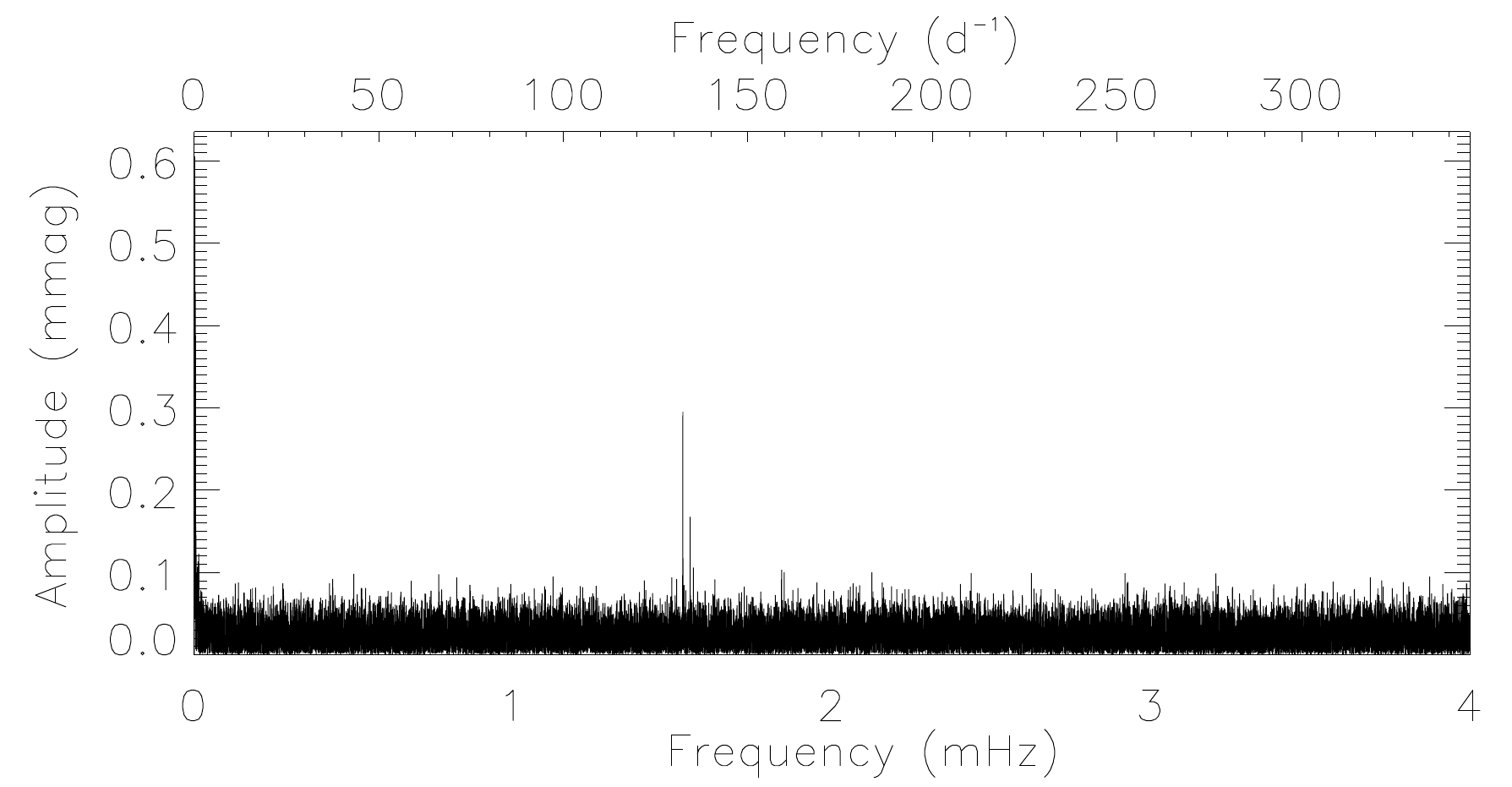}
\includegraphics[width=0.9\columnwidth]{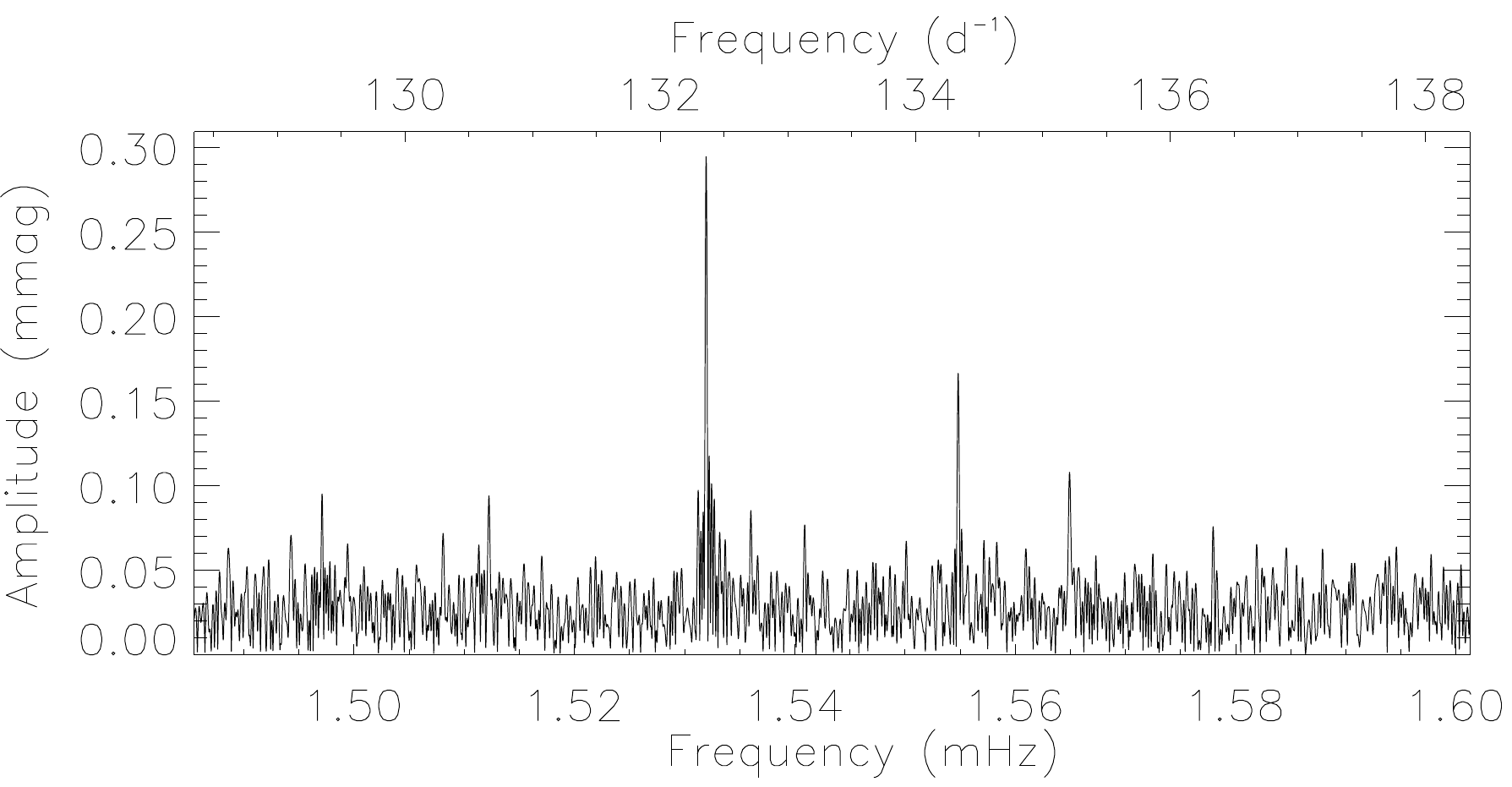}
\includegraphics[width=0.9\columnwidth]{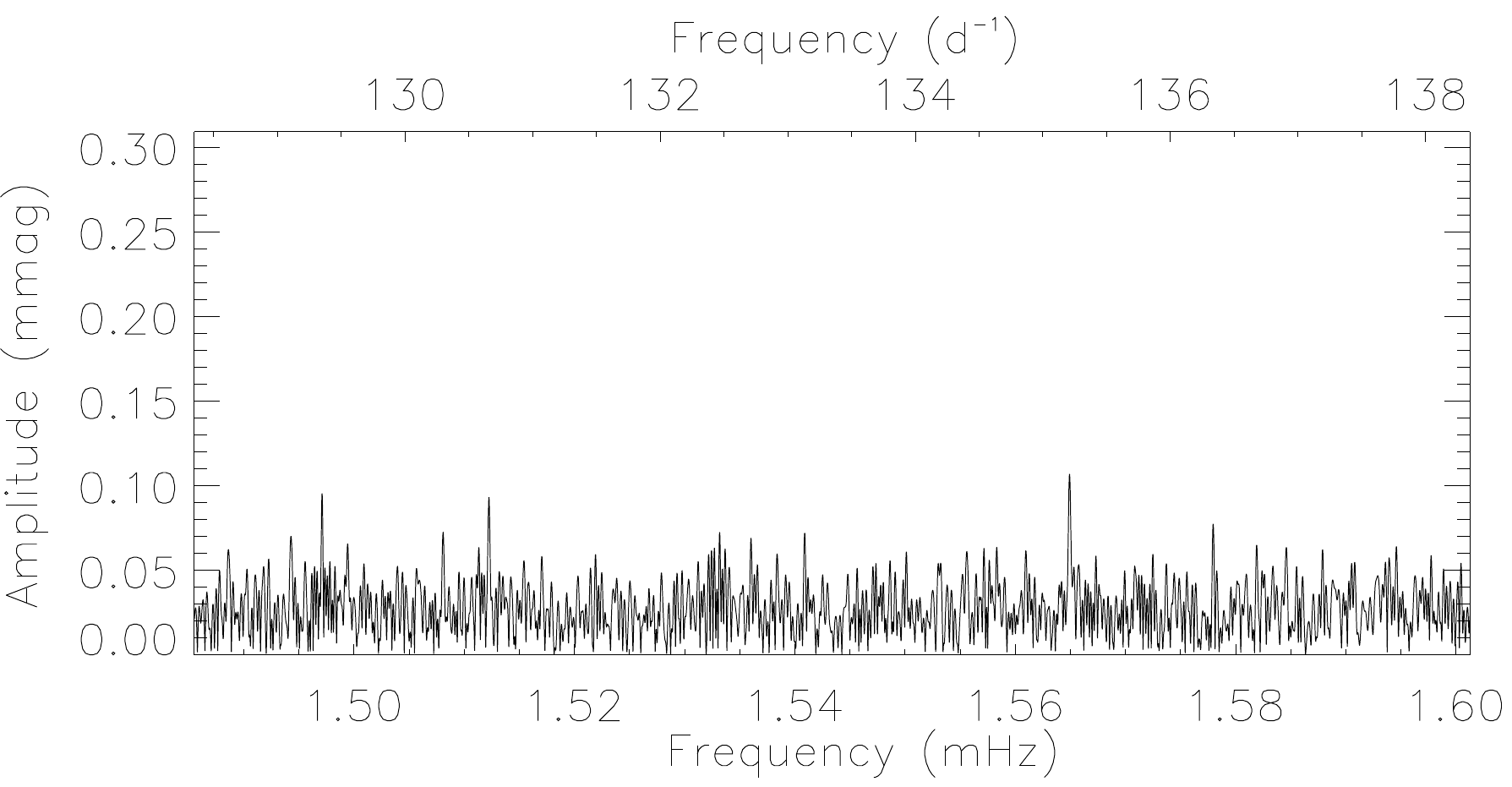}
\caption{Top: amplitude spectrum of TIC\,167695608 to nearly the Nyquist frequency. The roAp pulsations are clear at about 1.55\,mHz. Middle: detailed view of the pulsation modes in this star. Bottom: amplitude spectrum of the residuals after removing all of the frequencies shown in Table\,\ref{tab:167695608}. There are potentially further modes present in this star, however the current data are inconclusive.}
\label{fig:167695608_ft}
\end{figure}

There is no clear signal of rotation in TIC\,167695608, which is consistent with the longer time-base observations from SuperWASP. The middle panel of Fig.\,\ref{fig:167695608_ft} shows a detailed view of the pulsation modes after a highpass filter has been applied. The dominant mode, at 1.53\,mHz, is already known, {but additional, lower amplitude modes have been revealed by these TESS data}. Furthermore, there may be additional modes present in this star. For the modes we are confident of, we fit them by non-linear least-squares and show the results in Table\,\ref{tab:167695608}.

\begin{table}
\centering
\caption{Details of the pulsation modes found in TIC\,167695608. The zero-point for phases is BJD\,2458381.51320.}
\label{tab:167695608}
\begin{tabular}{lccrr}
\hline
ID & Frequency & Amplitude & \multicolumn{1}{c}{Phase} & \multicolumn{1}{c}{S/N} \\
    & (mHz)  		 & (mmag) & \multicolumn{1}{c}{(rad)}&\\
                               &  &  \multicolumn{1}{c}{$\pm 0.022$} & &\\
\hline
$\nu_1$ & $1.531950\pm0.000008$ & $0.294$ & $1.62\pm0.14$ & 13.4 \\
$\nu_2$ & $1.554816\pm0.000015$ & $0.164$ & $-2.023\pm0.26$ & 7.5 \\
$\nu_3$ & $1.564918\pm0.000022$ & $0.107$ & $3.10\pm0.40$ & 4.9 \\
\hline
\end{tabular}
\end{table}

The frequencies found at $\nu_1$ and $\nu_2$ in this star are separated by $22.9\,\umu$Hz which could represent half of the large frequency separation. The difference between $\nu_2$ and $\nu_3$ could be the small separation. However, as there are many more data to come for this star, we will not claim that here.

\subsection{TIC\,211404370}
\label{hd203932}

TIC\,211404370 (HD\,203932) was classified as Ap\,SrEu by \citet{houk1982}. \citet{martinez1993} measured Str\"omgren and H$\beta$ indices for this star: $V = 8.820; b-y = 0.175; m_1 = 0.196; c_1 = 0.742; \beta = 2.694$, from which the parameters $\delta m_1 = 0.004$ and $\delta c_1 = -0.020$.  The H$\beta$ index indicates an equivalent spectral type near mid-F, so this is one of the coolest roAp stars. The TIC temperature and gravity, $T_{\rm eff} = 7500$\,K,  $\log g = 4.1$, suggest an equivalent spectral type around F0. \citet{gelbmann1997} performed an abundance analysis on TIC\,211404370 and showed it to be an Ap star with abundances similar to $\alpha$\,Cir. They derived $T_{\rm eff} = 7540 \pm 100$\,K and $\log g = 4.30 \pm 0.15$, putting the star close to the zero-age main-sequence. Their metallicity, $\rm{[M/H]} = 0.0 \pm 0.1$ is consistent with the Str\"omgren parameters. Here we derive an effective temperature of $7370 \pm 160$\,K, even lower than the previous value, albeit consistent given the errors. \citet{hubrig2004} derived a longitudinal magnetic field strength of $\left<B_z\right> = -267 $\,G from FORS1 data for TIC\,211404370 and later \cite{ryabchikova2008} have set a limit of 1\,kG to the star's magnetic field modulus.

TIC\,211404370 was discovered to be an roAp star by \citet{kurtz1984}. It was studied further photometrically by \citet{kurtz1988}, who found it to be multiperiodic, with pulsation mode frequencies in the range of $2.8 - 2.9$\,mHz ($P$ = 5.9\,min), making this, at the time, the shortest known period for an roAp star. However, later multi-site observations by \citet{martinez1990} found only the dominant pulsation frequency, leading them to suggest that the other frequencies found by \citet{kurtz1988} were possibly transient. 

\begin{figure}
\centering
\includegraphics[width=0.9\columnwidth]{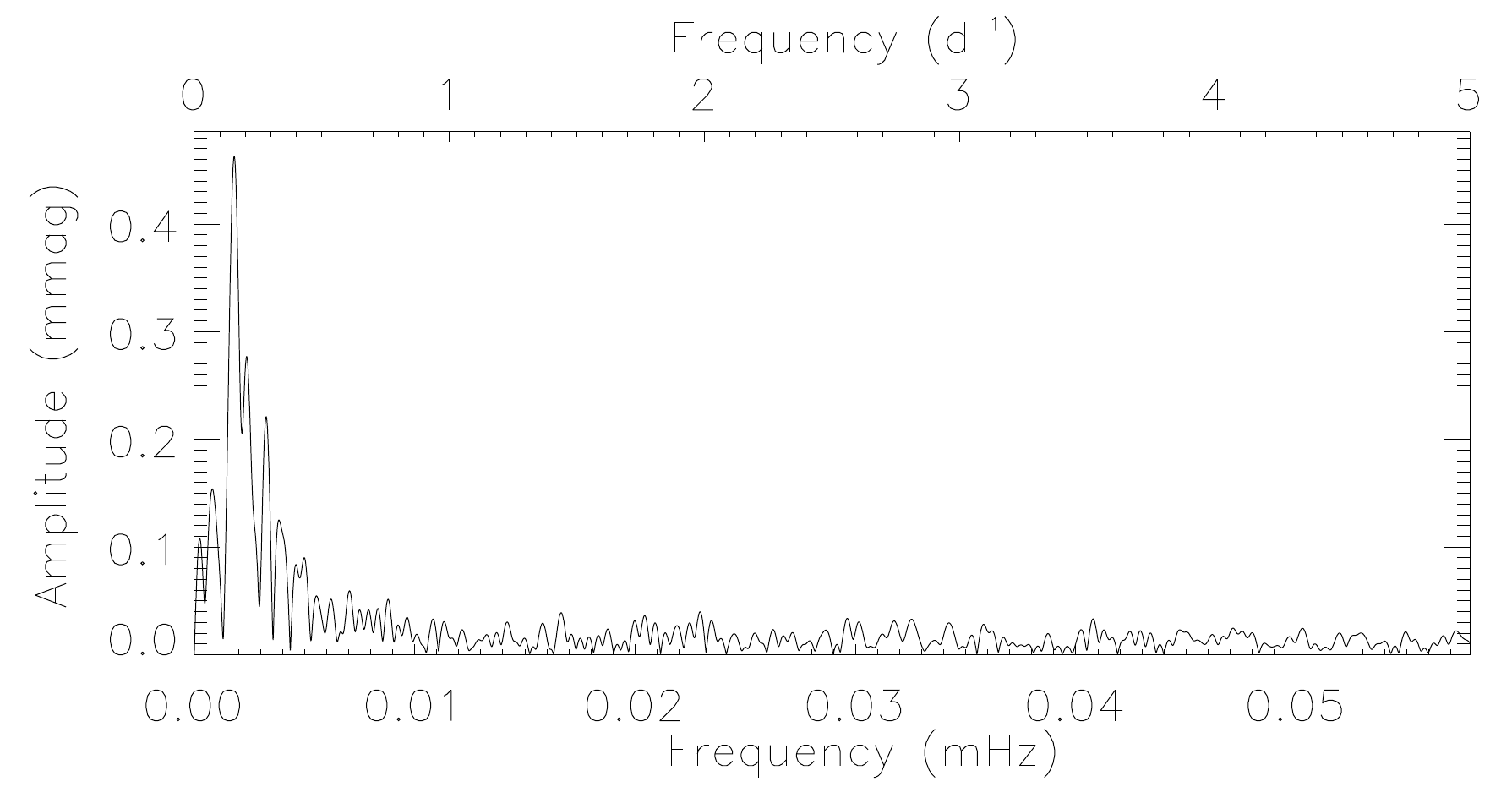}
\includegraphics[width=0.9\columnwidth]{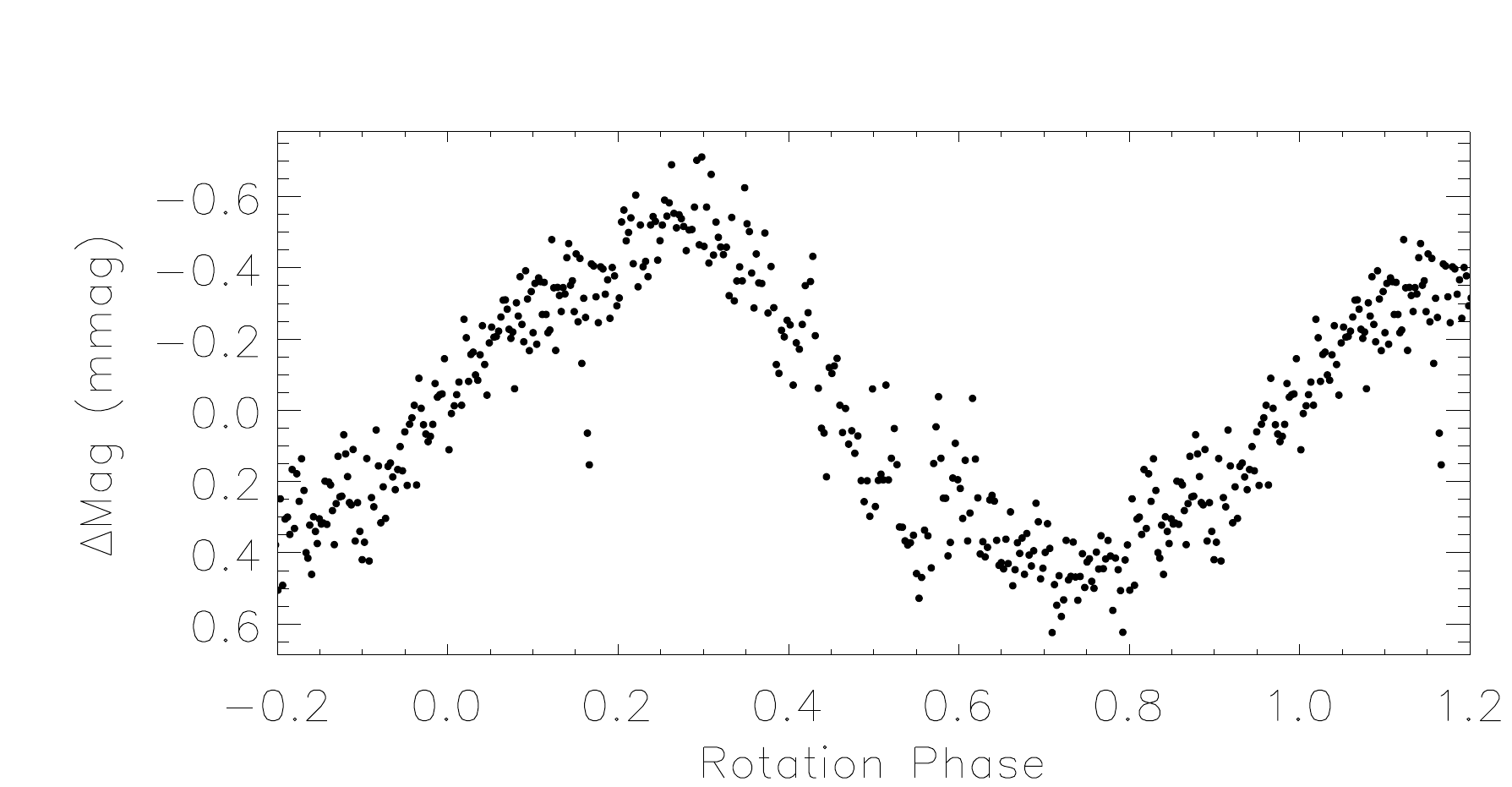}
\includegraphics[width=0.9\columnwidth]{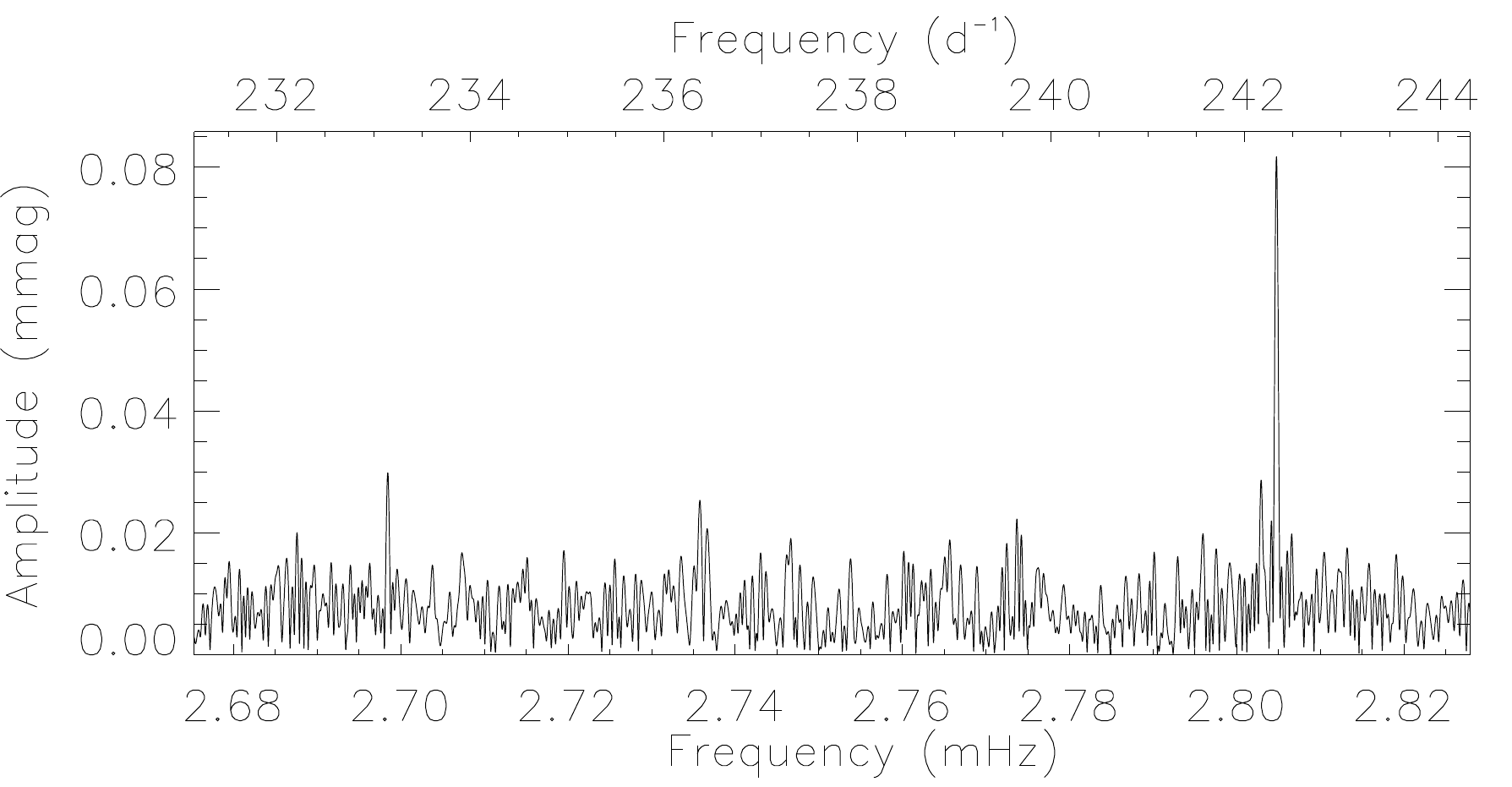}
\includegraphics[width=0.9\columnwidth]{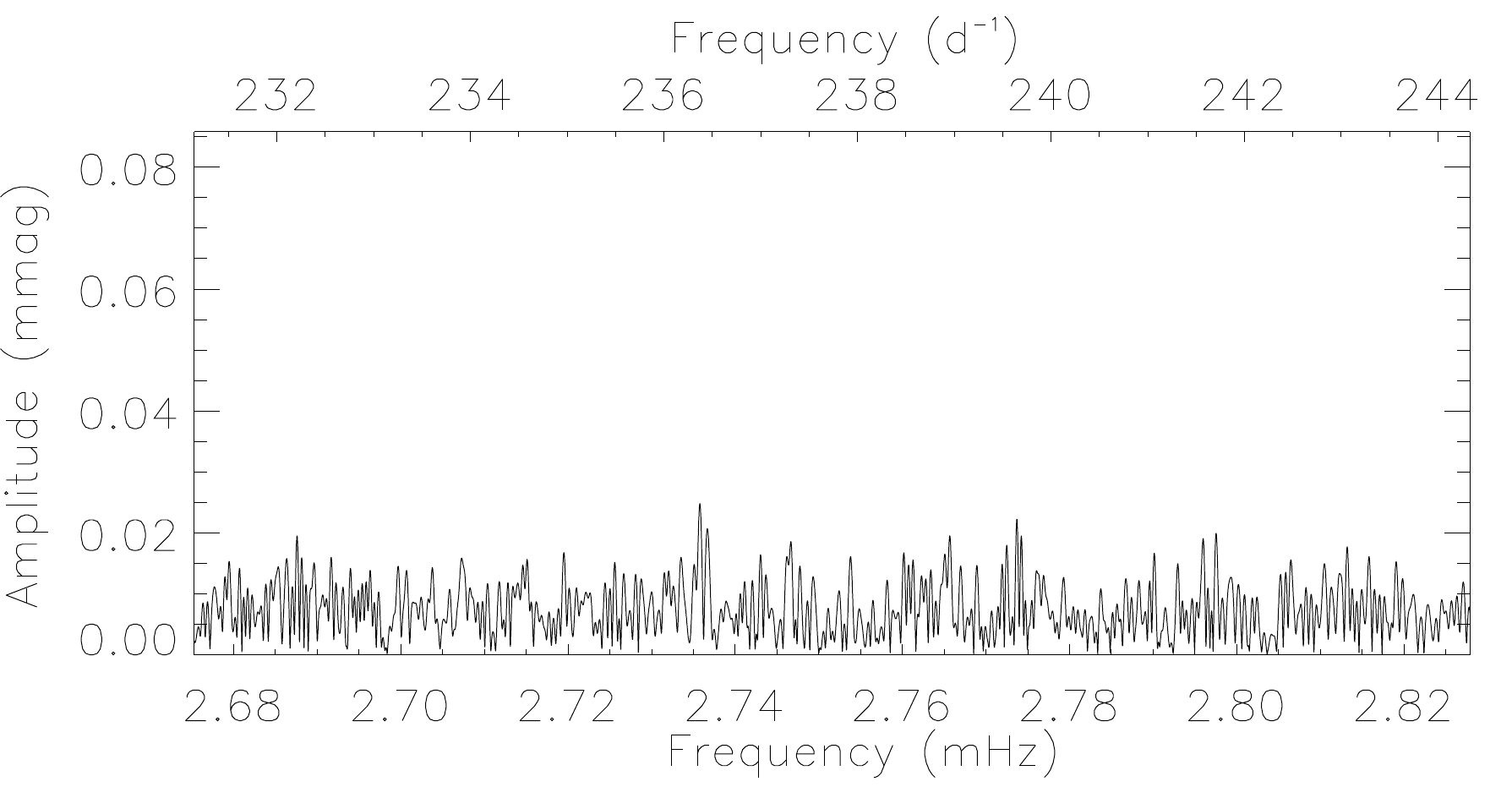}
\caption{Top: low-frequency amplitude spectrum of TIC\,211404370. The dominant peak suggests a rotation period of about $6.4$\,d. There are also harmonics and instrumental artefacts present. Second: phase-folded light curve on the determined rotation period. The single wave nature suggests only one magnetic pole is seen. Third: amplitude spectrum showing the pulsation modes in the star. Bottom: amplitude spectrum of the residuals after removing the three frequencies listed in Table\,\ref{tab:211404370_OPM}.}
\label{fig:211404370_ft}
\end{figure}

TIC\,211404370 was observed during sector 1 of TESS's primary mission. The observations reveal the star to be an $\alpha^2$~CVn variable, with a rotation period of $6.442\pm0.012$\,d. This is the first measurement of the rotation period of this star. The top panel of Fig.\,\ref{fig:211404370_ft} shows the low-frequency amplitude spectrum of the TESS light curve. Clearly visible is the rotation signal, with some harmonics present. Also mixed into the stellar signal are some instrumental artefacts {which are evidenced by the broad power at low-frequency and deviations from a smooth curve in the phase-folded light curve at phase 0.55.} 

We detected two pulsation modes, with a third frequency being a rotationally split sidelobe. The frequencies are given in Table\,\ref{tab:211404370_OPM}. An inspection of the $\nu_2$ peak in Fig.\,\ref{fig:211404370_ft} gives an indication of a second sidelobe to this mode at $\nu_2+\nu_{\rm rot}$. To investigate this further, and to apply the oblique pulsator model to TIC\,211404370, we fitted a triplet with sidelobes split by exactly the rotation frequency to the data by linear least-squares. The results are shown in Table\,\ref{tab:211404370_OPM}.

\begin{table}
\centering
\caption{A linear least-squares fit to the presumed triplet at $\nu_2$ in TIC\,211404370. The frequencies are split by exactly the rotation frequency, and the zero-point in time is chosen so the sidelobe phases are equal. The zero-point is BJD\,2458341.53111.}
\label{tab:211404370_OPM}
\begin{tabular}{llcrr}
\hline
ID & \multicolumn{1}{c}{Frequency} & Amplitude & \multicolumn{1}{c}{Phase} & \multicolumn{1}{c}{S/N}\\
    & \multicolumn{1}{c}{(mHz)}  		 & (mmag) & \multicolumn{1}{c}{(rad)} & \\
                                  &  &  \multicolumn{1}{c}{$\pm 0.006$} & &\\
\hline

$\nu_1$ & $2.698436\pm0.000046$ & $0.030$ & $-0.89\pm0.20$ & 5.0\\
$\nu_2-\nu_{\rm rot}$ & $2.802956$ & $0.029$ & $-0.75\pm0.21$ & 4.8\\
$\nu_2$ & $2.804756\pm0.000016$ & $0.082$ & $-2.84\pm0.07$ & 13.7\\
$\nu_2+\nu_{\rm rot}$ & $2.806556$ & $0.016$ & $-0.75\pm0.38$ & 2.7\\

\hline
\end{tabular}
\end{table}

The $\nu_2+\nu_{\rm rot}$ sidelobe has a significance of $2.7\sigma$. With the sidelobe present, we can apply the dipole form of the oblique pulsator model, as shown in equation\,(\ref{eq:OPM_trip}). We find that $\tan i\tan\beta=0.54\pm0.11$. Although we cannot disentangle $i$ and $\beta$, we  illustrate their relationship in Fig.\,\ref{ibeta}. We find that $i+\beta\leq90^\circ$ which is consistent with the observed single-wave light curve shown Fig.\,\ref{fig:211404370_ft}, i.e. only one pulsation {pole} and {one} magnetic pole are seen.

Previous ground-based observations of TIC\,211404370 have often reported {that it pulsates in multiple frequencies}, but with the presence of different modes at different epochs. In both the work by \citet{kurtz1988} and \citet{martinez1990}, the authors report separations between modes of about $35\,\umu$Hz, or multiples thereof. Here, the difference between $\nu_1$ and $\nu_2$ is {$106.3\,\umu$Hz}, or about $3\times35\,\umu$Hz. Therefore, the TESS observations of TIC\,211404370 are consistent with previous studies of this star, showing transient frequencies with separations of 35\,$\umu$Hz which could represent the large frequency separation, or half of it. From the parameters in Table\,\ref{properties}, we estimate $\Delta\nu = 73 \pm 32 \, \umu$Hz, the large uncertainty resulting from the uncertainty in the Gaia DR2 parallax for this star, {$\pi=5.2 \pm 0.9$~mas}. Despite that, the value obtained from the scaling indicates that 35\,$\umu$Hz most likely represents half of the large frequency separation in this star. 

\subsection{TIC\,237336864}
\label{hd218495}

TIC\,237336864 (HD\,218495) was classified as Ap\,EuSr by \citet{houk1975}. \citet{martinez1993} measured Str\"omgren and H$\beta$ indices for this star: $V = 9.356; b-y = 0.114; m_1 = 0.252; c_1 = 0.812; \beta = 2.870$, from which the parameters $\delta m_1 = -0.049$ and $\delta c_1 = -0.098$ can be derived; both of these indices are indicative of a strong Ap star. The H$\beta$ index indicates an equivalent spectral type near mid-A. The effective temperature derived here, $7950 \pm 160$\,K, is consistent with that. A longitudinal magnetic field strength of $\left<B_z\right> = -912$\,G has been detected in this star by \citet{hubrig2004} based on FORS1 data.

Rapid pulsations were discovered in TIC\,237336864 by \citet{martinezkurtz1990} who found an oscillation at 2.24\,mHz. No further photometric observations of this star have been published since that time. 

\begin{figure}
\centering
\includegraphics[width=0.9\columnwidth]{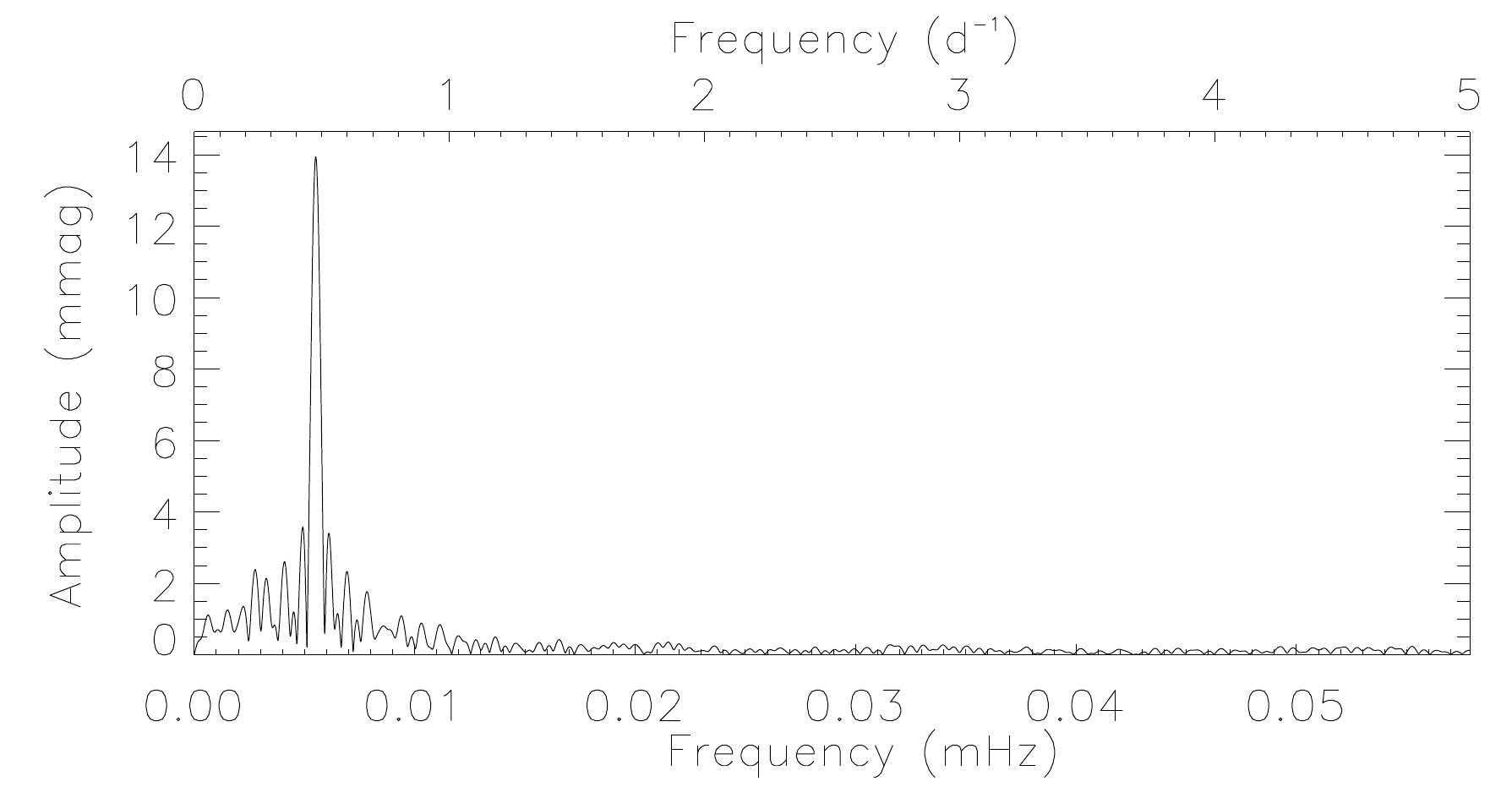}
\includegraphics[width=0.9\columnwidth]{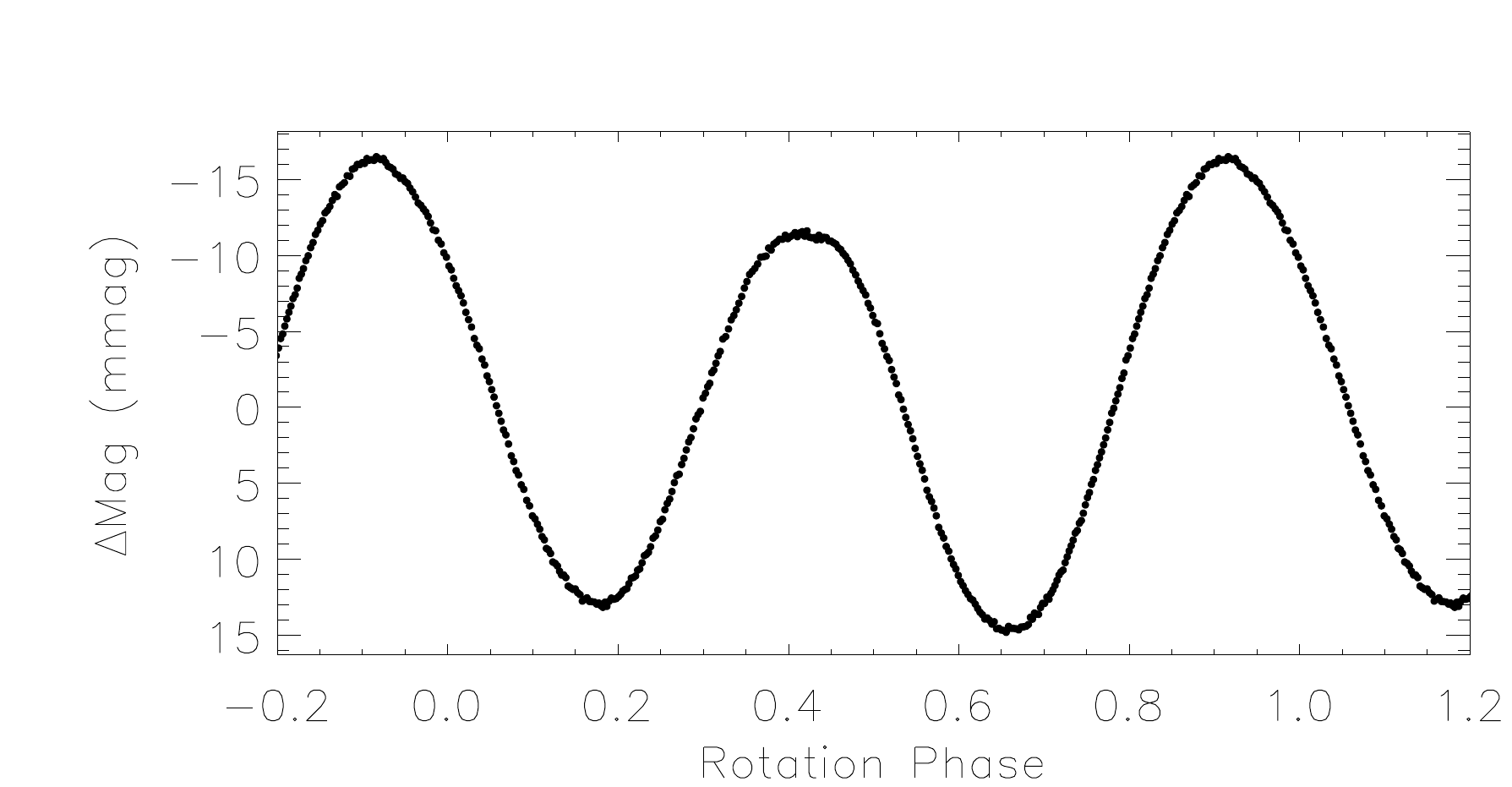}
\caption{Top: low-frequency amplitude spectrum of TIC\,237336864. The dominant peak suggests a rotation period of about $2.1$\,d, however we find this to be the first harmonic of the rotation frequency. Bottom: phase-folded light curve, folded on a period of $4.2006\pm0.0001$\,d. The unequal maxima and minima show this to be correct period. The light curve has been binned 50:1.}
\label{fig:237336864_rot}
\end{figure}

TESS observed TIC\,237336864 only during sector 1. The top panel of Fig.\,\ref{fig:237336864_rot} shows the amplitude spectrum of these data. There is a significant peak which corresponds to a period of 2.1\,d. However, this is a case where the harmonic of the rotation has a much more significant signal than the true rotation frequency. To prove we have determined the correct period (as shown in Table\,\ref{properties}) we present a phase-folded light curve in the bottom panel of Fig.\,\ref{fig:237336864_rot} which shows the double-wave nature of the rotational modulation. We extracted the rotation frequency and processed an 8-harmonic series using both linear and non-linear least-squares to derive the rotation frequency, $\nu_{\rm rot} = 0.00275536\pm0.00000007$\,mHz, which gives a rotation period of $P_{\rm rot} = 4.2006\pm0.0001$\,d.

\begin{figure}
\centering
\includegraphics[width=0.9\columnwidth]{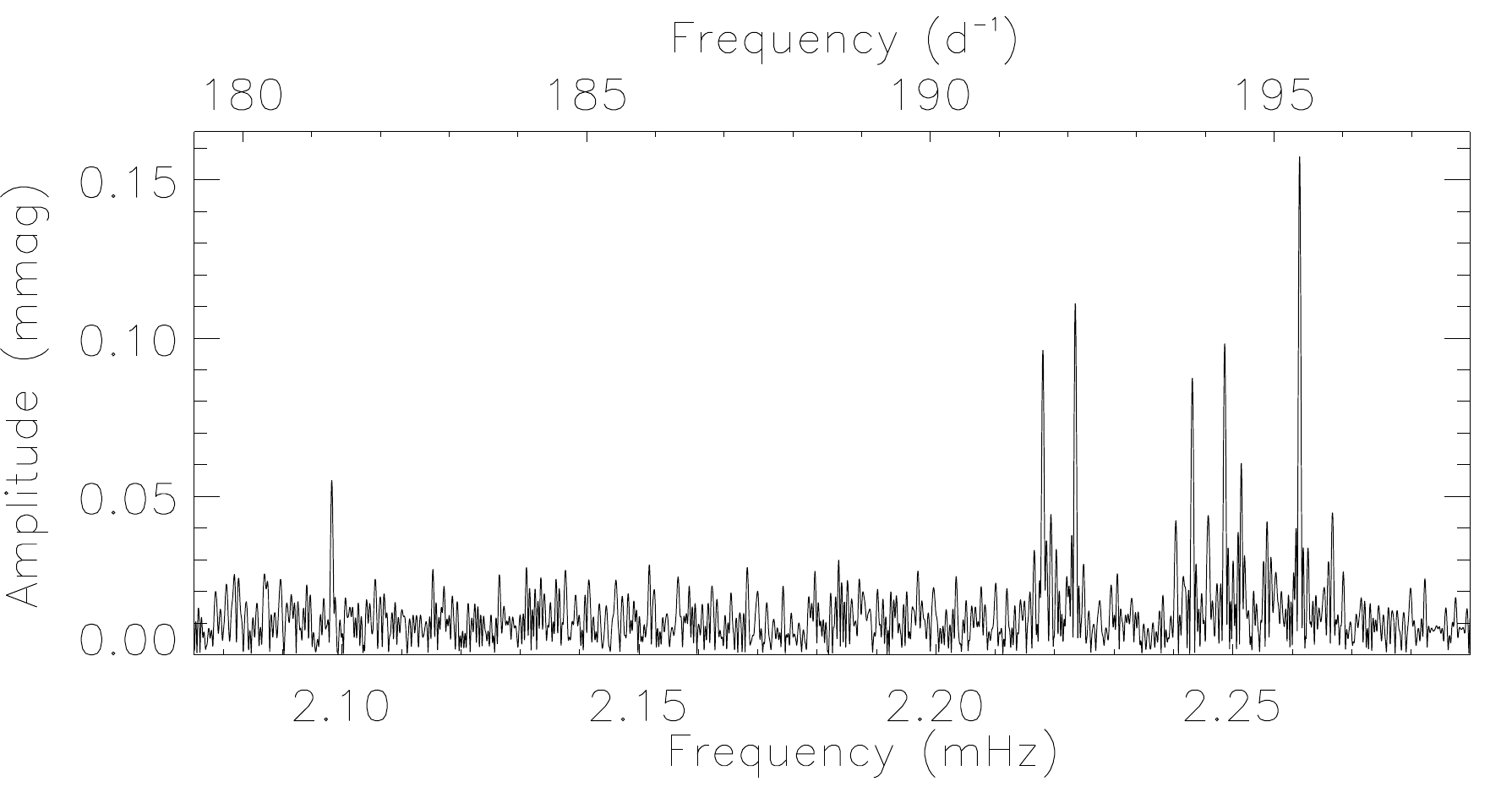}
\includegraphics[width=0.9\columnwidth]{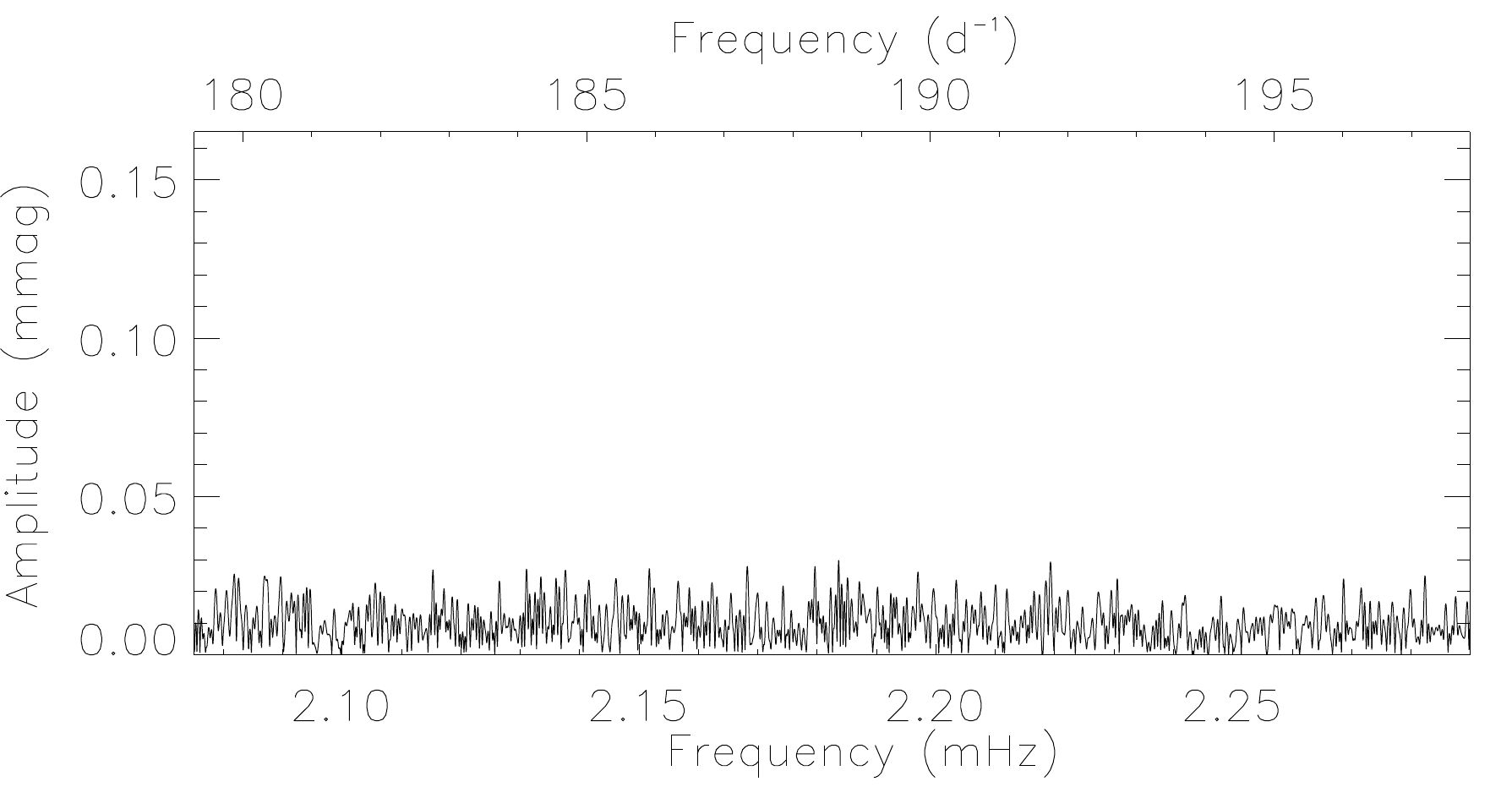}
\caption{Top: amplitude spectrum of TIC\,237336864 showing the pulsations detected in this star. Bottom: amplitude spectrum of the residuals after removing all of the frequencies shown in Table\,\ref{tab:237336864_forced}. There are no remaining peaks with significant amplitude.}
\label{fig:237336864_ft}
\end{figure}

As outlined earlier, we pre-whitened the rotational variations and some low-frequency artefacts to create a highpass filtered data set to study the pulsation frequencies. This did not affect those pulsation frequencies, and provides a better estimate of errors in least-squares fitting by removing the variance at low frequency. The top panel of Fig.\,\ref{fig:237336864_ft} shows an amplitude spectrum of the pulsation modes in TIC\,237336864; this is a much richer spectrum than previously reported from ground-based observations. The plot shows a singlet, a doublet, a quintuplet, and a triplet of frequencies. The multiplets are all split by either $\nu_{\rm rot}$ or 2$\nu_{\rm rot}$. We optimised the frequencies by non-linear least-squares fitting, {then forced the multiplets to be equally split by $\nu_{\rm rot}$ or 2$\nu_{\rm rot}$}. Table\,\ref{tab:237336864_forced} shows the derived frequencies. We found that this fitted the data as well as the non-linear least-squares determined frequencies. The bottom panel of Fig.\,\ref{fig:237336864_ft} shows the amplitude spectrum of the residuals with only noise with highest peaks less than 40\,$\umu$mag. The purpose of force fitting the multiplets is to test the phase relationships of the multiplet members, which is informative about the mode geometry within the oblique pulsator model. 

Table\,\ref{tab:237336864_forced} shows those results, where we have chosen the zero-point in time to give equal phases for the doublet.  We note that the phases of the two highest amplitude components of the quintuplet and the outer components of the triplet are equal to within 1$\sigma$. This is what is expected for oblique pulsation, but the different amplitudes for the multiplets suggest either different degrees, $\ell$, or different pulsation axes, or both (see, e.g., \citealt{bigot2011}, \citealt{kurtz2011}). A more detailed examination of this is deferred to a future paper. 

\begin{table}
\centering
\caption{A linear least-squares fit of the frequencies extracted for TIC\,237336864. The zero-point for the phases has been chosen to be BJD\,2458338.92557 to set the phases of the doublet to be equal.}
\label{tab:237336864_forced}
\begin{tabular}{llcrr}
\hline
ID & \multicolumn{1}{c}{Frequency} & Amplitude & \multicolumn{1}{c}{Phase} & \multicolumn{1}{c}{S/N}\\
    & \multicolumn{1}{c}{(mHz)}  		 & (mmag) & \multicolumn{1}{c}{(rad)} & \\
                                      &  &  \multicolumn{1}{c}{$\pm 0.008$} & &\\
\hline
$\nu_1$ & $2.098269\pm0.000032$	& $0.056$ & $1.23\pm0.14$ & 7.1\\
$\nu_2-\nu_{\rm rot}$ & $2.217975$	& $0.099$ & $0.76\pm0.08$ & 12.5\\
$\nu_2+\nu_{\rm rot}$ & $2.223486$	& $0.114$ & $0.76\pm0.07$ & 14.4\\
$\nu_3-2\nu_{\rm rot}$& $2.240365$	& $0.045$ & $1.67\pm0.18$ & 5.7\\
$\nu_3-1\nu_{\rm rot}$& $2.243121$	& $0.090$ & $1.89\pm0.09$ & 11.4\\
$\nu_3$& $2.245876\pm0.000039$	& $0.046$ & $1.22\pm0.17$ & 5.9\\
$\nu_3+1\nu_{\rm rot}$& $2.248631$	& $0.104$ & $2.01\pm0.08$ & 13.1\\
$\nu_3+2\nu_{\rm rot}$& $2.251387$	& $0.058$ & $-0.86\pm0.14$ & 7.3\\
$\nu_4-2\nu_{\rm rot}$& $2.255722$	& $0.047$ & $-2.21\pm0.17$ & 5.9\\
$\nu_4$& $2.261233\pm0.000011$	& $0.159$ & $-1.98\pm0.05$ & 20.1\\
$\nu_4+2\nu_{\rm rot}$& $2.266744$	& $0.048$ & $-1.65\pm0.17$ & 6.1\\
\hline
\end{tabular}
\end{table}

Looking at the quintuplet, $\nu_3$, we were able to constrain the geometry of the star with the oblique pulsator model under equation\,(\ref{eq:OPM_quad}). We obtained $\tan i\tan\beta = 2.11\pm0.26$. We graphically show the relationship in Fig.~\ref{ibeta}, and see that $i+\beta$ is always $>90\degr$ implying we see both pulsation poles, consistent with the double-wave nature of the light curve seen in Fig.\,\ref{fig:237336864_rot}.

We now look at the mode frequency separation, assuming that each multiplet arises from oblique pulsation, so that the central frequency is the actual pulsation frequency. That gives $\nu_{1} = 2.098269$\,mHz; $\nu_{2} = 2.220730$\,mHz, where we have taken the average of the two components to obtain the central frequency of an assumed triplet;  $\nu_{3} = 2.245876$\,mHz; and $\nu_{\rm 4} = 2.261233$\,mHz. We then look at the frequency separations for the mode frequencies of these multiplets: $\nu_{\rm 2} - \nu_{\rm 1} = 122$\,$\umu$Hz; $\nu_{\rm 3} - \nu_{\rm 2} = 25$\,$\umu$Hz; $\nu_{\rm 4} - \nu_{\rm 3} = 15$\,$\umu$Hz. {These are plausibly in the range of the large separation, or half or a multiple of that.} However, scaling from the Sun with the parameters provided in Table\,\ref{properties}, we estimate $\Delta\nu = 91 \pm 16 \, \umu$Hz. Further insight clearly requires detailed modelling of this star. 

\subsection{TIC\,348717688}
\label{hd19918}

TIC\,348717688 (HD\,19918) was classified as Ap\,SrCrEu by \citet{houk1975}. \citet{martinez1993} measured Str\"omgren and H$\beta$ indices for this star: $V = 9.336; b-y = 0.169; m_1 = 0.216; c_1 = 0.822; \beta = 2.855$, from which the parameters $\delta m_1 = -0.010$ and $\delta c_1 = -0.058$ can be derived; both of these indices are indicative of a mild Ap star. The H$\beta$ index indicates an equivalent spectral type near mid-A. Here we derive a temperature of $T_{\rm eff} = 7480 \pm$ 160\,K, lower than {$T_{\rm eff}=8110$\,$\pm$\,150\,K} derived by \citet{ryabchikova2007}. {A mean longitudinal magnetic field strength of {$-625\pm 87$\,G} has been derived from FORS1 data for this star \citep{hubrig2006}. A mean field modulus of 1.6\,kG was obtained from the analysis of high-resolution spectra by \citet{ryabchikova2007}.}

Rapid pulsations were discovered in TIC\,348717688 by \citet{martinez1991} and studied in more detail by \citet{martinez1995}, who found two pulsation modes separated by either the large separation or half of that, depending on mode identification, in observations obtained in the years 1990, 1991, 1992 and 1994. {\citet{martinez1995}} also found a strong harmonic to the principal pulsation frequency.

TIC\,348717688 was observed during TESS's sector 1 and will be revisited during sectors 12 and 13. The large gap between the current data set and subsequent observations will introduce a window pattern {which may interfere with the pulsation analysis}, but with two consecutive sectors, the future data will allow for a more precise analysis than we present here.

\begin{figure}
\centering
\includegraphics[width=0.9\columnwidth]{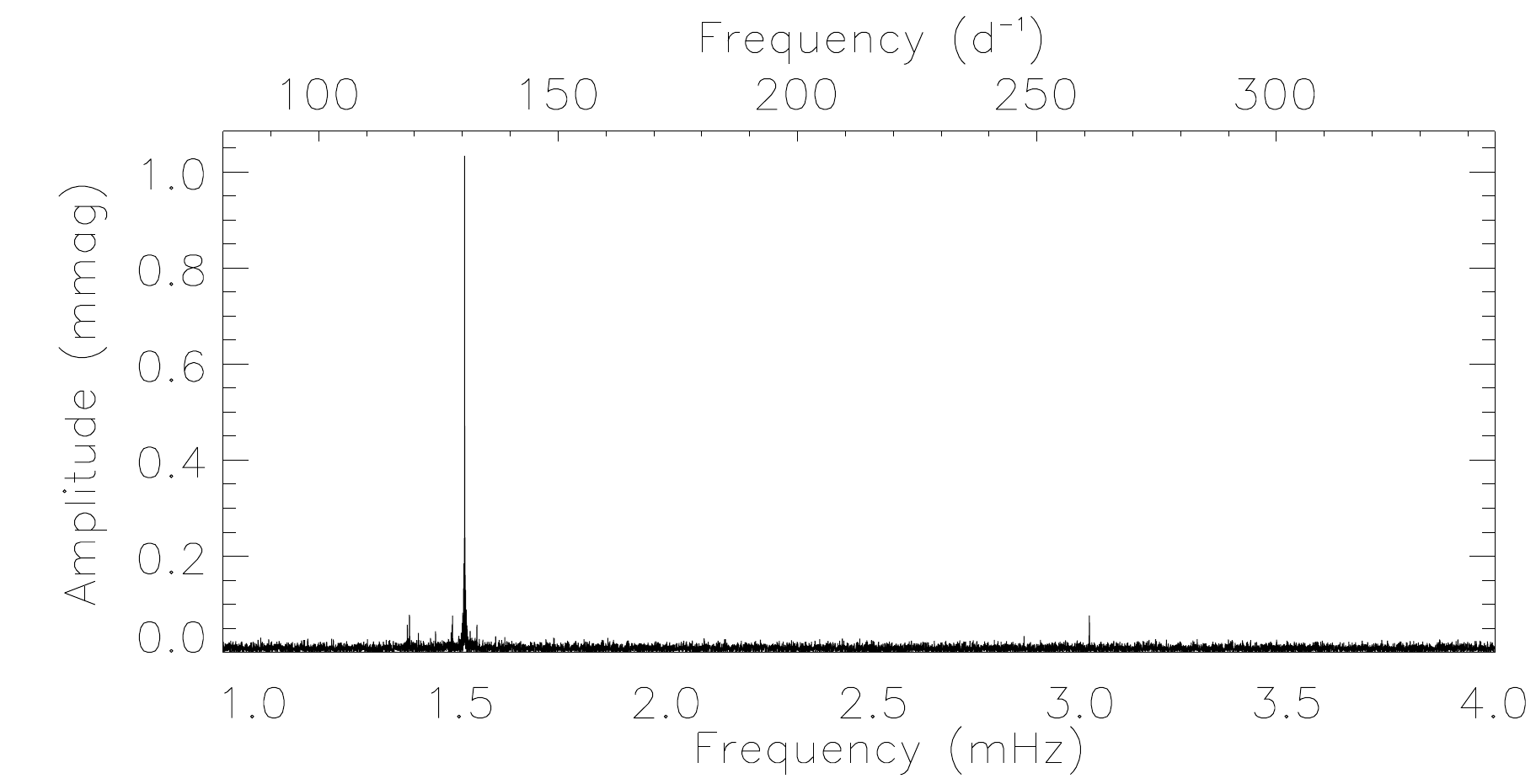}
\includegraphics[width=0.9\columnwidth]{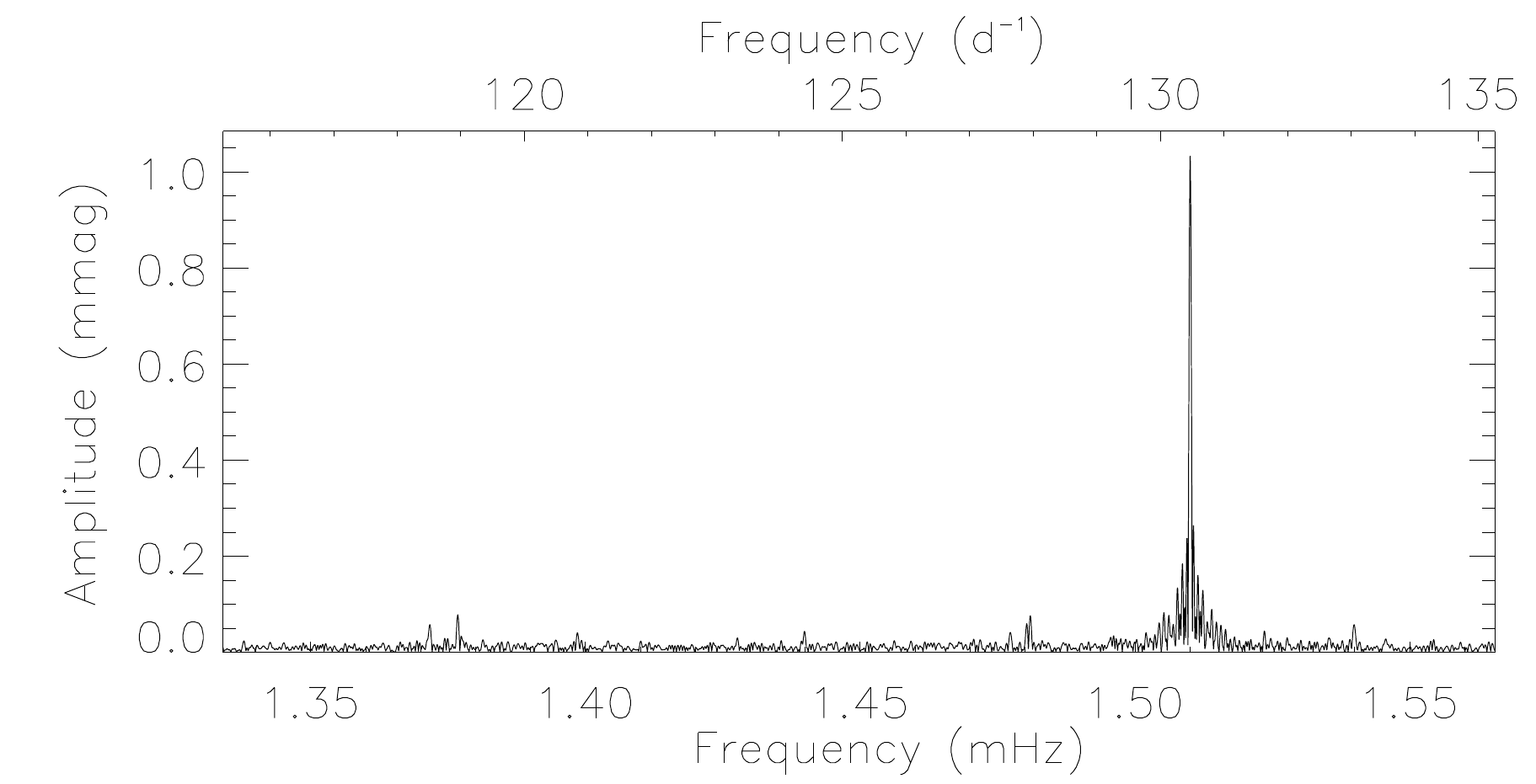}
\includegraphics[width=0.9\columnwidth]{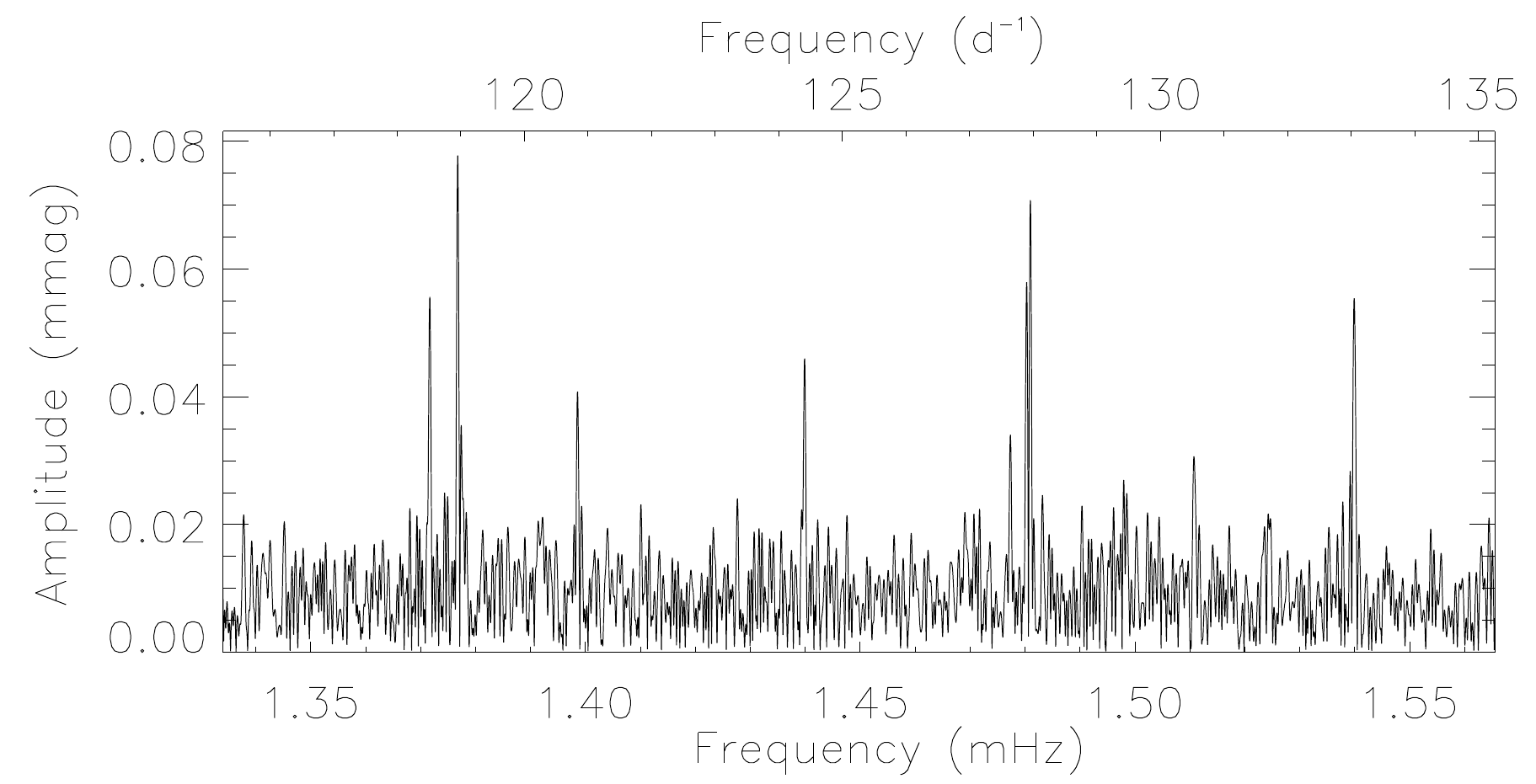}
\includegraphics[width=0.9\columnwidth]{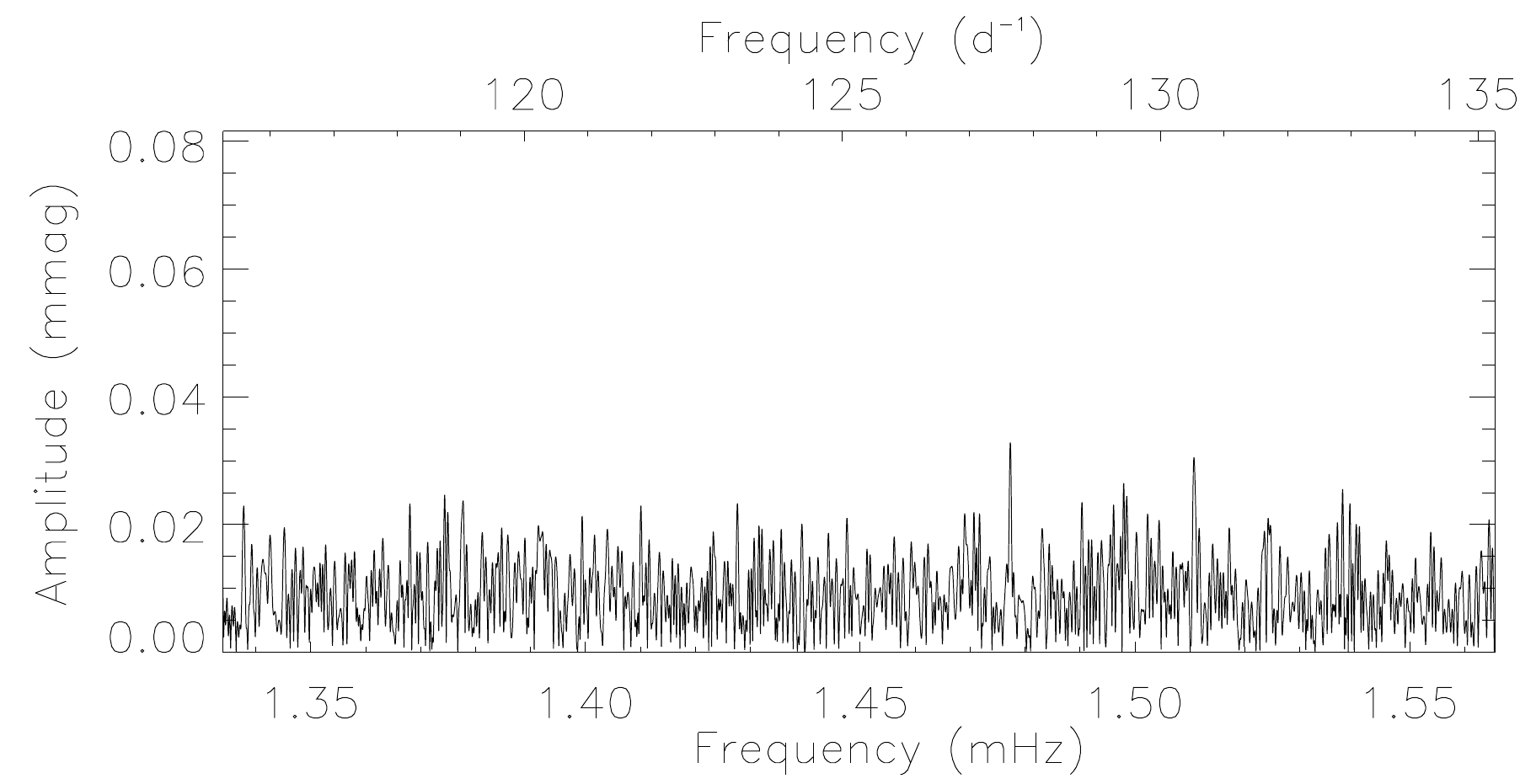}
\caption{Top: amplitude spectrum of TIC\,348717688 to almost the Nyquist frequency. Evident is the principal pulsation mode and its harmonic. Second: zoom of the pulsation modes in this star. Third: the amplitude spectrum after removing the principal mode to show the presence of further, low-amplitude, modes. Note the change in the ordinate scale. Bottom: amplitude spectrum of the residuals after removing all of the frequencies shown in Table\,\ref{tab:348717688}. There are no remaining peaks with significant amplitude.}
\label{fig:348717688_ft}
\end{figure}

The top panel of Fig.\,\ref{fig:348717688_ft} shows an amplitude spectrum of the sector 1 data, where the principal pulsation frequency, $\nu_7 = 1.510057 \pm 0.000002$\,mHz, its harmonic at $2\nu_7$ and other significant peaks can be seen. Those are clearer in the second panel which has a higher frequency resolution. The third panel shows the amplitude spectrum after pre-whitening by $\nu_7$, where other pulsation mode frequencies are seen better. The bottom panel shows an amplitude spectrum of the residuals after removing all of the frequencies shown in Table\,\ref{tab:348717688}.

\begin{table}
\centering
\caption{Details of the pulsation mode found in TIC\,348717688. The zero-point for phases is BJD\,2458339.23761.}
\label{tab:348717688}
\begin{tabular}{lccrr}
\hline
ID & Frequency & Amplitude & \multicolumn{1}{c}{Phase} & \multicolumn{1}{c}{S/N} \\
    & (mHz)  		 & (mmag) & \multicolumn{1}{c}{(rad)} &\\
                                          &  &  \multicolumn{1}{c}{$\pm 0.007$} & &\\
\hline
$\nu_1$ & $1.371648\pm0.000029$ & $0.055$ & $-0.471\pm0.129$ & 7.9 \\
$\nu_2$ & $1.376747\pm0.000021$ & $0.077$ & $2.345\pm0.093$ & 11.0\\
$\nu_3$ & $1.398550\pm0.000040$ & $0.040$ & $2.111\pm0.179$ & 5.7\\
$\nu_4$ & $1.439870\pm0.000034$ & $0.047$ & $1.419\pm0.152$ & 6.7\\
$\nu_5$ & $1.480236\pm0.000034$ & $0.050$ & $2.975\pm0.147$ & 7.1\\
$\nu_6$ & $1.480999\pm0.000025$ & $0.066$ & $0.512\pm0.111$ & 9.4\\
$\nu_7$ & $1.510057\pm0.000002$ & $1.031$ & $-0.090\pm0.007$ & 147.3\\
$\nu_8$ & $1.539902\pm0.000029$ & $0.055$ & $1.219\pm0.128$ & 7.9\\
$2\nu_7$ & $3.020152\pm0.000021$ & $0.076$ & $2.652\pm0.094$ & 10.9\\
\hline
\end{tabular}
\end{table}

In total, we extracted nine frequencies from the amplitude spectrum seen in the third panel of Fig.\,\ref{fig:348717688_ft}. The results are given in Table\,\ref{tab:348717688}. The principal mode frequency $\nu_7 = 1.510057 \pm 0.000002$\,mHz is very close to that found by \citet{martinez1995}, $1.510208 \pm 0.000004$\,mHz. The small difference between these two measurements of $\nu_7$ is significant, hence probably indicating some change of frequency between these data sets; this is not uncommon for roAp stars. The second frequency found by \citet{martinez1995} was at $1.48061 \pm 0.00001$\,mHz and is between two of the peaks listed in Table\,\ref{tab:348717688}, hence may be unresolved. This suggests that care is called for in any attempt to model these frequencies.

{With a large number of frequencies, we search for repeating frequency separations to find the large and/or small frequency separations. We find that $\nu_8-\nu_7=\nu_7-\nu_5=29.8\,\umu$Hz and $\nu_5-\nu_4\simeq\nu_4-\nu_3\simeq40.8\,\umu$Hz. Both of these values are plausible for the large frequency separation, or half of it. From the parameters in Table\,\ref{properties}, we estimate $\Delta\nu=59 \pm 10\,\umu$Hz, thus pointing towards the observed 29.8~$\umu$Hz spacing being half of the true large frequency separation.} 

{Furthermore, the difference between $\nu_1$ and $\nu_2$, {of $5.1\mu$Hz}, could be the rotation frequency, or two times the rotation frequency, under the assumption of a triplet with a missing sidelobe, or a triplet with no central component. The derived rotation period would either be $2.27$\,d or $4.54$\,d, both of which are feasible. With future sector 12 and 13 observations, we hope to be able to confirm all of the separations discussed here.}
	
\citet{kurtz2006} obtained 2\,h of high time resolution, high spectral resolution observations of TIC\,348717688 with the UVES spectrograph on the ESO VLT. They also found two frequencies, $1.510$\,mHz and $1.383$\,mHz in radial velocity variations of lines of Pr\,{\sc{iii}} and H$_\alpha$. Of course, with only 2\,h time span, their second frequency is not resolved in those spectroscopic data. \citet{ryabchikova2007} performed a more detailed analysis of the UVES data for this star and 9 other roAp stars and discussed the pulsation radial velocity amplitude as a function of atmospheric height. {A detailed analysis of the pulsational line profile variability as a function of atmospheric height, using the same UVES data, was presented by \citet{kochukhov07}.} The new results here from the TESS data show the dominance of $\nu_7$ to the variations in this star, hence indicate that the spectroscopic results are secure, as they are not likely to be affected by the additional, {significantly lower amplitude,} pulsation modes. 

\citet{kurtz2006} comment on a relatively small amplitude ratio of $A_{2\nu_7}/A_{\nu_7} =0.10$ for $\nu_7$ and its harmonic in the spectroscopic data compared to the $B$ photometric data, for which they found a ratio of $A_{2\nu_7}/A_{\nu_7} =0.36$\footnote{Again, note that we have changed the labeling of the modes presented by \citet{kurtz2006} to fit the convention we use in this work.}. As can be seen in Table\,\ref{tab:348717688}, $A_{2\nu_7}/A_{\nu_7} =0.07$ for the TESS filter data, which is similar to the spectroscopic results. This is not a surprise, since the amplitudes of the pulsation modes and their harmonics in roAp stars are very sensitive to atmospheric height and photometric bandpass, as well as to particular spectral lines or depths in their line profiles, sample different atmospheric heights. There is information here, but it may be difficult to exploit it in modelling the pulsations. 

\subsection{TIC\,394124612}
\label{cont2}

TIC\,394124612 (HD\,218994) was classified with spectral type A3\,Sr by \citet{houk1975}. \citet{martinez1993} measured Str\"omgren and H$\beta$ indices for this star: $V = 8.565 ; b-y = 0.154; m_1 = 0.196 ; c_1 = 0.826; \beta = 2.807$, from which the parameters $\delta m_1 = 0.008$ and $\delta c_1 = 0.032$ were derived, indicative of a normal main sequence A-type star. 

The star is part of a visual binary system with a separation of 1.2\,arcsec \citep{renson91}. At least one of the components of the binary system is a $\delta$\,Sct star, according to \cite{kurtz2008}. In the same work, a longitudinal magnetic field strength of $440 \pm 23$\,G was determined.

TIC\,394124612 was observed during the Cape Survey, but not found to pulsate \citep{martinez1994}. Nevertheless, high time-resolution spectroscopic observations later revealed pulsational variability in the Nd\,{\sc{iii}} and Pr\,{\sc{iii}} lines of this star \citep{gonzalez08}, with a frequency of 1.17\,mHz (a period of 14.2\,min). 

\begin{figure}
\centering
\includegraphics[width=0.9\columnwidth]{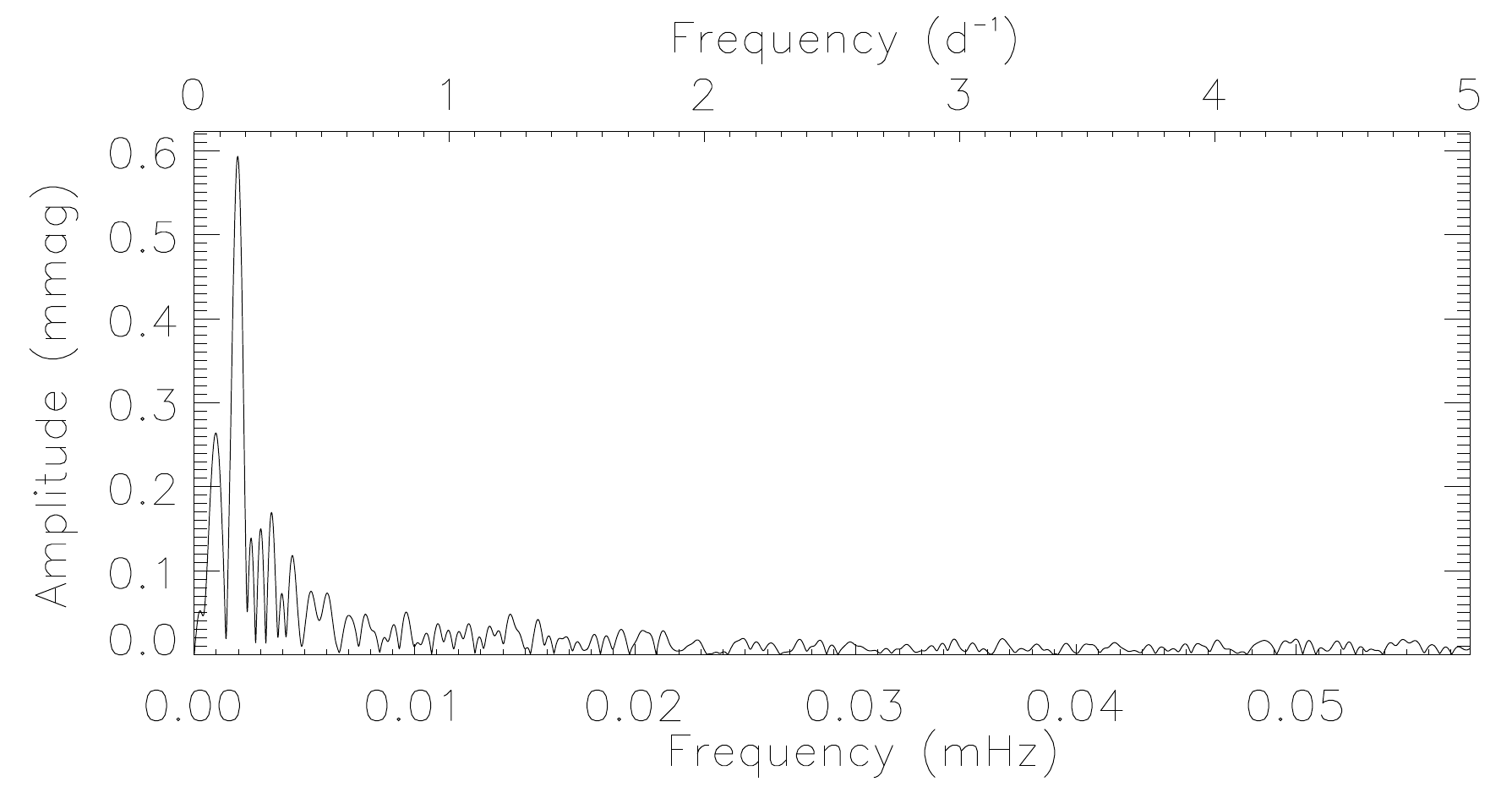}
\includegraphics[width=0.9\columnwidth]{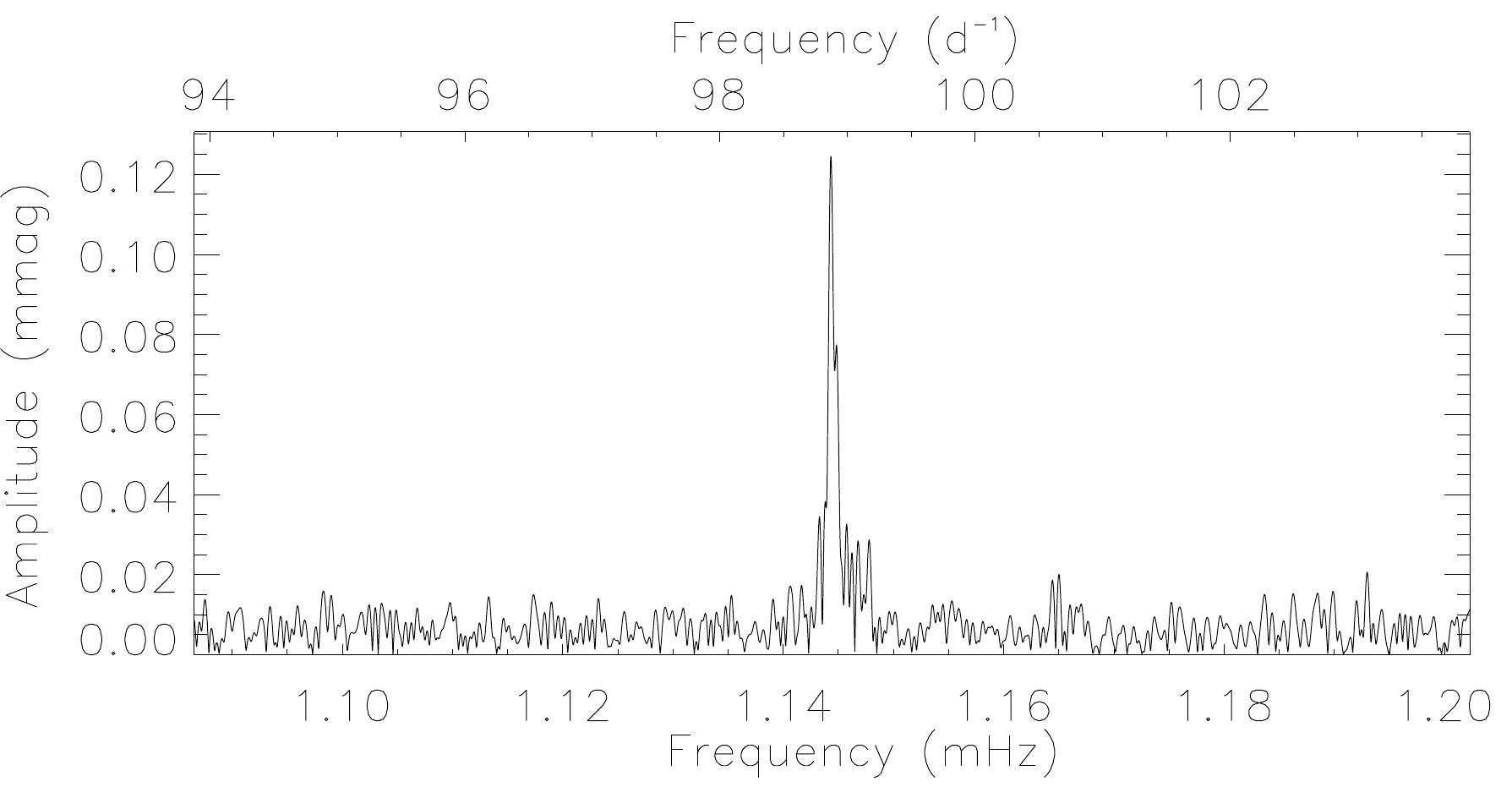}
\includegraphics[width=0.9\columnwidth]{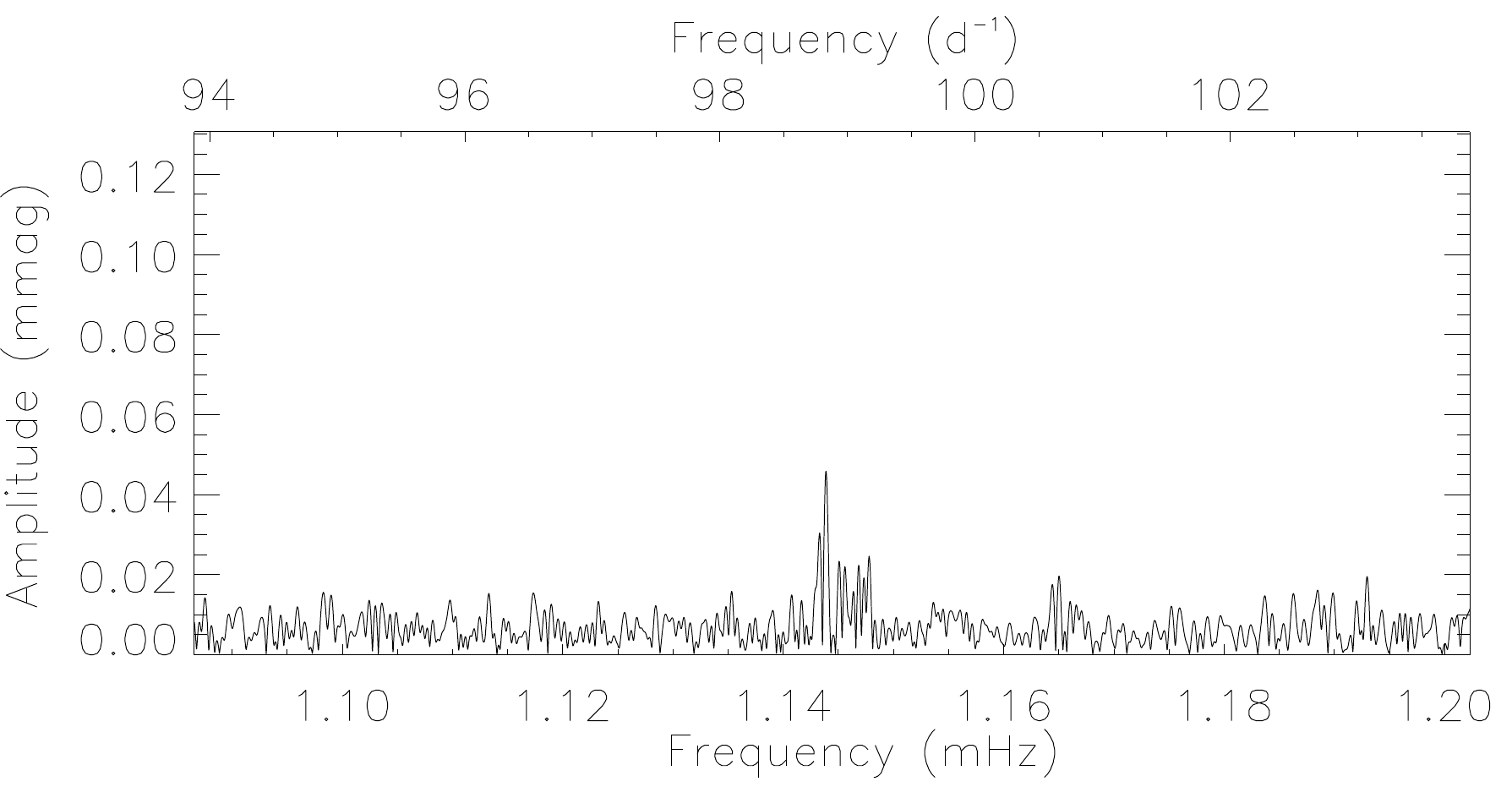}
\caption{Top: amplitude spectrum at low-frequency showing the rotation signal of TIC\,394124612. Middle: zoom of the pulsation mode found in this star. Bottom: the amplitude spectrum of the residuals after removing the frequency shown in Table\,\ref{tab:394124612}. There is clearly some remaining signal, but this is not resolved in the 27.88-d data set.}
\label{fig:394124612_ft}
\end{figure}

TIC\,394124612 was observed by TESS during sector 1 only. The data show a low frequency signal which we take to be the rotation period of the Ap star (top panel of Fig.\,\ref{fig:394124612_ft}). Such a strong low frequency peak is not expected to be present in a non-peculiar A star. Under this assumption, we determine the rotation period for TIC\,394124612 as $5.855\pm0.008$\,d, {which is significantly different from the value found in the literature, as discussed in Sec.~\ref{res:rot}}.

As stated above, TIC\,394124612 is a visual double star. Given the small separation of the two components, and the large pixel size of TESS, the data contain pulsation signals from both stars. The $\delta$\,Sct frequencies are found in the range $0.040-0.515$\,mHz ($3.46-44.54$\,d$^{-1}$), and are not analysed here. Rather, we remove all of these frequencies before we study the variability in the Ap star.

The middle panel of Fig.\,\ref{fig:394124612_ft} shows the roAp pulsation signal in TIC\,394124612. Obviously, the peak is not represented by a clean sinc function, which implies frequency/amplitude modulation, or further unresolved modes. We fitted the signal by non-linear least-squares and show the results in Table\,\ref{tab:394124612}. The bottom panel of Fig.\,\ref{fig:394124612_ft} shows the amplitude spectrum of the residuals after removing the pulsation. The remaining power supports our previous point. {We caution that the frequency we fitted may be perturbed by unresolved modes in the vicinity -- a longer time-base is needed to fully resolve this pulsation.}

\begin{table}
\centering
\caption{Details of the pulsation mode found in TIC 394124612. The zero-point for phase is BJD\,2458339.23972.}
\label{tab:394124612}
\begin{tabular}{lccrr}
\hline
ID & Frequency & Amplitude & \multicolumn{1}{c}{Phase} & \multicolumn{1}{c}{S/N} \\
    & (mHz)  		 & (mmag) & \multicolumn{1}{c}{(rad)}\\
                                              &  &  \multicolumn{1}{c}{$\pm 0.008$} & &\\
\hline
$\nu_1$ & $1.14444\pm0.00004$ & $0.123$ & $-1.61\pm0.16$ &15.4 \\
\hline
\end{tabular}
\end{table}

{Work by \citet{gonzalez08} suffered from alias effects and the authors exercised caution in their frequency determination. The TESS data however confirm their frequency determination was correct, i.e. $1.14444\pm0.00004$\,mHz.}

\section{Mode amplitude comparisons}
\label{sec:comp}

It is well known that the amplitudes of the pulsation modes in the roAp stars change dramatically depending on which photometric bandpass is used to observe them \citep{medupe1998}. {This is primarily because the photometric variations are dominated by the temperature fluctuations during the pulsation cycle. Since the A stars have peak flux towards the blue, the pulsation amplitude with this temperature variation drops to the red of the Rayleigh-Jeans tail, simplisitically thinking of the spectral energy distribution as a black-body. In addition, the different filters probe different heights in the atmosphere of the star, where the pulsation amplitude, which has contributions from both acoustic and Alfven components, varies significantly (see, e.g., \citealt{quitral2018}).}  The optimum filters for roAp observations, to maximise the measured pulsation amplitude, are bluewards of about $4400$\,\AA, with Johnson $B$ often providing the best signal-to-noise as it samples near to the flux maximum. TESS's broad red filter ($6000-10\,000$\,\AA) is weighted towards the Rayleigh-Jeans tail, and is therefore not optimal for the study of roAp stars. 

To provide a quantitative measure of the reduction in the pulsation amplitudes due to the TESS filter, we made a comparison of some of the known roAp stars where $B$ data are available. We chose stars which show stable modes, have clearly dominant or well resolved modes, and which are not known to be rotationally modulated to remove the risk of comparing amplitudes at different rotation phases. Table\,\ref{tab:amp_comp} shows the results of this exercise.

\begin{table*}
\centering
\caption{Comparison of the mode amplitudes for TESS and ground-based $B$ observations of known roAp stars.}
\label{tab:amp_comp}
\begin{tabular}{lccccc}
\hline
Star & Mode & TESS Amplitude & $B$ Amplitude & $A_B/A_{\rm TESS}$ & Reference \\
(TIC) & & (mmag) & (mmag) & & \\
\hline
69855370    & $\nu_1$ & $0.172\pm0.007$ & $0.96\pm0.07$ & $5.58\pm0.47$ & \citet{martinez1998}\\
            & $\nu_2$ & $0.120\pm0.007$ & $0.50\pm0.07$ & $4.17\pm0.63$ & \citet{martinez1998}\\
167695608   & $\nu_1$ & $0.294\pm0.022$ & $2.31\pm0.15$ & $7.86\pm0.78$ & Unpublished data\\
211404370   & $\nu_2$ & $0.083\pm0.006$ & $0.64\pm0.03$ & $7.71\pm0.66$ & \citet{martinez1990}\\
348717688   & $\nu_1$ & $1.031\pm0.007$ & $1.24\pm0.04$ & $1.20\pm0.04$ & \citet{martinez1995}\\
\hline
\end{tabular}
\end{table*}

Typically we are seeing a reduction in the amplitudes as seen by TESS of about a factor of 6. The comparison shows that the factor is different from star to star, thus making the transform from TESS to $B$ amplitudes uncertain for the newly discovered roAp stars. Having these $B$ and TESS amplitudes side by side demonstrates that although amplitudes are reduced, the S/N is just as good, or better, in the case of TESS. This is a result of near-continuous, space-based photometry, which are two great advantages for the study of roAp stars.

The reduction factor for TIC\,348717688 is implausibly small, however (see \citealt{medupe1998}). We conclude that the amplitude was significantly higher at the time of the TESS observations than it was during the three years of observations by \citet{martinez1995}. Whether this is a change in pulsation amplitude, or a change in aspect for an oblique pulsator with a very long rotation period is not known. The lack of any rotational photometric variability found so far would be consistent with a very long rotation period (years), and such long periods are known among the Ap stars.

{The amplitude reduction in TIC\,139191168 is about a factor of 20 compared to the $B$ observations. This is a striking difference. However, we do not include this star in our comparison table as it is multiperiodic.}

\section{The well characterised noAp stars}
\label{no}

\subsection{TIC\,277688819}
\label{hd208217}

TIC\,277688819 (HD\,208217) was classified as A0p\,(pSrEuCr) by \citet{houk1975}. \citet{manfroid1983} and \citet{mathys1985} studied the photometric variability of this star in the Str\"omgren system, finding a rotational period of 8.35\,d. \citet{mathys1997} discovered resolved magnetically split lines in the spectra of TIC\,277688819 and deduced that the field strength of this star changes by about $\pm$1\,kG about the mean value of $\langle B \rangle=7.8$\,kG. Using these magnetic measurements and new photometric observations, \citet{manfroid1997} derived an improved value of the rotational period, {8.44475~$\pm$~0.00011\,d}. New mean field modulus, longitudinal field, crossover, and quadratic magnetic field measurements were published for TIC\,277688819 by \citet{mathys2017}. He also noted that this star exhibits a long-term radial velocity variation indicative of motion in a binary system but could not determine the orbital period nor identify a contribution of the secondary in the spectra. The star is known to be an astrometric binary \citep{makarov2005}, but it is unknown if the astrometric companion is the same as the spectroscopic one.

\citet{landstreet2000} fitted all magnetic field measurements available to them with an axisymmetric low-order multipolar field geometry model. According to the authors, the surface magnetic field geometry of TIC\,277688819 is dominated by a 13.1\,kG dipole field with $\beta=86\degr$ and $i=15\degr$. On the other hand, \citet{bagnulo2002} derived a different, quadrupole-dominated magnetic field configuration by applying an alternative non-axisymmetric multipolar field parametrisation to the same set of magnetic observations.

\citet{hubrig2000} investigated the evolutionary state of a sample of Ap stars, including TIC\,277688819. They reported a photometric effective temperature of $7900\pm300$\,K and obtained $M=2.00\pm0.11\,$M$_\odot$, $\log L/{\rm L}_\odot = 1.38\pm0.13$. A similar study of a larger sample of Ap stars by \citet{kochukhov2006} also included TIC\,277688819, for which these authors found $T_{\rm eff} = 8000\pm200$\,K, $M=1.93\pm0.10\,$M$_\odot$, $\log L/{\rm L}_\odot = 1.29\pm0.12$. Spectroscopic analysis by \citet{freyhammer2008} indicated $T_{\rm eff} = 7500-8000$\,K from the H$_\alpha$ line. Here we derive $T_{\rm eff}=8320\pm\,170$~K, {which is consistent with previous determinations.}

\citet{martinez1994} reported null results in the search of high-overtone pulsations based on photometric observations of TIC\,277688819 on seven different nights. \citet{freyhammer2008} did not detect rapid oscillations above $\sim$10\,m\,s$^{-1}$ with 2.5-h of time-resolved high-resolution spectroscopic observations with UVES at VLT. An independent search of radial velocity pulsations with the HARPS spectrograph \citep{kochukhov2008} also yielded null results. Thus, TIC\,277688819 is a well-established noAp star in our sample.

\begin{figure}
\centering
\includegraphics[width=0.9\columnwidth]{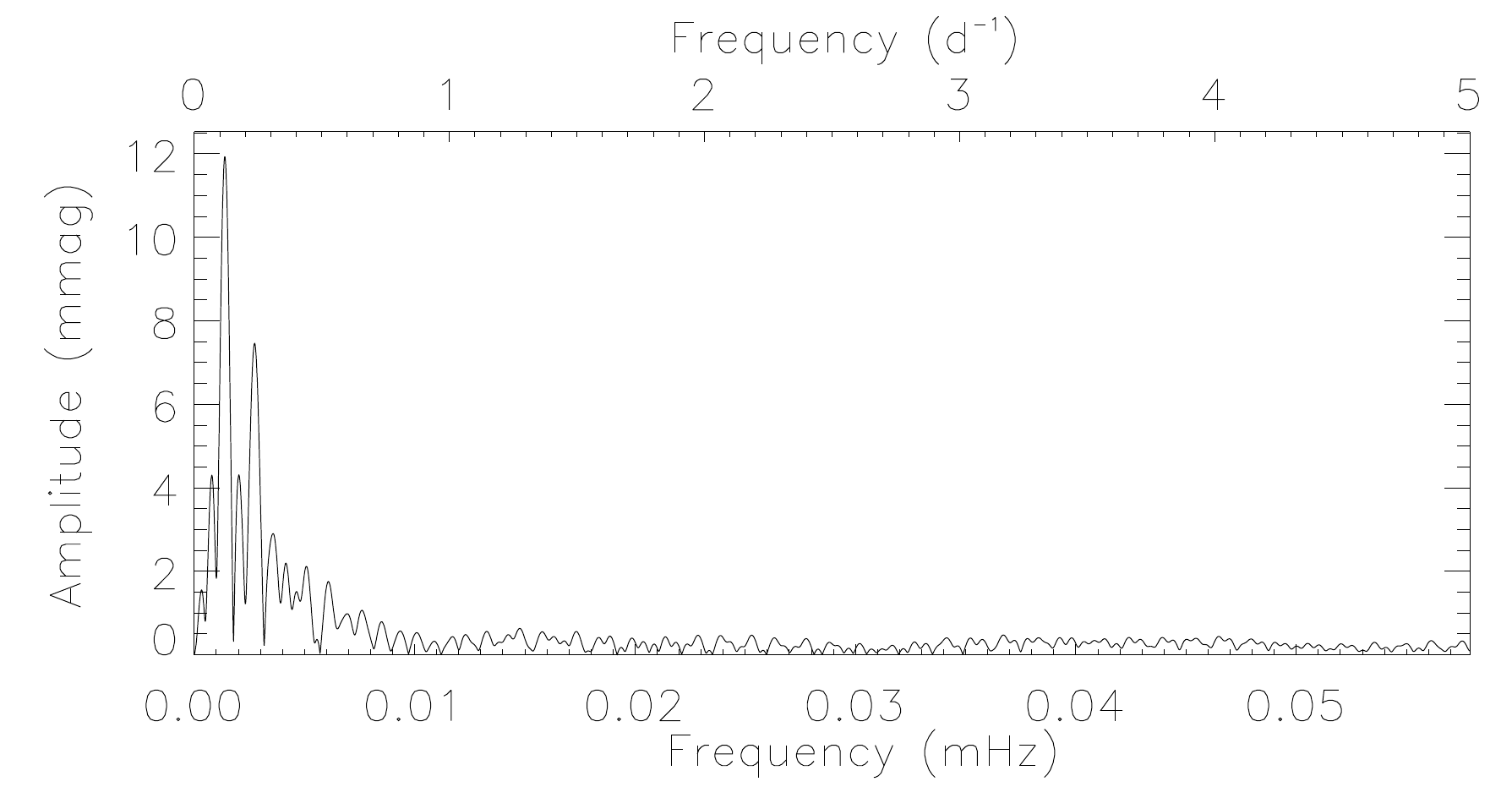}
\includegraphics[width=0.9\columnwidth]{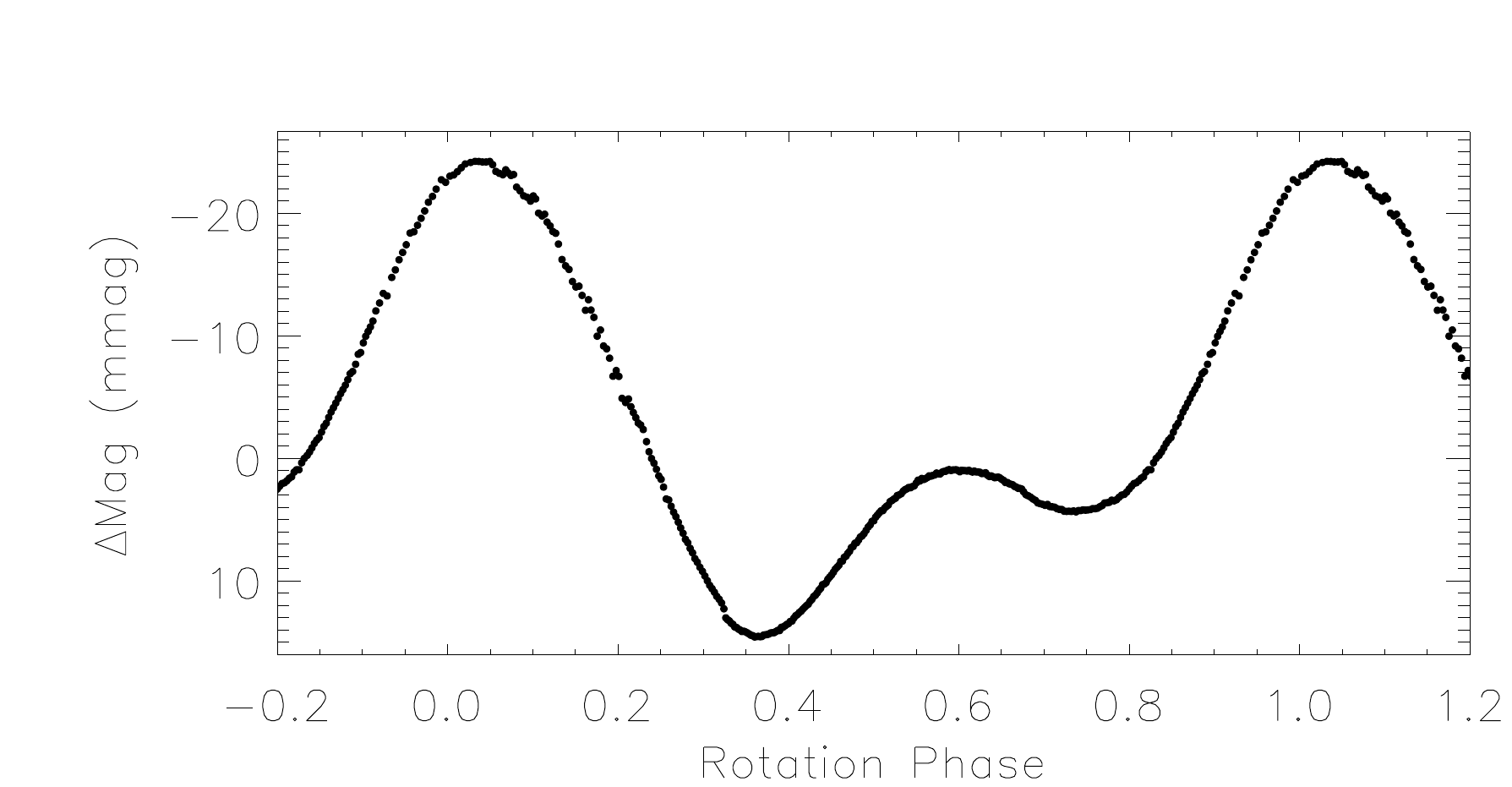}
\includegraphics[width=0.9\columnwidth]{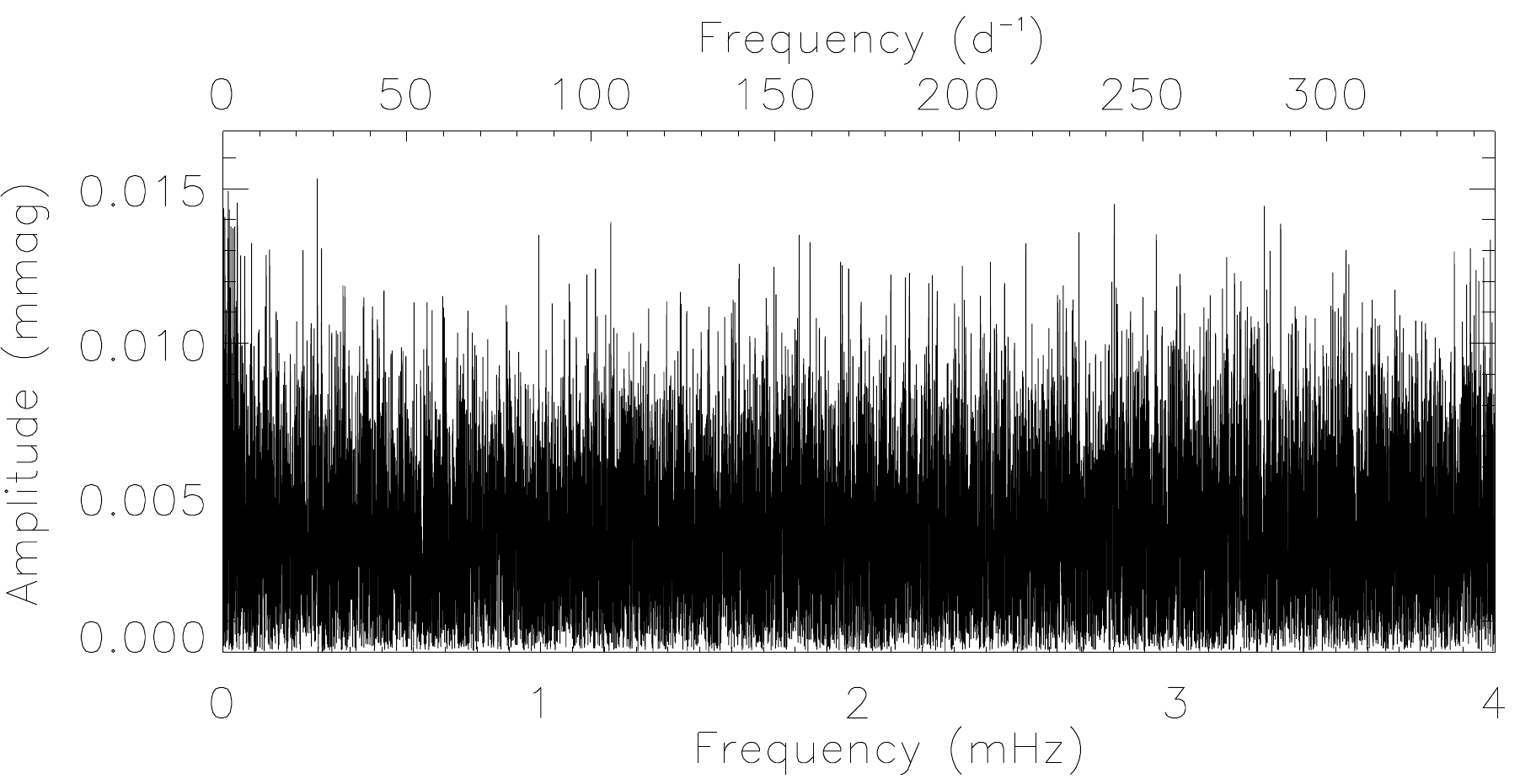}
\caption{Top: amplitude spectrum at low-frequency showing the rotation signal of TIC\,277688819. Middle: phase-folded light curve showing a double-wave nature. The light curve is folded on a period of $8.3200\pm0.0084$\,d and is binned 50:1.  Bottom: an amplitude spectrum to almost the Nyquist frequency of the TESS data showing a distinct lack of pulsational variability {down to a limit of $13\,\umu$mag}. The rotation signal, its harmonics, and low-frequency noise have been removed.}
\label{fig:277688819_ft}
\end{figure}

The TESS sector 1 data for TIC\,277688819 show low-frequency signals of rotation. The amplitude spectrum showing the rotation signal, and the phase-folded light curve are shown in the top two panels of Fig.\,\ref{fig:277688819_ft}. The bottom panel of Fig.\,\ref{fig:277688819_ft} shows the amplitude spectrum of the data set after applying a highpass filter. There are no signs of high-frequency variability in this star, to a limit of $13\,\umu$mag. This limit was estimated from the top of the noise peaks (i.e. the Fourier grass) and four times the error on the highest amplitude noise peak. These TESS data, therefore, support the case that TIC\,277688819 is a noAp star.

\subsection{TIC\,281668790}
\label{hd3980}

TIC\,281668790 (HD\,3980) is a bright southern Ap star, classified as F0\,(pSrEuCr) by \citet{bidelman1955} and as A3\,V\,(pSrCr) by \citet{abt1995}. \citet{renson1979} discussed photometric variability of this star, suggesting a period of 2.13\,d. A revised analysis by \citet{maitzen1980} established a {rotational period of 3.9516~$\pm$ 0.0003\,d} with an outstanding 0.13\,mag double-wave variation in the Str\"omgren $v$ passband. These authors also reported coarse photographic measurements of the longitudinal magnetic field with an amplitude of about 2\,kG, but no clear changes with the rotational phase. \citet{catalano1991} and \citet{catalano1998} confirmed the 3.9\,d rotational period of TIC\,281668790 with near-IR {\it JHK} photometric observations.

Definitive longitudinal field measurements were made for TIC\,281668790 by \citet{hubrig2006} and \citet{kochukhov2006} with the use of observations collected with the FORS1 low-resolution spectropolarimeter at the ESO VLT. The latter authors also derived $T_{\rm eff} = 8260\pm200$\,K, $\log L/{\rm L}_\odot = 1.24\pm0.04$, $M=1.91\pm0.03\,$M$_\odot$ using photometric temperature calibrations, theoretical stellar evolutionary models, and Hipparcos parallax. \citet{elkin2008} determined $T_{\rm eff} = 8000\pm200$\,K, $\log g=4.0\pm0.2$ by fitting theoretical spectra to the hydrogen Balmer lines. They also reported several additional longitudinal field measurements, which together with the literature data allowed them to establish a clear sinusoidal variation of $\langle B_z \rangle$ between about $-2$ and $+2$\,kG. 

Chemical abundances derived by \citet{elkin2008} for TIC\,281668790 revealed a large heavy element overabundance and a discrepancy between abundances of singly and doubly ionised rare-earth elements, typical of roAp stars \citep{ryabchikova2004}. \citet{nesvacil2012} carried out a detailed phase-resolved spectroscopic study of TIC\,281668790, deriving surface abundance distributions of 13 elements with the Doppler imaging technique. They also analysed all available magnetic field measurements, finding a dipolar field strength of $B_{\rm p}=6.9\pm1.0$\,kG and an obliquity of $\beta=88\pm12\degr$. The atmospheric parameters, $T_{\rm eff} = 8300\pm250$\,K, $\log g=4.0\pm0.2$, obtained by \citet{nesvacil2012} are compatible with previous determinations.

Null results of photometric searches of rapid oscillations in TIC\,281668790 were reported by \citet{weiss1979} and \citet{martinez1994}. Furthermore, \citet{elkin2008} could not find radial velocity oscillations above a few tens of m\,s$^{-1}$ in the period range typical of roAp stars using two 1-2\,h-long timeseries observations with the UVES spectrograph at the ESO VLT. These ground-based observations established TIC\,281668790 as a noAp star.

\begin{figure}
\centering
\includegraphics[width=0.9\columnwidth]{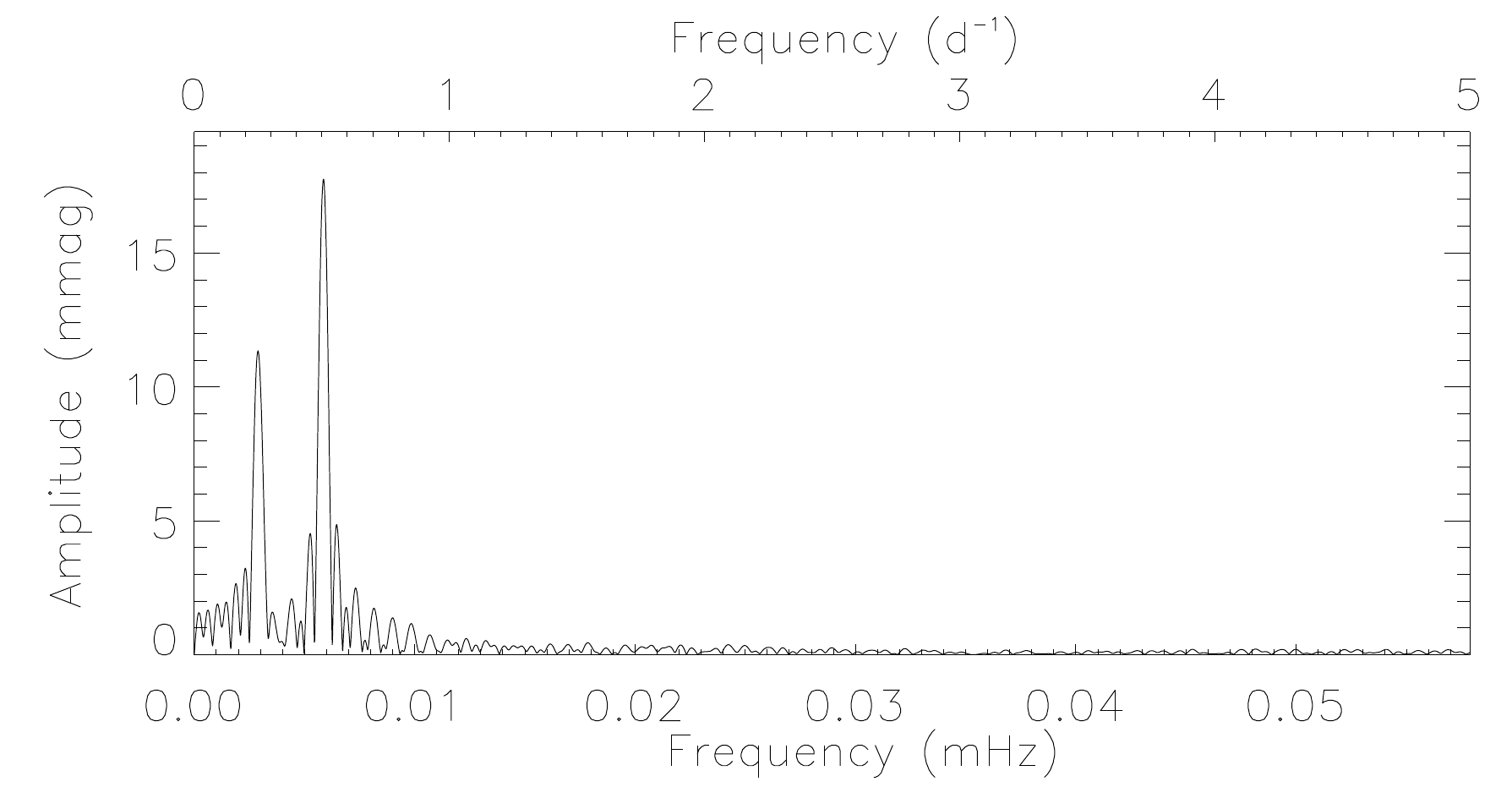}
\includegraphics[width=0.9\columnwidth]{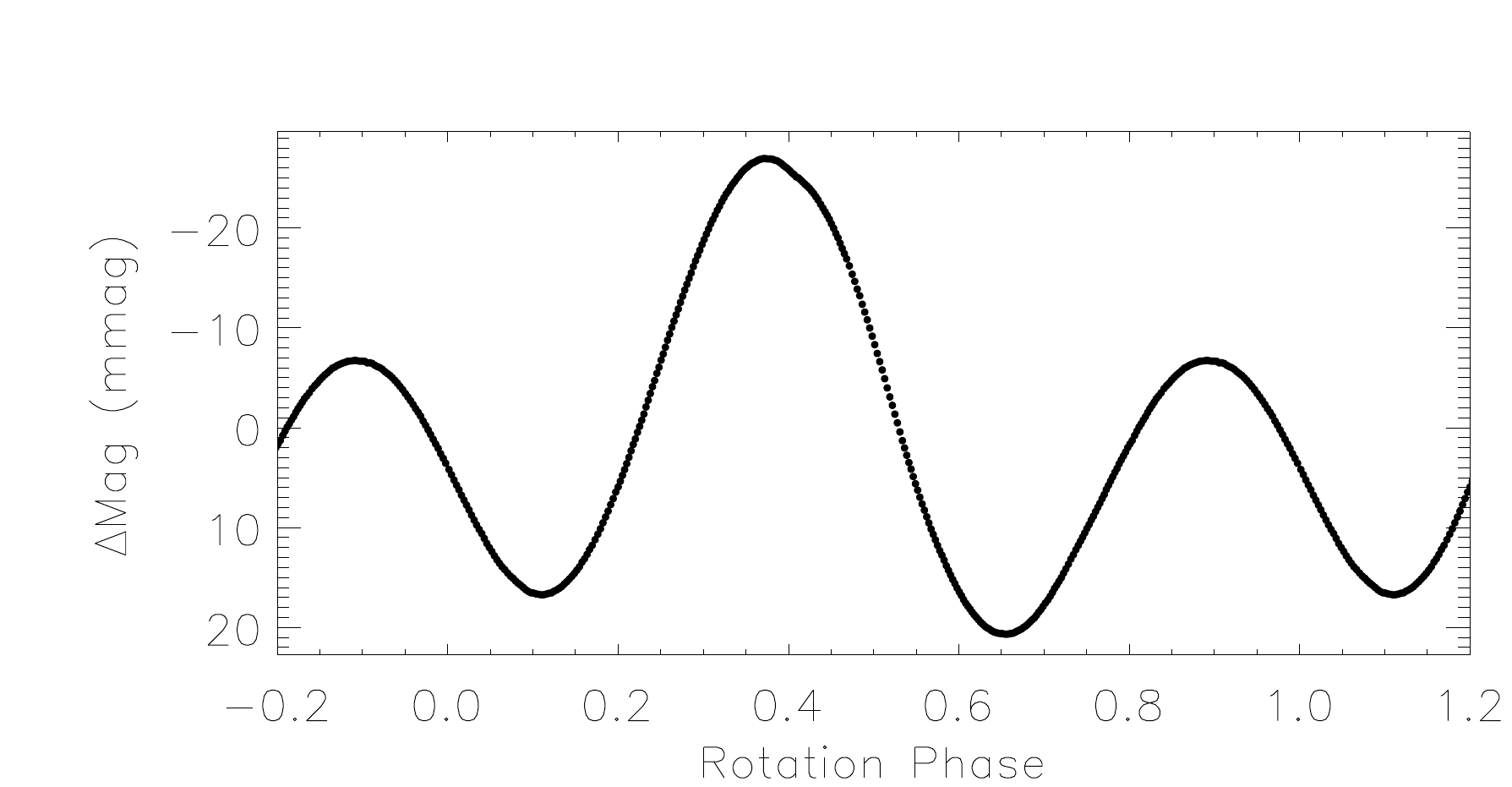}
\includegraphics[width=0.9\columnwidth]{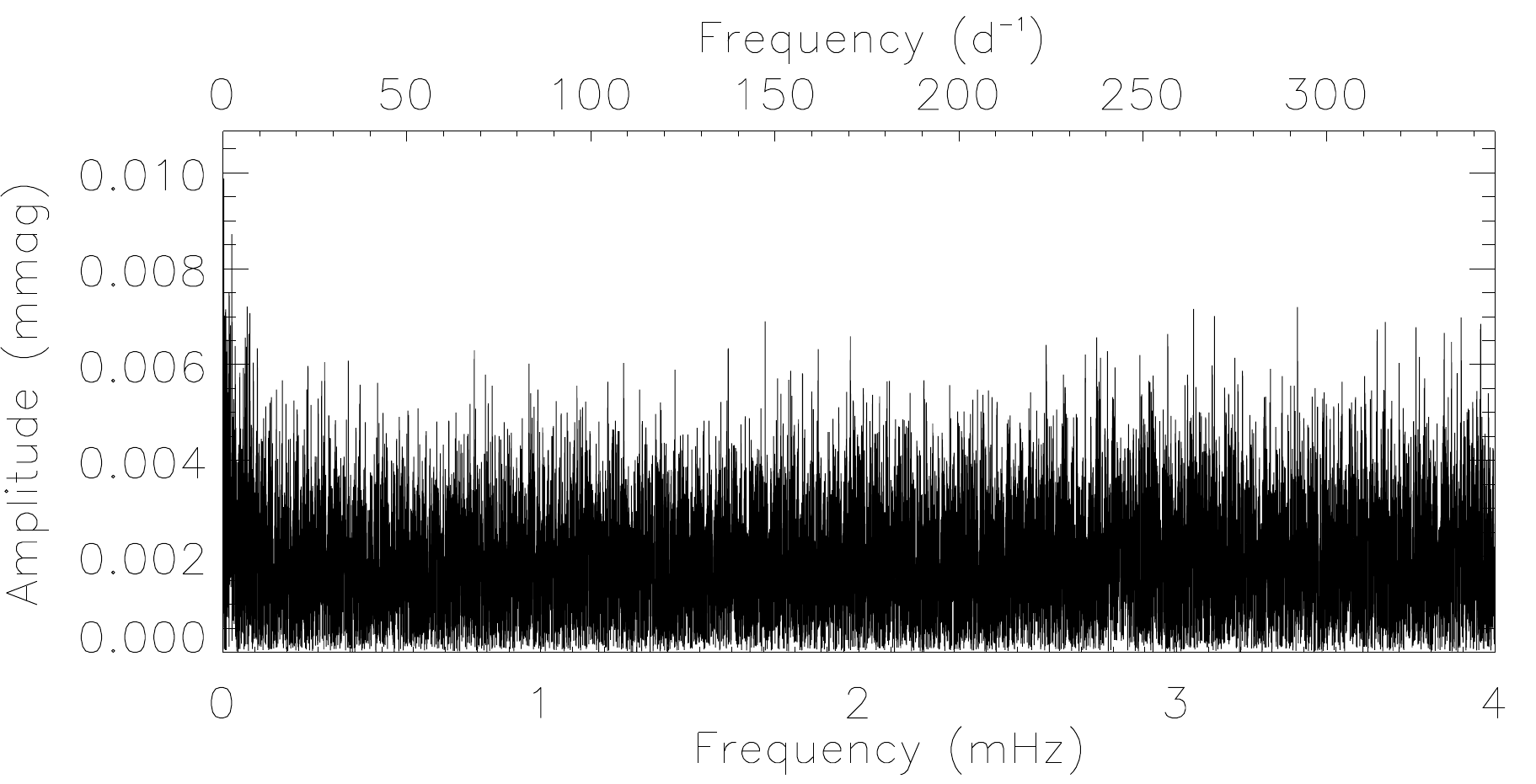}
\caption{Top: amplitude spectrum at low-frequency showing the rotation signal of TIC\,281668790. Middle: phase-folded light curve showing a double-wave. The light curve is folded on a period of $3.9517\pm0.0001$\,d and is binned 50:1.  Bottom: an amplitude spectrum to almost the Nyquist frequency of the TESS data showing a distinct lack to pulsational variability to a limit of $6\,\umu$mag. The rotation signal, its harmonics, and low-frequency noise have been removed.}
\label{fig:281668790_ft}
\end{figure}

The TESS observations in sector 2 support the ground-based observations of TIC\,281668790. The amplitude spectrum at low frequency of the TESS data shows the rotation frequency and its first harmonic (Fig.\,\ref{fig:281668790_ft}). This indicated a double-wave nature to the light curve, which is evidenced by the phase-folded light curve shown in Fig.\,\ref{fig:281668790_ft}. The rotation period is the same as that given in the literature, within the errors.

The bottom panel of Fig.\,\ref{fig:281668790_ft} shows the full amplitude spectrum of TIC\,281668790, demonstrating the lack of a detection of pulsations in the TESS data. The detection limit to pulsational variability, at 6$\,\umu$mag, is estimated from the top of the noise peaks which is approximately 4 times the error of the highest amplitude noise peak. These results strongly support the case that TIC\,281668790 is a noAp star.

\section{Conclusions}
\label{conclusion}

In this work we present the results from the analysis of the first data collected by the TESS satellite on Ap stars, {acquired during the obsevation of sectors 1 and 2, with a 2-min cadence. The sample studied is composed of 83 stars, of which 80 were previously classified as chemically peculiar,  7 of which were known to be roAp stars prior to the TESS launch.} The main outcomes from this analysis can be summarized as follows:
\begin{itemize}
    \item {Five} new roAp stars were discovered, including one hosting the highest frequency pulsations of any known roAp star (TIC\,350146296; highest pulsation frequency of 3.562\,mHz, corresponding to a period of 4.68\,min). {The new roAp stars TIC\,152808505, TIC\,350146296, and TIC\,431380369 show multiple principal pulsation frequencies and rotational multiplets. In addition, TIC\,12968953 was found to be multiperiodic and TIC\,41259805 was found to have a single pulsation frequency split by rotation.} {Also, one of the new roAp stars, TIC 41259805, has the shortest rotation period measured for this class of objects, {$P_{\rm rot} = 1.71$\,d, although other Ap stars not known to be roAp have shorter rotation periods. }} Considering that {three} of these stars were found among {the sample of 73 Ap stars not previously known to pulsate}, we {estimate a $4$~per~cent} incidence of the roAp phenomenon among Ap stars. {An important result of this study is the knowledge that roAp stars are rare, rather than just difficult to detect. Previous ground-based  studies were all more limited by S/N than the uniform, high-precision TESS data.}
    
      \item  Seven previously known roAp stars, members of the sample, were also analysed. In five of them, TIC\,167695608, TIC\,237336864, TIC\,348717688, TIC\,139191168 and  TIC\,211404370, either additional or different pulsation frequencies were discovered. Moreover, in another two, TIC\,69855370 and TIC\,394124612, it was possible to distinguish the correct pulsation frequency from possible aliases present in the ground-based data.
    \item The new roAp star TIC\,350146296 and the previously known roAp star TIC\,237336864 show particularly interesting and challenging pulsation spectra, with the presence of several multiplets. {These stars are promising candidates for further asteroseismic studies.}
    \item Two of the best characterised noAp stars known to date, TIC\,281668790 and TIC\,277688819, with properties very similar to those of typical roAp stars were also analysed. The analyses set pulsation amplitude limits in these stars of 6 and 13\,$\umu$mag, respectively, in the TESS filter.
    \item We found a typical reduction factor of 6 in the pulsation amplitudes observed through the TESS filter, compared with the $B$ filter usually used in ground-based surveys. This corresponds to an amplitude suppression of 83~per~cent. To determine this value, we considered only stars showing  stable  modes, with a clearly dominant mode or well resolved modes, and with no sign of rotational modulation. Still, a definitive value requires simultaneous observations from TESS and ground-based observatories.
    \item We have identified {27} new rotational variables in our sample. Moreover, we found that the rotation period has been misidentified in {five} of the previously known rotational variables, TIC\,307642246, TIC\,309148260, TIC\,336731635, TIC\,348898673 {and TIC\,394124612}; we provided corrected values for them. {For another two stars, we found rotation periods that are slightly, but significantly, different to the values published in the literature (Table\,\ref{rot})}.
\end{itemize}

The results from this study provide important insights to the state-of-the-art theoretical models of chemically peculiar, magnetic A stars. In particular, concerning pulsational variability, we find puzzling frequency separations, possible additional cases of stars with frequencies higher than those predicted to be excited by the opacity mechanism, and additional pulsating stars located outside the theoretical instability strip. {These new observations are challenges to theory.} Moreover, we continue to witness a lack of roAp stars located close to the theoretical blue edge and continue to fail to find pulsations in stars that are, from the photometric and spectroscopic point-of-view, twins of roAp stars. All of these findings reinforce the urgency to revisit the theory of pulsations in chemically peculiar, strongly magnetic, stars.

\section*{acknowledgements}
We thank the referee for very detailed and useful comments to the original manuscript. This work was supported by FCT - Funda\c c\~ao para a Ci\^encia e a Tecnologia  through national funds and by FEDER through COMPETE2020 - Programa Operacional Competitividade e Internacionaliza\c c\~ao by these grants: UID/FIS/04434/2019, PTDC/FIS-AST/30389/2017 \& POCI-01-0145-FEDER-030389. MC is supported in the form of work contract funded by national funds through FCT. Funding for the Stellar Astrophysics Centre is provided by The Danish National Research Foundation (Grant agreement no.: DNRF106). DLH and DWK acknowledge financial support from the Science and Technology Facilities Council (STFC) via grant ST/M000877/1. This work has made use of data from the European Space Agency (ESA) mission
{\it Gaia} ({https://www.cosmos.esa.int/gaia}), processed by the {\it Gaia}
Data Processing and Analysis Consortium (DPAC,
{https://www.cosmos.esa.int/web/gaia/dpac/consortium}). Funding for the DPAC
has been provided by national institutions, in particular the institutions
participating in the {\it Gaia} Multilateral Agreement. LFM acknowledges support from the UNAM by the way of DGAPA project PAPIIT IN100918. The research leading to these results has received funding from the European Research Council (ERC) under the European Union's Horizon 2020 research and innovation programme (grant agreement N$^\circ$670519: MAMSIE) and from the Fonds Wetenschappelijk Onderzoek - Vlaanderen (FWO) under the grant agreement G0H5416N (ERC Opvangproject). 
%The research leading to these results has received funding from the European Research Council (ERC) under the European Union's Horizon 2020 research and inno- vation programme (grant agreement No.670519: MAMSIE).
MS acknowledges the financial support of Postdoc@MUNI project CZ.02.2.69/0.0/0.0/16 027/0008360. EN acknowledges the Polish National Science Center grants no.2014/13/B/ST9/00902. JCS acknowledges funding support from Spanish public funds for research under projects ESP2017-87676-2-2 and ESP2015-65712-C5-5-R, and from project RYC-2012-09913 under the `Ram\'on y Cajal' program of the Spanish Ministry of Science and Education. AGH acknowledges funding support from Spanish public funds for research under projects ESP2017-87676-2-2 and ESP2015-65712-C5-5-R of the Spanish Ministry of Science and Education. \'AS, ZsB, and RSz acknowledge the financial support of the GINOP-2.3.2-15-2016- 00003, K-115709, K-113117, K-119517 and PD-123910 grants of the Hungarian National Research, Development and Innovation Office (NKFIH), and the Lend\"ulet Program of the Hungarian Academy of Sciences, project No. LP2018-7/2018. GH has been supported by the Polish NCN grant 2015/18/A/ST9/00578. MLM acknowledges funding support from Spanish public funds for research under project ESP2015-65712-C5-3-R. JPG acknowledges funding support from Spanish public funds for research under project ESP2017-87676-C5-5-R. MLM and JPG also acknowledges funding support from the State Agency for Research of the Spanish MCIU through the "Center of Excellence Severo Ochoa" award for the Instituto de Astrofísica de Andalucía (SEV-2017-0709). IS acknowledges funding support of NSF under projects DN 08-1/2016 and DN 18/13-12.12.2017. P. Ko{\l}aczek-Szyma\'nski acknowledges support from the NCN grant no. 2016/21/B/ST9/01126. This paper includes data collected by the TESS mission. Funding for the TESS mission is provided by the NASA Explorer Program.

\bibliography{tess.bib}
\appendix
\section{Author Affiliations}
\label{sec:affiliations}
{\small
$^1$ Instituto de Astrof\'\i sica e Ci\^encias do Espa\c co, Universidade do
Porto, CAUP, Rua das Estrelas, PT4150-762 Porto, Portugal\\
$^2$ Stellar Astrophysics Centre, Aarhus University, Ny Munkegade 120, 8000, Aarhus, Denmark \\
$^3$ Jeremiah Horrocks Institute, University of Central Lancashire, Preston PR1 2HE, United Kingdom\\ 
$^4$ South African Astronomical Observatory, P.O. Box 9, Observatory, Cape Town, South Africa\\
$^5$ Konkoly Observatory, MTA CSFK, H-1121, Konkoly Thege Mikl\'os \'ut 15-17, Budapest, Hungary \\
$^6$ MTA CSFK Lend\"ulet Near-Field Cosmology Research Group\\
$^7$ Institute of Astronomy, KU Leuven, Celestijnenlaan 200D, 3001 Leuven, Belgium\\
$^8$ Center for Exoplanets and Habitable Worlds, 525 Davey Laboratory, The Pennsylvania State University, University Park, PA 16802, USA \\
$^9$ Copernicus Astronomical Center, Polish Academy of Sciences, Bartycka 18, 00-716 Warsaw, Poland\\
$^{10}$ Astronomical Institute of University of Wroc\l{}aw, ul. Kopernika 11, 51-622 Wroc\l{}aw, Poland\\
$^{11}$ Instituto de Astrof\'\i sica de Andaluc\'\i a~(CSIC), Glorieta de la Astronom\'\i a s/n, E-18008 Granada, Spain\\
$^{12}$ Department of Theoretical Physics and Astrophysics, Masaryk University, Kotl\'{a}\v{r}sk\'{a} 2, 61137 Brno, Czech Republic \\
$^{13}$ Astronomical Institute, Czech Academy of Sciences, Fri\v{c}ova 298, 25165, Ond\v{r}ejov, Czech Republic\\
$^{14}$ Astrophysics Group, Keele University, Staffordshire ST5 5BG, United Kingdom \\
$^{15}$ Department of Physics and Astronomy, Uppsala University, Box 516, 75120 Uppsala, Sweden \\
$^{16}$ Department of Physics, Lehigh University, 16 Memorial Drive East, Bethlehem, PA 18015, USA \\
$^{17}$ School of Earth and Space Exploration, Arizona State University, Tempe, AZ 85281, USA \\
$^{18}$ Department of Physics, and Kavli Institute for Astrophysics and Space Research, Massachusetts Institute of Technology, Cambridge,MA 02139, USA\\
$^{19}$ Earth and Planetary Sciences, MIT, 77 Massachusetts Avenue, Cambridge, MA 02139, USA \\
$^{20}$ Department of Chemistry and Physics, Florida Gulf Coast University, 10501 FGCU Blvd. S., Fort Myers, FL 33965 USA \\
$^{21}$ Instituto de Astronom\'{\i}a--Universidad Nacional Aut\'onoma de M\'exico, Ap. P. 877, Ensenada, BC 22860, Mexico\\
$^{22}$ Department of Physics, University of Zanjan, Zanjan , Iran \\
$^{23}$ Departamento de F\'{\i}sica e Astronomina, Faculdade de Ci\^encias da Universidade do Porto, Portugal\\
$^{24}$ Institute of Astronomy and NAO, Bulgarian Academy of Sciences, blvd. Tsarigradsko chaussee 72, Sofia 1784, Bulgaria\\
$^{25}$ Royal Observatory of Belgium, Ringlaan 3, 1180 Brussels, Belgium\\
$^{26}$ Dept. F\'{\i}sica Te\'orica y del Cosmos, Universidad de Granada, Campus de Fuentenueva s/n, E-18071, Granada, Spain\\
$^{27}$ Department of Physics, Institute for Advanced Studies in Basic Sciences (IASBS), Zanjan 45137-66731, Iran\\
$^{28}$ Sydney Institute for Astronomy, School of Physics, University of Sydney 2006, Australia \\
$^{29}$ Department of Physic $\&$ Astronomy, University of British Columbia, Vancouver, Canada\\
$^{30}$ Dept. of Physics $\&$ Astronomy, Camosun College, Victoria, British Columbia, Canada\\
$^{31}$ Department of Astrophysics, University of Vienna, Tuerkenschanzstarsse 17, 1180 Vienna, Austria\\

}
\end{document}